\documentclass[10pt, aps, amsmath, twocolumn, superscriptaddress, showpacs, prb, floatfix]{revtex4-2}
  
\usepackage{graphicx}
\usepackage{multirow}
\usepackage{verbatim}
\usepackage{bm}
\usepackage{latexsym}
\usepackage{inputenc}
\usepackage{natbib}
\usepackage[normalem]{ulem}

\usepackage{dcolumn}
\usepackage{braket}
\usepackage{dsfont}
\usepackage{bbold}
\usepackage{xcolor}
\usepackage{booktabs}
\usepackage{siunitx}
\usepackage{makecell}
\usepackage[colorlinks=true, allcolors=blue]{hyperref}
\usepackage{comment}
\usepackage{orcidlink}
\usepackage{mathtools}

\begin{document}

\title{Anatomy of the modern theory of orbital magnetism from first-principles:
\\
term-by-term analysis in the gauge-covariant formalism
}

\author{Hojun Lee\,\orcidlink{0000-0002-7406-1936}}
\thanks{These authors contributed equally to this work}
\affiliation{Department of Physics, Pohang University of Science and Technology, Pohang 37673, Korea}
\affiliation{Center for Quantum Dynamics of Angular Momentum, Pohang 37673, Korea}

\author{Insu Baek\,\orcidlink{0000-0002-7475-2876}}
\thanks{These authors contributed equally to this work}
\affiliation{Department of Physics, Pohang University of Science and Technology, Pohang 37673, Korea}
\affiliation{Center for Quantum Dynamics of Angular Momentum, Pohang 37673, Korea}

\author{Mirco Sastges\,\orcidlink{0009-0008-2403-7365}}
\thanks{These authors contributed equally to this work}
 \affiliation{Peter Grünberg Institut, Forschungszentrum Jülich, 52425 Jülich, Germany}
 \affiliation{Institute of Physics, Johannes Gutenberg University Mainz, 55099 Mainz, Germany}
 \affiliation{Department of Physics, RWTH Aachen University, 52056 Aachen, Germany}

\author{Yuriy Mokrousov\,\orcidlink{0000-0003-1072-2421}}
\affiliation{Peter Grünberg Institut, Forschungszentrum Jülich, 52425 Jülich, Germany}
\affiliation{Institute of Physics, Johannes Gutenberg University Mainz, 55099 Mainz, Germany}

\author{Hyun-Woo Lee\,\orcidlink{0000-0002-1648-8093}}
\affiliation{Department of Physics, Pohang University of Science and Technology, Pohang 37673, Korea}
\affiliation{Center for Quantum Dynamics of Angular Momentum, Pohang 37673, Korea}

\author{Dongwook Go\,\orcidlink{0000-0001-5740-3829}}
\thanks{Corresponding author: \href{mailto:dongwookgo@korea.ac.kr}{dongwookgo@korea.ac.kr}}
\affiliation{Department of Physics, Korea University, Seoul 02841, Republic of Korea}
\affiliation{Center for Quantum Dynamics of Angular Momentum, Pohang 37673, Korea}

\begin{abstract}
We present an in-depth analysis of the orbital magnetism by means of the so-called modern theory based on the Berry phase across distinct classes of materials--$d$ transition-metals, $sp$ metals, and transition metal dichalcogenides--highlighting the importance of the microscopic nature of band structure characteristics on the orbital magnetization. We adopt a gauge-covariant formulation of the modern theory proposed in [\href{https://journals.aps.org/prb/abstract/10.1103/PhysRevB.85.014435}{Lopez {\it et al.} Phys. Rev. B {\bf 85}, 014435 (2012)}], which enables the calculation of orbital magnetism in a controlled manner in any chosen gauge of Wannier functions and gives the total contribution as a gauge-invariant measurable. This captures consistently the contributions due to the anomalous position, velocity, and orbital angular momentum of Wannier basis, as well as the contributions due to Hamiltonian such that their sum is gauge-invariant. For $d$ transition metals, we find that the atom-centered approximation captures the majority of the total contribution given by modern theory, which we attribute to localized nature of $d$ electrons. However, $5d$ metals tend to exhibit relatively larger deviation between the two methods than $3d$ metals do, as $5d$ electrons are more delocalized than $3d$ electrons. On the other hand, $sp$ metals exhibit a strong deviation between the two methods, where large kinetic energy of $sp$ electrons is important. Finally, in 1H-MoS$_2$, we find that the valley orbital moment at each valley far exceeds the atomic limit orbital moment of $d$ electrons due to coherent hybridization between valence and conduction bands in direct band gaps. Our work elucidates the interplay of the chemical nature of electronic orbitals and the effect of band structures in a consistent manner and highlights the role of Berry phase in orbital magnetism. The results suggest a promising direction of orbitronics beyond controlling atomic orbitals, in which the orbital magnetism can be greatly enhanced by exploiting Berry phase.
\end{abstract}

\maketitle

\section{Introduction}\label{Introduction}

\subsection{Orbital magnetism in solids}

Traditionally, studies on magnetism have focused on the spin contribution as it is spontaneously ordered by electron correlations and thus dominates magnetism. Another contribution due to the orbital angular momentum (OAM) of electrons did not receive much attention until recently as it is weakly induced from the spin contribution via the relativistic spin-orbit coupling (SOC) in most solids, such as $3d$ magnets Fe, Co, and Ni~\cite{Vleck32Book}, and thus smaller than the spin contribution. However, interest in the OAM is rising with the recent emergence of {\it orbitronics}, which aims to exploit the orbital degree of freedom of electrons as an information carrier in quantum transport and information processing and extends beyond the charge- and spin-based paradigms~\cite{Go21EPL,Jo24NPJS,Wang25AElecM,Cysne25NPJS,Atencia24APX}. Although the OAM in the ground state is quenched in most materials, transient nonequilibrium quantum states may break the time-reversal symmetry and carry large OAM or orbital currents. For instance, an external electric field can induce an accumulation of nonequilibrium OAM and the current dependent on the sign of the OAM, the phenomena known as the orbital Edelstein effect~\cite{Osumi21ComPhys,Park11PRL,Go17SR,Yoda18NanoLett,Anas23NatPhys} and the orbital Hall effect~\cite{Bernevig05PRL,Kontani09PRL,Go18PRL,Jo18PRB,Salemi22PRM,Choi23NAT,Lyalin23PRL,Liu24PRL,Liu24PRL2,Park25AdvMat}, respectively. It has been found that the OAM-carrying electron excitations often dominate over the spin counterpart, regardless of the quenching of the OAM in the ground state. These nonequilibrium OAM-polarized carriers are highly susceptible to {\it real} space symmetries and also enable efficient manipulation of spontaneously broken orders, for example via the orbital torque~\cite{Go20PRR,Lee21NatCom,Kim21PRB,Sala22PRR,Hayashi23CP,Go23PRL,Fukunaga23PRR,Ding24PRL,Lee24CAP,Moriya24NanoLett,Chiba24NanoLett} to control magnetization. Conversely, dynamics of order parameters such as lattice and magnetization dynamics can excite OAM-polarized carriers in the phenomenon called orbital pumping~\cite{Hayashi24NatElec,Han25PRL,Go24arXiv}.

In low-symmetry materials, the OAM may dominate over the spin even in equilibrium, where orbital hybridizations are pronounced by the lack of symmetry protection. For example, at a surface or interface, parity-mixing of orbitals develops an electric polarization, which in turn leads to chiral OAM textures in $\mathbf{k}$ space~\cite{Park11PRL,Park12PRL,Hagiwara25AdvMat}. Similarly, the broken inversion symmetry in two-dimensional materials results in pronounced OAM in the direction out of the plane, i.e., valley-dependent OAM in~e.g.~hexagonal MoS$_2$~\cite{Bhowal20PRB,Bhowal20PRB2,Cysne21PRL,Cysne22PRB,Costa23PRL}. Broken symmetries by magnetic orders also lead to large OAM, such as chiral and topological textures in real space and orbitally driven weak moment in altermagnets. In kagome superconductors AV$_3$Sb$_5$ (A$=$K, Rb, Cs), a spontaneously ordered loop current has been found~\cite{Wilson24NatRevMat}.

\begin{figure*}[t!]
\includegraphics[width=0.8\textwidth]{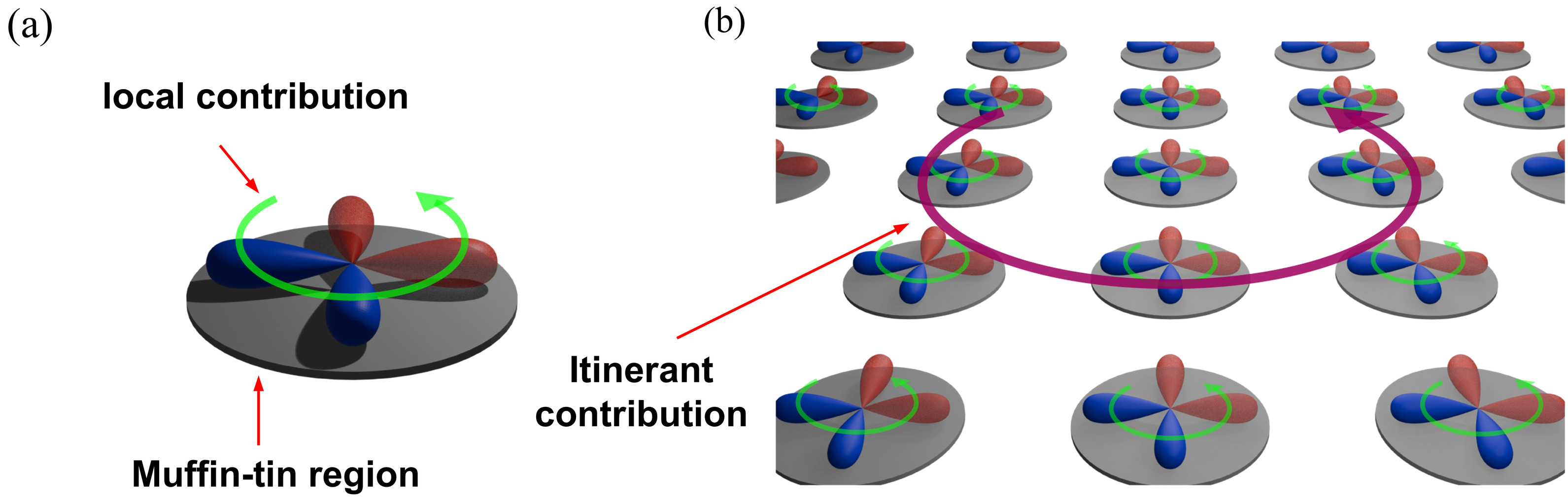}
\caption{\label{fig1}\textbf{Schematic illustration of the atom-centered approximation (ACA) and the modern theory of orbital magnetism.} (a) ACA accounts only for the orbital magnetism arising from the local orbital angular motion of electrons inside each MT sphere. (b) The modern theory evaluates not only the local contribution within the MT sphere but also the contribution from the interstitial region and the itinerant motion of electrons.}
\end{figure*}


The main difficulty in theoretically treating the OAM lies in the position operator appearing in the definition of the OAM operator, 
\begin{equation}\label{eq:OAM}
    \hat{\mathbf{L}} = \frac{m}{2}
    \left( 
    \hat{\mathbf{r}}\times \hat{\mathbf{v}} - \hat{\mathbf{v}}\times \hat{\mathbf{r}}
    \right),
\end{equation}
where $m$ is the bare mass of an electron, and $\hat{\mathbf{r}}$ and $\hat{\mathbf{v}}$ are the position and velocity operators, respectively. Here, a problem arises from the ambiguity with respect to the choice of the coordinate origin, and moreover, the position operator is intrinsically ill-defined in the Bloch representation defined for the periodic boundary conditions~\cite{Resta98PRL,Si25EPL}.

\subsection{Atom-centered approximation}


A commonly used method to compute the OAM in electronic structure calculations is the \textit{atom-centered approximation} (ACA). As shown in Fig.~\ref{fig1}(a), it fixes the coordinate origin as the center of an atomic nucleus in the unit cell and integrates matrix elements of the OAM operator in the so-called muffin-tin (MT) sphere,
\begin{eqnarray} \label{eq:ACA}
    {\mathbf{L}}_\mathrm{\mu,nn^{\prime}}^\mathrm{ACA} (\mathbf{k})
    &=&
    -\frac{i\hbar}{2} 
    \int d^3r 
    \Theta (R_\mu - |\mathbf{r}-\boldsymbol{\tau}_\mu|)
   \\
    & &
    \times \Big[
    u_{n\mathbf{k}}^* (\mathbf{r})
    (\mathbf{r} - \boldsymbol{\tau}_\mu)\times 
    \left\{
    \partial_\mathbf{r} u_{n'\mathbf{k}} (\mathbf{r})
    \right\}
    \nonumber
    \\
    & & 
    -
    \left\{ 
    \partial_\mathbf{r} u_{n\mathbf{k}}^* (\mathbf{r}) \right\}
    \times
    (\mathbf{r} - \boldsymbol{\tau}_\mu)  u_{n'\mathbf{k}} (\mathbf{r})
    \Big],
    \nonumber
\end{eqnarray}
where $\hbar$ is the reduced Planck constant, $u_{n\mathbf{k}}$ is a cell-periodic Bloch state with band index $n$ and crystal momentum $\mathbf{k}$, $\mu$ is the atom index, and $\boldsymbol{\tau}_\mu$ is the center of the $\mu$-th nucleus in the unit cell. Note that the integral is cut off by the Heaviside step function such that the contribution outside the MT sphere of radius $R_\mu$ is ignored. The total orbital magnetization within the ACA is given by
\begin{equation}\label{eq:ACA-sum}
    \mathbf{M}^\mathrm{ACA} = -\frac{\mu_\mathrm{B}}{\hbar} \sum_{\mu} \sum_n \int [dk] f_{nk} \mathbf{L}_{nn}^\mathrm{ACA} (\mathbf{k}),
\end{equation}
where $\mu_\mathrm{B}$ is the Bohr magneton, $f_{n\mathbf{k}}$ is the Fermi-Dirac distribution function for $u_{n\mathbf{k}}$, and $[dk]=d^Nk/(2\pi)^N$ is the integral measure of $N$-dimensional $\mathbf{k}$ space.

The ACA serves as a convenient and widely used method for evaluating the OAM in solids~\cite{Sharma07PRB}. Also, the construction of the OAM {\it operator} is straightforward as in Eq.~\eqref{eq:ACA}, which can easily be applied to calculate nonequilibrium properties, such as current-induced OAM and the orbital Hall effect~\cite{Bernevig05PRL,Kontani09PRL,Go18PRL}. Although the ACA captures only the local orbital angular motion of electrons around each atom, and therefore has a clear limitation. It gives a reasonable estimation of the orbital magnetization for systems where electrons are localized near atomic nuclei, such as $d$ transition metals and $4f$ rare-earth metals~\cite{Hanke16PRB}. However, the result may depend on the muffin-tin radius $R_\mu$, so the consistency--whether the ACA saturates as $R_\mu$ increases--must be checked. Furthermore, the ACA fails to describe the OAM carried by highly delocalized states, such as the itinerant contribution across several atomic sites~\cite{Hanke16PRB}.

\subsection{Modern theory}


On the other hand, within the Berry phase theory, the position operator is explicitly dealt with in the Bloch representation in periodic systems. This is now known as the {\it modern theory of orbital magnetism}~\cite{Thonhauser12IJMPB,Vanderbilt18Book,Aryasetiawan19JPCS} and shares common features with the theory of electric polarization~\cite{Kingsmith93PRB,Vanderbilt93PRB}. It was developed by two independent approaches, one from the representation by localized Wannier functions~\cite{Resta05CPC,Thonhauser05PRL,Ceresoli06PRB,Ceresoli10PRB} and the other from the effective theory for wave packet dynamics~\cite{Chang96PRB,Sundaram99PRB,Xiao05PRL,Xiao06PRL}. Despite the difference in the approaches, both give the same expression for the orbital magnetization. Later a rigorous quantum mechanical calculation of the free energy in the presence of a magnetic field has proved the validity of both approaches~\cite{Shi07PRL}. According to the modern theory, the total orbital magnetization is given by 
\begin{eqnarray}\label{eq:modern}
    \mathbf{M} &=& \frac{e}{2\hbar}\sum_{n}\int [dk]f_{n\mathbf{k}}
    \\
    & &
    \times \mathrm{Im}
    \left[ 
    \bra{\partial_{\mathbf{k}} u_{n\mathbf{k}}} (\hat{H}_{\mathbf{k}} + \mathcal{E}_{n\mathbf{k}} - 2\mathcal{E}_{\text{F}})
    \times 
    \ket{\partial_{\mathbf{k}}u_{n\mathbf{k}}}
    \right],
    \nonumber
\end{eqnarray}
where $H_{\mathbf{k}} = e^{-i\mathbf{k}\cdot\mathbf{r}} H e^{i\mathbf{k}\cdot\mathbf{r}}$ is the cell-periodic Hamiltonian, $\mathcal{E}_{n\mathbf{k}}$ is the energy of the $n$-th band, and $\mathcal{E}_\text{F}$ is the Fermi energy. The modern theory captures {\it all} contributions to the orbital magnetization, including the local circulation within the unit cell and the itinerant circulation across neighboring cells, as shown in Fig.~\ref{fig1}(b). Thus, the modern theory goes beyond the ACA and incorporates the nonlocal nature of the orbital motion of electrons. 

Although Eq.~\eqref{eq:modern} describes the OAM from all the sources, the $\mathbf{k}$-space formula makes it difficult to understand how electrons circulate microscopically in {\it real} space. Bianco and Resta have introduced a `local marker' formulation for the orbital magnetization such that the local contribution in real space can be probed and its macroscopic average agrees with Eq.~\eqref{eq:modern}. This formulation was originally developed for insulating systems~\cite{Bianco13PRL}, but has been successfully applied to both insulating and metalic systems~\cite{Marrazzo16PRL,Wang22PRB}. Recently, Ref.~\cite{Vidarte25arXiv} has taken an approach to evaluating Eq.~\eqref{eq:OAM} explicitly in real space for any finite-size system. The real-space approaches are useful for treating disordered, inhomogeneous, and finite systems, and enable a separation between bulk and surface contributions~\cite{Marrazzo16PRL,Wang22PRB,Vidarte25arXiv}. On the other hand, the real-space picture is straightforward in the ACA.

Another issue to be resolved in the modern theory is how the ACA is automatically included within it and which terms encode nonlocal circulations. Because many orbitronic phenomena have been calculated assuming the ACA, it is important to establish the limitations of the ACA and to what extent the ACA is valid compared to the modern theory. Hanke {\it et al.} performed a comparative study on various materials ranging from transition metals to Chern insulators and noncollinear magnets, revealing the crucial role of the Berry phase~\cite{Hanke16PRB}. However, to the best of our knowledge, it has not yet been clarified which term in the modern theory makes the difference, because Eq.~\eqref{eq:modern} itself does not explicitly make a difference between atomic and interstitial contributions.

Meanwhile, often in theoretical studies~\cite{Canonico20PRB,Bhowal21PRB,Pezo22PRB,Cysne22PRB,Busch23PRR,Gobel24PRL}, Eq.~\eqref{eq:modern} is calculated by inserting the identity,
\begin{align}\label{eq:the_identity}
    \hat{\mathbb{1}}_{\mathbf{k}}= \sum_{n'} \ket{u_{n'\mathbf{k}}} \bra{u_{n'\mathbf{k}}},
\end{align}
right before $\ket{\partial_{\mathbf{k}}u_{n\mathbf{k}}}$ and replacing the Berry connection by
\begin{equation}\label{eq:connection-naive}
    i\braket{u_{n\mathbf{k}} | \partial_\mathbf{k} u_{m\mathbf{k}}}
    =
    \frac{i
    \bra{u_{n\mathbf{k}}}
    \partial_\mathbf{k} \mathcal{H}_\mathbf{k}
    \ket{u_{m\mathbf{k}}}
    }{\mathcal{E}_{n\mathbf{k}} - \mathcal{E}_{m\mathbf{k}}}, \ \ n \neq m.
\end{equation}
Although the above approach is mathematically precise, there are technical subtleties that make the numerical evaluation based on this approach flawed for the following reason. First, essentially all numerical schemes do not deal with a complete set of basis functions, the number of which is infinite. For example, in most schemes, even in {\it ab initio} methods, a limited number of basis states are employed, and thus the identity in Eq.~\eqref{eq:the_identity} does not hold. Practically, by naively inserting `$\hat{\mathbb{1}}$', one actually projects onto the reduced Hilbert space where an effective theory is defined, and ignores the contributions arising from the space outside the reduced Hilbert space. This could be a reasonable approximation if the latter contributions are negligible. This is indeed the case for some quantities such as the anomalous conductivity. However, it is not the case for the orbital magnetization~\cite{Thonhauser12IJMPB,Ceresoli10PRB,Lopez12PRB} unless the reduced Hilbert space contains all eigenstates in a very wide energy window near the Fermi energy. Such a wide energy window makes numerical calculations costly. Therefore, one needs a formulation that enables one to compute Eq.~\eqref{eq:modern} in terms only of much smaller number of states within a few or at best a few tens of eV energy window near the Fermi energy. Second, the $\mathbf{k}$-derivative in Eq.~\eqref{eq:connection-naive} is usually applied only to a matrix representation of the Hamiltonian and \textit{not} to the basis of the representation. However, as shown in Ref.~\cite{Go24PRB}, the latter derivative is crucial for evaluating orbital-related quantities. As we will show in Sec.~\ref{sec:theory}, naively using Eqs.~\eqref{eq:the_identity} and~\eqref{eq:connection-naive} to evaluate Eq.~\eqref{eq:modern} fails to capture the contribution of the ACA consistently within the modern theory and misses other important contributions. Importantly, this approach breaks the gauge invariance of Eq.~\eqref{eq:modern}, that is, Eq.~\eqref{eq:modern} being independent of the choice of basis functions used for its evaluation. To restore the gauge invariance, it is crucial to include the $\mathbf{k}$-derivative acting on basis states, which produces anomalous-position-like terms, related to the Berry phase inherent in the basis states regardless of the details of the Hamiltonian~\cite{Go24PRB}. This must be evaluated explicitly at the level of {\it ab initio} without inserting the `$\hat{\mathbb{1}}$' which is incomplete in practice~\cite{Lopez12PRB}.

\subsection{Outline of the paper}

Our work aims to resolve the above-mentioned critical issues on the modern theory. We start from the gauge-invariant expression of the orbital magnetization based on the ground-state projector, introduced by Ceresoli {\it et al.}~\cite{Ceresoli06PRB} and adopt the Wannier-based description introduced in the seminal work by Lopez {\it et al.}~\cite{Lopez12PRB}, which bridges the real-space description encoded in the position and higher moments evaluated in Wannier basis and connects to the $\mathbf{k}$-space formula of Eq.~\eqref{eq:modern}. We scrutinize the gauge-covariant formulation of the modern theory, which ensures the same form of the equations that need to be implemented (gauge-covariance) and the same value of the resulting orbital magnetization (gauge-invariance), irrespective of uncontrolled gauge effects arising from the choice of Wannier functions and diagonalization of the effective Hamiltonian. Importantly, we construct a quantum mechanical operator for the orbital moment based on the {\it occupation-weighted covariant derivative}, which reproduces all the results of the gauge-covariant implementation. Note that this is more general than the Ansatz of the orbital moment operator written in terms of ordinary derivative~\cite{Bhowal21PRB,Cysne22PRB,Busch23PRR,Gobel24PRL,Pezo22PRB,Go24NanoLett,Lee25PRB}. Our approach naturally captures both the contributions from localized Wannier functions and the contributions due to band hybridizations, whose sum is gauge-invariant. Based on this, we analyze band-resolved contributions to the orbital moment systematically, which is essential for understanding microscopic origins of the orbital moment.

For detailed analysis, we decompose the orbital magnetization into a few separate contributions, which differ from each other with regard to the nature of the electron's circulating motion responsible for the orbital magnetization, that is, whether the circulation is fundamentally originating from localized Wannier basis or whether it is induced by coherent band hybridizations due to the gauge difference between the Wannier bases and the Hamiltonian's eigenstates, which, roughly speaking, corresponds to real-space or reciprocal-space effects, respectively. The latter is related to the contribution captured by the naive evaluation of Eq.~\eqref{eq:modern} in effective models with a finite number of basis. On the other hand, the former is related to what is known as the `atomic' OAM operator in models. In particular, the former agrees \textit{quantitatively} with the ACA contribution in many materials when the Wannier functions are chosen to resemble atomic orbitals. For different choices of the Wannier functions, however, the agreement declines, implying that the former alone is not gauge-invariant. We prove that only the sum of all these contributions makes the orbital magnetization gauge-invariant, and each contribution does not individually satisfy the gauge invariance. Note that this scheme is different from the local versus itinerant circulations in Wannier-based descriptions.
We denote this scheme of decomposition by {\it $J$-decomposition}, following the notation introduced in Refs.~\cite{Wang06PRB, Lopez12PRB}, where $J$ stands for the term in Eq.~\eqref{eq:connection-naive}. The terms proportional to $J^0$, $\mathbf{M}^{(0)}$, are solely described by matrix elements between Wannier states and thus describe atomic-like contributions. The terms proportional to $J^2$, $\mathbf{M}^{(2)}$, are the most affected by the coherent hybridizations; they tend to diverge to infinity as the energy difference approaches zero, and thus capture the most nonlocal contributions. The rest of the contributions are the terms proportional to $J^1$, $\mathbf{M}^{(1)}$, describing an intermediate-level circulation, mixed between intra-atomic and inter-atomic motions. We remark that the $J$-decomposition differs from the decomposition into the wave packet self-rotation versus the center-of-mass circulation of the wave packet (see Sec.~\ref{subsec:gauge_invariance}). Each of the self-rotation contribution and the center-of-mass circulation contribution contains the $J^{0}$, $J^{1}$, and $J^{2}$ contributions in the $J$-decomposition scheme, as demonstrated below.

Our first-principles-based term-by-term analysis on different classes of materials generally shows the tendency that the orbital magnetism in transition metals has a mostly atomic character, dominated by $\mathbf{M}^{(0)}$, and the ACA constitutes more than 70\% of the total contribution. However, deviations among different elements are worth noticing. Elements with relatively more localized $d$ orbitals (e.g. Fe, Co, Ni) have the orbital magnetism dominated by the $\mathbf{M}^{(0)}$, while elements like W exhibit sizable $\mathbf{M}^{(1)}$ and $\mathbf{M}^{(2)}$. In Bi, a prototypical example of the $sp$ metal, we find that the ACA captures only 42\% of the total orbital magnetism given by the modern theory, where nonlocal $\mathbf{M}^{(1)}$ and $\mathbf{M}^{(2)}$ are important. We observe a similar tendency for a large deviation between the ACA and the modern theory in a light metal Al. We attribute this behavior to the larger spread of $sp$ orbitals. Finally, we investigate transition metal dichalcogenides (TMDs), 1H-MoS$_{2}$ and T$_d$-WTe$_2$ monolayer, as two prototypical examples of a semiconductor with the valley degree of freedom and a topological insulator. In MoS$_{2}$, we find that $\mathbf{M}^{(2)}$ terms near the direct band gaps at $\mathrm{K}$ and $\mathrm{K}'$ are highly dominant, leading to pronounced valley-dependent orbital moment, whereas the ACA yields only a small portion at these points. Furthermore, the valley-dependent orbital moment evaluated by the modern theory varies across the insulating energy gap, whereas the orbital moment obtained from the ACA remains fixed. In WTe$_2$, we observe that the momentum-resolved orbital moment tends to take on very large values near avoided band crossings, where the Berry curvature is pronounced due to strong band hybridizations.

The rest of the article is organized as follows. In Secs.~\ref{subsec:gauge_invariance}-\ref{subsec:gauge_covariant_expression}, we revisit the gauge-invariant formulation of the modern theory of orbital magnetization and its Wannier representation, respectively, which were first presented by Lopez {\it et al.}~\cite{Lopez12PRB}. In Sec.~\ref{subsec:orbital_moment_operator}, we introduce a gauge-covariant expression of the modern theory, which is implemented in our first-principles code as it is. Section~\ref{sec:first-principles_calculation} is devoted to first-principles calculation of real materials. After briefly explaining computational methods (Sec.~\ref{subsec:computation_method_outline}) and how to analyze term by term (Sec.~\ref{subsec:term-by-term_analysis}), we show detailed results on $d$ transition metals, $sp$ metals, and TMDs in Sec.~\ref{subsec:d_transition_metals}-\ref{subsec:TMDs}, respectively, and discuss implications of the results. Finally, Sec.~\ref{sec:concluding_remarks} summarizes and concludes the article with several remarks.

\section{Theoretical Formalism}\label{sec:theory}

\subsection{Gauge invariance of orbital magnetization}\label{subsec:gauge_invariance}

One can show that the modern theory expression for the orbital magnetization can be written to be invariant with respect to $U(N)$ gauge transform, which is thus measurable. To show the gauge invariance more explicitly, Ceresoli {\it et al.}~\cite{Ceresoli06PRB} have shown that the orbital magnetization can be written in terms of the ground-state projector for each $\mathbf{k}$
\begin{align}\label{eq:projection_P}
    \hat{P}_{\mathbf{k}} = \sum_{n}^\infty \ket{u_{n\mathbf{k}}} f_{n\mathbf{k}} \bra{u_{n\mathbf{k}}}.
\end{align}
In the zero temperature limit, the ground-state projector $\hat{P}_{\mathbf{k}}$ satisfies the property of a projection operator. We assume this limit and set the occupation function $f_{n\mathbf{k}}$ to one or zero. For convenience, we define the complementary projector
\begin{eqnarray}\label{eq:projection_Q}
    \hat{Q}_\mathbf{k} &=& \hat{\mathbb{1}}_\mathbf{k} - \hat{P}_\mathbf{k}
    =
    \sum_{n}^\infty \ket{u_{n\mathbf{k}}} g_{n\mathbf{k}} \bra{u_{n\mathbf{k}}}
\end{eqnarray}
where 
\begin{equation}\label{eq:identity}
    \hat{\mathbb{1}}_\mathbf{k} = \sum_n^\infty \ket{u_{n\mathbf{k}}} \bra{u_{n\mathbf{k}}}
\end{equation}
is the complete identity for each $\mathbf{k}$, and 
\begin{equation}\label{eq:occupation_g}
g_{n\mathbf{k}} = 1  - f_{n\mathbf{k}}.
\end{equation}
In terms of ground-state projector and its complementary, the modern theory orbital magnetization can be written as
\begin{align}\label{eq:modern2}
    M_{\gamma}(\mathbf{k}) &= \frac{\varepsilon_{\alpha \beta \gamma}e}{2\hbar} \text{Im} 
    \left[
    \text{Tr}\left( \hat{G}_{\mathbf{k}, \alpha \beta} + \hat{K}_{\mathbf{k}, \alpha \beta} - 2\mathcal{E}_{\text{F}}\hat{F}_{\mathbf{k}, \alpha \beta} \right)
    \right] ,
\end{align}
where
\begin{subequations}\label{eq:modern_FGK}
\begin{eqnarray}
\label{eq:modern_F}
    \hat{F}_{\mathbf{k}, \alpha \beta} &=& (\partial_{\alpha} \hat{P}_{\mathbf{k}}) \hat{Q}_{\mathbf{k}} (\partial_{\beta} \hat{P}_{\mathbf{k}}),
\\[6pt]
\label{eq:modern_G}
    \hat{G}_{\mathbf{k}, \alpha \beta} &=& (\partial_{\alpha} \hat{P}_{\mathbf{k}}) \hat{Q}_{\mathbf{k}} \hat{H}_{\mathbf{k}} \hat{Q}_{\mathbf{k}} (\partial_{\beta} \hat{P}_{\mathbf{k}}),
\\[6pt]
\label{eq:modern_K}
    \hat{K}_{\mathbf{k}, \alpha \beta} &=& \hat{H}_{\mathbf{k}} (\partial_{\alpha} \hat{P}_{\mathbf{k}}) \hat{Q}_{\mathbf{k}} (\partial_{\beta} \hat{P}_{\mathbf{k}}).
\end{eqnarray}
\end{subequations}
Here, $\alpha$, $\beta$, $\gamma$ $\in$ $\{\hat{x}, \hat{y}, \hat{z}\}$, and $\partial_{\alpha}$ denotes $\partial_{k_{\alpha}}$. The summation over $\alpha$ and $\beta$ is implied through the Levi-Civita tensor $\epsilon_{\alpha \beta \gamma}$. These three objects are gauge-covariant under any unitary transformation mixing the occupied bands (see Appendix~\ref{appendix:gauge-invariant-objects}). Therefore, Eq.~\eqref{eq:modern2} yields the same values independent of the gauge choice and is known as the \textit{gauge-invariant orbital magnetization}.
The orbital magnetization [Eq.~\eqref{eq:modern2}] can be decomposed, for which commonly used conventions are introduced below.

The first way of classifying the orbital magnetization is based on the semi-classical wave packet dynamics~\cite{Chang96PRB,Sundaram99PRB,Xiao06PRL}. According to it, self-rotation (SR) of a wave packet generates the orbital magnetization
\begin{subequations}\label{eq:modern_wavepacket}
\begin{align}\label{eq:modern_SR}
    M_{\gamma}^{\text{SR}}(\mathbf{k}) &= \frac{\varepsilon_{\alpha \beta \gamma}e}{2\hbar} \text{Im} 
    \left[ \text{Tr}\left( \hat{G}_{\mathbf{k}, \alpha \beta} - \hat{K}_{\mathbf{k}, \alpha \beta} \right) \right].
\end{align}
On the other hand, the Berry phase correction of the density of state leads to another contribution to the thermodynamic energy~\cite{Xiao05PRL}, given by 
\begin{align}\label{eq:modern_CM}
    M_{\gamma}^{\text{CM}}(\mathbf{k}) &= \frac{\varepsilon_{\alpha \beta \gamma}e}{\hbar} \text{Im} \left[ \text{Tr}\left( \hat{K}_{\mathbf{k}, \alpha \beta} - \mathcal{E}_{\text{F}}\hat{F}_{\mathbf{k}, \alpha \beta} \right) \right] .
\end{align}
\end{subequations}
Here, the superscript stands for the center-of-mass (CM) motion of a wave packet for its physical meaning.

There is an alternative decomposition of the orbital magnetization from the Wannier-based description~\cite{Ceresoli06PRB,Lopez12PRB}, by which $M_{\gamma}(\mathbf{k})$ is expressed as the sum of local circulation (LC) $M^{\text{LC}}_{\gamma}(\mathbf{k})$ and itinerant circulation (IC) $M^{\text{IC}}_{\gamma}(\mathbf{k})$ contributions. They are given by 
\begin{subequations}\label{eq:modern_wannier}
\begin{eqnarray}\label{eq:modern_LC}
    M_{\gamma}^{\text{LC}}(\mathbf{k}) &=& 
    M_{\gamma}^{\text{SR}}(\mathbf{k}) + \frac{1}{2}M_{\gamma}^{\text{CM}}(\mathbf{k})
    \\
    &=&
    \frac{\varepsilon_{\alpha \beta \gamma}e}{2\hbar} \text{Im} \left[ \text{Tr}\left( \hat{G}_{\mathbf{k}, \alpha \beta} - \mathcal{E}_{\text{F}}\hat{F}_{\mathbf{k}, \alpha \beta} \right) \right] , \nonumber 
\end{eqnarray}
and
\begin{eqnarray}\label{eq:modern_IC}
    M_{\gamma}^{\text{IC}}(\mathbf{k}) &=& 
    \frac{1}{2}M_{\gamma}^{\text{CM}}(\mathbf{k})
    \\
    &=&
    \frac{\varepsilon_{\alpha \beta \gamma}e}{2\hbar} \text{Im} \left[ \text{Tr}\left( \hat{K}_{\mathbf{k}, \alpha \beta} - \mathcal{E}_{\text{F}}\hat{F}_{\mathbf{k}, \alpha \beta} \right) \right] , \nonumber 
\end{eqnarray}
\end{subequations}
respectively. In the Wannier representation, $M^{\text{LC}}_{\gamma}$ presents the interior contribution to the orbital magnetization of the Wannier functions, while $M^{\text{IC}}_{\gamma}$ is the itinerant contribution of the Wannier functions flowing across the boundary of a finite sample~\cite{Thonhauser05PRL,Ceresoli06PRB,Souza08PRB}.

The combinations in Eqs.~\eqref{eq:modern_wavepacket} and \eqref{eq:modern_wannier} are gauge invariant and energy-origin independent, and can be connected to physically observable quantities. For example, $M^{\text{SR}}_{\gamma}$ is proportional to the MCD spectrum~\cite{Souza08PRB,Resta20PRR,Mahfouzi25PRB}. Note that $M^{\text{SR}}_{\gamma} = M^{\text{LC}}_{\gamma} - M^{\text{IC}}_{\gamma}$, which implies that the value of each term can be determined by measuring both $M_{\gamma}$ and $M^{\text{SR}}_{\gamma}$. In the remainder of this paper, we adopt the decomposition $M_{\gamma} = M^{\text{SR}}_{\gamma} + M^{\text{CM}}_{\gamma}$ to present the calculating results for real materials. This approach facilitates the comparison of the values obtained from experiments measuring $M_{\gamma}$, such as gyromagnetic experiments, with experiments measuring $M^{\text{SR}}_{\gamma}$, such as MCD experiments. Finally, we note that the gauge covariance of the individual building blocks [Eq.~\eqref{eq:modern_FGK}] does not by itself imply that each of them is a separately measurable quantity.

\subsection{Wannier representation}\label{subsec:Wannier_reresentation}

\begin{figure*}[t]
\includegraphics[width=0.7\textwidth]{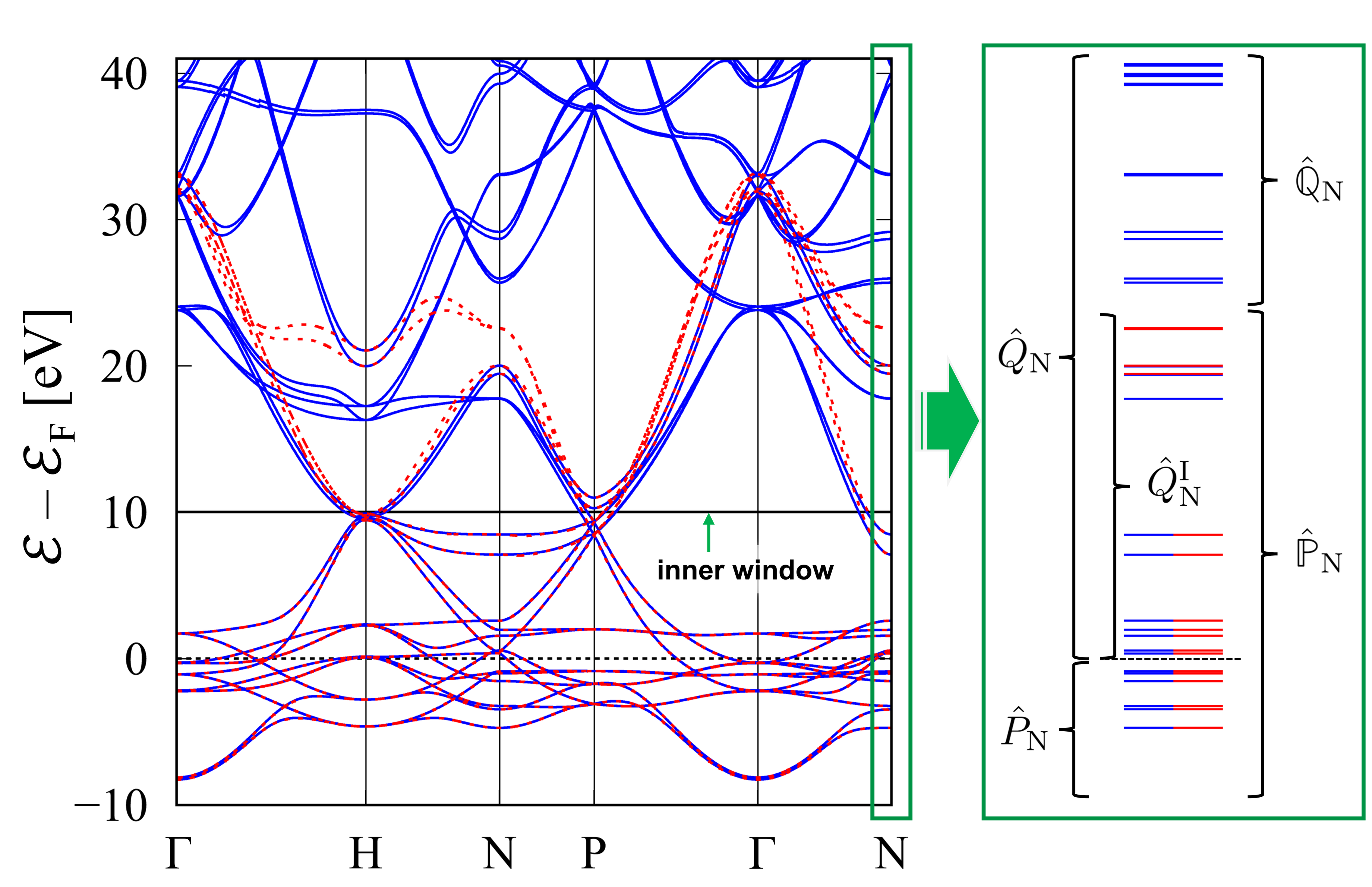}
\caption{\label{fig2}\textbf{Schematic illustration of the space selection.} The band structure of bcc Fe is presented with the inner window (black dotted line). The energy eigenvalues of the Bloch states and the Bloch-like states are plotted as the blue solid lines and red dotted lines, respectively. The green box shows the spectrum of the bands at $\mathbf{k} = \text{N}$ point. The subspaces corresponding to the projection operators $\hat{P}_{\text{N}}$, $\hat{Q}_{\text{N}}$, and $\hat{\mathbb{P}}_{\text{N}}$, $\hat{\mathbb{Q}}_{\text{N}}$ are depicted.}
\end{figure*}

Although Eq.~\eqref{eq:modern2} is written in gauge-invariant form, actual calculation requires an implementation by taking properties of basis states into account. Here, we introduce a representation based on localized Wannier functions, which was first introduced by Lopez {\it et al.}~\cite{Lopez12PRB}. Note that this also shares many similar features with the calculation of the Berry curvature in Wannier basis~\cite{Wang06PRB}. This method ensures consistent implementation of formulas regardless of the choice of Wannier functions, whose effective Hamiltonians are not always identical and controlled depending on computation schemes.

To clarify this, we revisit the works of Vanderbilt \textit{et al.}~\cite{Marzari97PRB, Souza01PRB} on the construction of maximally localized Wannier functions. The Wannierization process is carried out in two steps: \textit{gauge selection} and \textit{space selection}. The gauge selection arises from the freedom in constructing Wannier functions,
\begin{align}\label{eq:Wannier_function}
    \ket{n\mathbf{R}} = \frac{1}{N} \sum_{\mathbf{k}} e^{i\mathbf{k}\cdot (\mathbf{r} - \mathbf{R})}\ket{u_{n\mathbf{k}}^\text{W}},
\end{align}
where $N$ is the number of unit cells or the number of $\mathbf{k}$-points in the first Brillouin zone. Here, the superscript `W' of $u_{n\mathbf{k}}$ stands for the {\it Wannier gauge}. Note that the {\it Wannier-gauge states} or {\it Bloch-like states} ${u_{n\mathbf{k}}^\text{W}}$ do not necessarily have to be eigenstates of the Hamiltonian, for which there is a gauge degree of freedom. Marzari and Vanderbilt developed a method to find the optimally connected gauge in $\mathbf{k}$-space in the smoothest variation as possible, that is called the {\it maximally localized Wannier functions}~\cite{Marzari97PRB,Marzari12RMP,Marrazzo24RMP}.

The space selection refers to choosing a set of bands for constructing Wannier functions. In contrast to the gauge selection, the space selection is a non-unitary process as the information in the bands that are not selected for constructing Wannier functions is dropped out. By this, a {\it reduced Hilbert space} $\bar{\mathcal{H}}$ is chosen from the full Hilbert space $\mathcal{H}$, in which an effective model is constructed. If the group of bands of interest is isolated (that is, the group of bands is separated from all other bands by finite gaps throughout the entire BZ), then the space selection becomes trivial. In such cases, we simply obtain the Wannier functions by Fourier transforming the Bloch states in that group. However, in general, the group of bands of interest is not isolated. For example, when dealing with metals or with conduction bands of insulators, the bands of interest are attached to other bands at points or lines of high symmetry~\cite{Souza01PRB}. In these cases, the bands of interest must be disentangled from the entire band structure. This process is generally called \textit{band disentanglement}. Optionally, a subset of the $\bar{\mathcal{H}}$ can be guaranteed to be included to keep certain band characters, which is known as the {\it inner space} or {\it frozen energy window}. Similar to the gauge selection, the space selection corresponds to choosing an optimal set of Bloch states which are optimally connected in $\mathbf{k}$-space.

Let us define the projection operator onto the $\bar{\mathcal{H}}$ as
\begin{align}\label{eq:projection_P_bb}
    \hat{\mathbb{P}}_{\mathbf{k}} = \sum_{n\in \bar{\mathcal{H}}} \ket{u_{n\mathbf{k}}^{\text{W}}} \bra{u_{n\mathbf{k}}^{\text{W}}},
\end{align}
and its complementary as
\begin{equation}\label{eq:projection_Q_bb}
    \hat{\mathbb{Q}}_\mathbf{k} = \hat{\mathbb{1}}_\mathbf{k} - \hat{\mathbb{P}}_\mathbf{k}.
\end{equation}
Note that the summation index runs only over $\bar{\mathcal{H}}$ ($\text{dim}\bar{\mathcal{H}} < \infty$), so $\hat{\mathbb{P}}_{\mathbf{k}}$ should not be regarded as the identity, for example, in tight-binding models. The Wannier gauge should be selected so that $\hat{\mathbb{P}}_{\mathbf{k}}$ varies smoothly with $\mathbf{k}$~\cite{Marzari97PRB}. The \textit{reduced Hamiltonian} $\mathbb{H}_{\mathbf{k}} = \hat{\mathbb{P}}_{\mathbf{k}}\hat{H}_{\mathbf{k}}\hat{\mathbb{P}}_{\mathbf{k}}$ is a $\text{dim} \bar{\mathcal{H}} \times \text{dim} \bar{\mathcal{H}}$ matrix,
\begin{align}\label{eq:Hamiltonian_Wannier_gauge}
    \mathbb{H}_{nm}^{\text{W}}(\mathbf{k}) &= \bra{u_{n\mathbf{k}}^{\text{W}}}\hat{H}_{\mathbf{k}}\ket{u_{m\mathbf{k}}^{\text{W}}}.
\end{align}
This is then diagonalized by finding the unitary transformation $V(\mathbf{k})$ such that
\begin{align}\label{eq:Hamiltonian_Hamiltonian_gauge}
    \mathbb{H}_{nm}^{\text{H}}(\mathbf{k}) 
    &= \left[ V^{\dagger}(\mathbf{k}) \mathbb{H}^{\text{W}}(\mathbf{k}) V(\mathbf{k}) \right]_{nm} = \bar{\mathcal{E}}_{n\mathbf{k}} \delta_{nm}.
\end{align}
The energy eigenvalue $\bar{\mathcal{E}}_{n\mathbf{k}}$ of the reduced Hamiltonian may not agree with the original energy eigenvalue ${\mathcal{E}}_{n\mathbf{k}}$ if either the space selection or the initial projection of Wannier functions, which requires a guess before the gauge selection, is not optimal. Note that the projected eigenenergies $\bar{\mathcal{E}}_{n\mathbf{k}}$ should have the same values as the true eigenenergies $\mathcal{E}_{n\mathbf{k}}$ for the group of bands of interest (e.g., occupied states), but they may differ for the bands outside the group (e.g., unoccupied states). In the disentanglement process, this can be controlled by setting the inner window. The bands included in the inner window remain invariant under the Wannier function choices, whereas the bands outside the window may vary. This is illustrated in Fig.~\ref{fig2}.
The corresponding eigenstate of the reduced Hamiltonian is
\begin{align}\label{eq:Bloch_Hamiltonian_gauge}
    \ket{u_{n\mathbf{k}}^{\text{H}}} = \sum_{m\in \bar{\mathcal{H}}} \ket{u_{m\mathbf{k}}^{\text{W}}}V_{mn}(\mathbf{k}),
\end{align}
where $V_{mn}(\mathbf{k}) = \braket{u^{\text{W}}_{m\mathbf{k}}|u^{\text{H}}_{n\mathbf{k}}}$. We call this gauge, which diagonalizes $\mathbb{H}_{\mathbf{k}}$, the \textit{Hamiltonian gauge} in the reduced Hilbert space $\bar{\mathcal{H}}$. If dominant band characters of {\it ab initio} Bloch states are preserved in the space selection and in the initial projection before the gauge selection, the {\it ab initio} Bloch state projected onto the $\bar{\mathcal{H}}$ may be faithfully represented by the Hamiltonian gauge state $u_{n\mathbf{k}}^\text{H}$, which is spanned by Wannier functions.

In the reduced Hilbert space $\bar{\mathcal{H}}$, the ground-state projection operator [Eq.~\eqref{eq:projection_P}] can be represented by 
\begin{eqnarray}\label{eq:projection_P2}
    \hat{P}_{\mathbf{k}} &=& 
    \sum_{n\in \bar{\mathcal{H}}}\ket{u_{n\mathbf{k}}^{\text{H}}}f_{n\mathbf{k}}^{\text{H}}\bra{u_{n\mathbf{k}}^{\text{H}}} \nonumber 
    \\
    &=& 
    \sum_{n,m\in \bar{\mathcal{H}}}\ket{u_{n\mathbf{k}}^{\text{W}}}f_{nm,\mathbf{k}}^{\text{W}}\bra{u_{m\mathbf{k}}^{\text{W}}},
\end{eqnarray}
where the occupation matrix in the Hamiltonian gauge $f_{n\mathbf{k}}^\mathrm{H}$ is band-diagonal, either $1$ or $0$ depending on whether $u_{n\mathbf{k}}^\text{H}$ is occupied or unoccupied, respectively. Note that the summation index is within the $\bar{\mathcal{H}}$ (the summation index goes only up to $\text{dim} \bar{\mathcal{H}}$), but Eq.~\eqref{eq:projection_P2} is still identical to Eq.~\eqref{eq:projection_P}. This is because the occupation is $0$ for unoccupied states, which do not contribute to the summation.
The occupation matrix in the Wannier gauge is generally non-diagonal and can be obtained by $f_{nm,\mathbf{k}}^\mathrm{W} =  [V  f^\text{H} V^\dagger]_{nm}$. The projection complementary to $\hat{P}$ [Eq.~\eqref{eq:projection_Q}] and that complementary to $\hat{\mathbb{P}}_\mathbf{k}$ [Eq.~\eqref{eq:projection_Q_bb}] are related by 
\begin{align}\label{eq:projection_Q2}
    \hat{Q}_{\mathbf{k}} = \hat{Q}^{\text{I}}_{\mathbf{k}} + \hat{\mathbb{Q}}_{\mathbf{k}},
\end{align}
where 
\begin{eqnarray}\label{eq:projection_QI}
    \hat{Q}_\mathbf{k}^\text{I} 
    &=&
    \sum_{n\in \bar{\mathcal{H}}} \ket{u_{n\mathbf{k}}^\text{H}} g_{n\mathbf{k}}^\text{H} \bra{u_{n\mathbf{k}}^\text{H}} \nonumber 
    \\
    &=&
    \sum_{nm \in \bar{\mathcal{H}}} \ket{u_{n\mathbf{k}}^\text{W}} g_{nm,\mathbf{k}}^\text{W} \bra{u_{n\mathbf{k}}^\text{W}}
\end{eqnarray}
is the projection operator onto the unoccupied states {\it in the reduced Hilbert space} $\bar{\mathcal{H}}$ described by Wannier basis.

\subsection{Gauge-related matrices: $\mathbb{A}$, $\mathbb{B}$, $\mathbb{C}$, and $J$}
\label{subsec:gauge_related_matrices}

The gauge-invariant form of the orbital magnetization [Eq.~\eqref{eq:modern2}] and its constituent terms [Eq.~\eqref{eq:modern_FGK}] involve differentiations with respect to $\mathbf{k}$. For example, one repeatedly encounters the derivative of the Hamiltonian-gauge Bloch state [Eq.~\eqref{eq:Bloch_Hamiltonian_gauge}]:
\begin{align}\label{eq:Bloch_Hamiltonian_gauge_derivative}
    \ket{\partial_{\mathbf{k}} u_{n}^{\text{H}}} &= \sum_{m}^{L} \left\{ \ket{\partial_{\mathbf{k}} u_{m}^{\text{W}}}V_{mn} (\mathbf{k}) - i\ket{u_{m}^{\text{H}}} \mathbf{J}_{mn}^{\text{H}} (\mathbf{k}) \right\},
\end{align}
where
\begin{align}\label{eq:J_H}
    \mathbf{J}^{\text{H}} (\mathbf{k}) = iV^{\dagger} (\mathbf{k})
    \left[ \partial_\mathbf{k} V (\mathbf{k}) \right]
\end{align}
is the connection arising from the gauge transform from the Wannier gauge to the Hamiltonian gauge.

\begin{figure*}[t!]
\includegraphics[width=440pt]{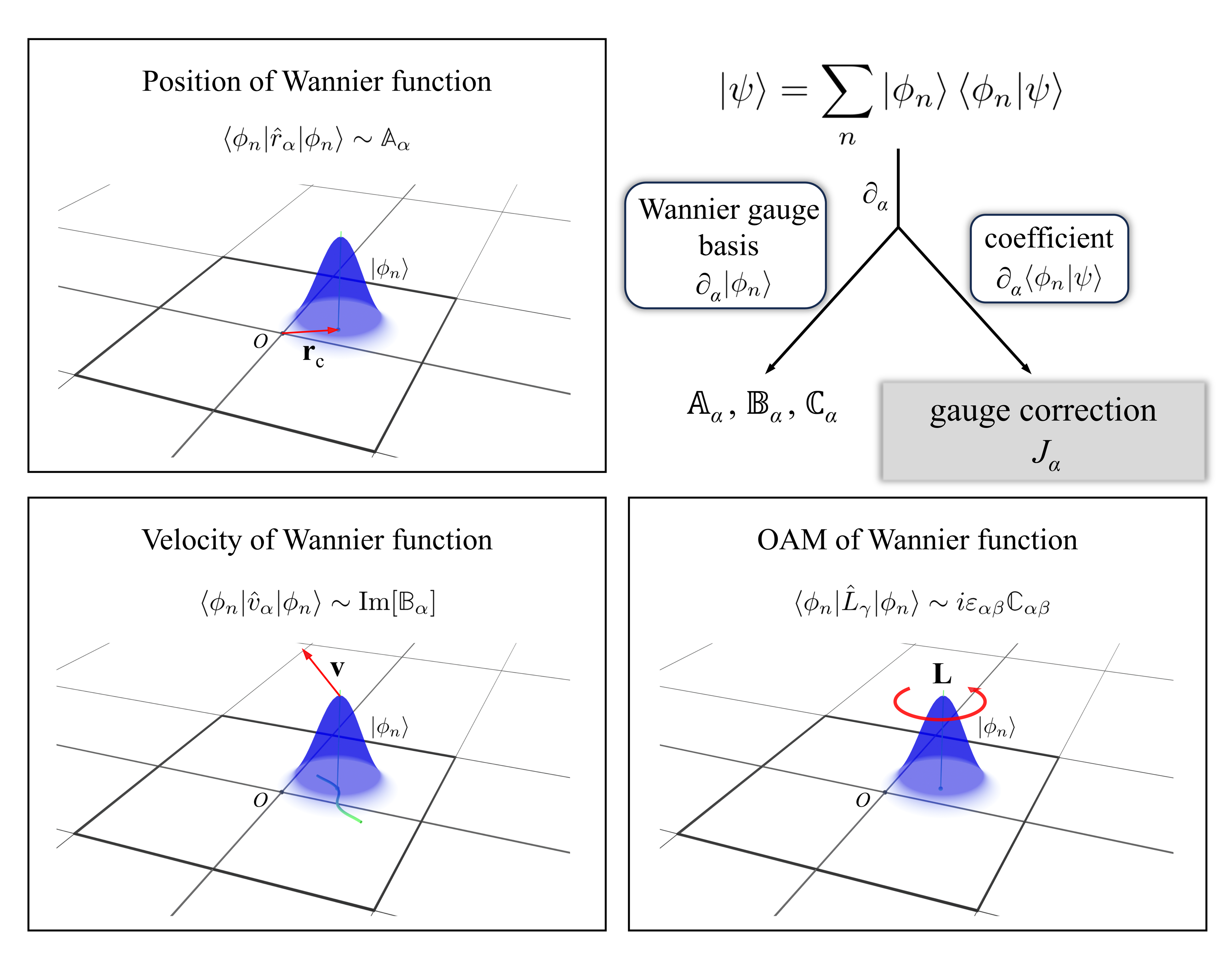}
\caption{\label{fig:Wannier_ABC}\textbf{Schematic illustration of the intuitive meaning of the Wannier gauge objects.} $\mathbb{A}_{\alpha}$, $\mathbb{B}_{\alpha}$, and $\mathbb{C}_{\alpha \beta}$ describe the position, velocity, and orbital angular motion of the Wannier basis, respectively, while the gauge correction $J_{\alpha}$ captures the inter-state mixing. Here, $\ket{\phi_{n}}$ and $\ket{\psi}$ denote Wannier-gauge states and Hamiltonian-gauge states, respectively.}
\end{figure*}

On the other hand, the derivative on the Wannier-gauge Bloch state requires the information on the basis states, Wannier functions. We remark that the information on the Hamiltonian alone is insufficient for evaluating the orbital magnetization [Eq.~\eqref{eq:modern2}], as well as the Berry curvature. For the orbital magnetization, one needs the so-called 'ABC' of gauge-related quantities {\it defined in the Wannier gauge},
\begin{subequations}\label{eq:gauge_ABC}
\begin{eqnarray}
\label{eq:gauge_A}
    \mathbb{A}_{\alpha,nm}^\text{W} &=& i\braket{u_{n}^\text{W}|\partial_{\alpha}u_{m}^\text{W}},
\\[6pt]
\label{eq:gauge_B}
    \mathbb{B}_{\alpha,nm}^\text{W} &=& i\bra{u_{n}^\text{W}}\hat{H}\ket{\partial_{\alpha}u_{m}^\text{W}},
\\[6pt]
\label{eq:gauge_C}
    \mathbb{C}_{\alpha \beta,nm}^\text{W} &=& \bra{\partial_{\alpha}u_{n}^\text{W}}\hat{H}\ket{\partial_{\beta}u_{m}^\text{W}},
\end{eqnarray}
\end{subequations}
which are $\text{dim} \bar{\mathcal{H}} \times \text{dim} \bar{\mathcal{H}}$ matrices defined in the reduced Hilbert space. Here, $\partial_{\alpha}$ denotes $\partial_{k_{\alpha}}$.

Meanwhile, the quantum geometric tensor is defined as
\begin{align}\label{eq:QGT}
    \mathbb{F}_{\alpha \beta, nm}^\text{W}  = \braket{\partial_{\alpha}u_{n}^\text{W}|\partial_{\beta}u_{m}^\text{W}},
\end{align}
whose imaginary part is known as the Berry curvature
\begin{align}\label{eq:Berry_curvature}
    i\mathbb{F}_{\alpha \beta}^\text{W} - i(\mathbb{F}_{\alpha \beta}^\text{W})^{\dagger} = \partial_{\alpha}\mathbb{A}_{\beta}^\text{W}  - \partial_{\beta}\mathbb{A}_{\alpha}^\text{W}.
\end{align}
These gauge-related quantities $\mathbb{A}_{\alpha}^{\text{W}}$, $\mathbb{B}_{\alpha}^{\text{W}}$, $\mathbb{C}_{\alpha\beta}^{\text{W}}$, and $\mathbb{F}_{\alpha\beta}^{\text{W}}$ can be calculated by the \textit{ab initio} code, such as \textsc{Quantum Espresso} and \textsc{Fleur}, via their interface to \textsc{Wannier90}~\cite{Urru25PRB}.

By the gauge transform between the Hamiltonian gauge and the Wannier gauge [Eq.~\eqref{eq:Bloch_Hamiltonian_gauge}], the so-called ABC of gauge-related quantities in Eq.~\eqref{eq:gauge_ABC} in the Hamiltonian gauge are written accordingly:
\begin{subequations}\label{eq:ABC_Hamiltonian_gauge}
\begin{eqnarray}\label{eq:A_Hamiltonian_gauge}
    \mathbb{A}_{\alpha,nm}^\mathrm{H} &=& 
    i \braket{u_{n}^\text{H} | \partial_\alpha u_m^\text{H}} \nonumber 
    \\
    &=&
    [V^\dagger \mathbb{A}_\alpha^\text{W} V]_{nm} + J_{\alpha,nm}^\mathrm{H},
    \\[6pt]
    \mathbb{B}_{\alpha,nm}^\text{H} &=& 
    i \bra{u_{n}^\text{H}} \hat{H} \ket{\partial_\alpha u_m^\text{H}} \nonumber 
    \\
    &=&
    \bar{\mathcal{E}}_n \mathbb{A}_{\alpha,nm}^\text{H},
    \\[6pt]
    \mathbb{C}_{\alpha\beta,nm}^\text{H} &=& 
    \bra{\partial_\alpha u_{n}^\text{H}} \hat{H} \ket{\partial_\beta u_{m}^\text{H}} \nonumber 
    \\
    &=&
    [V^\dagger \mathbb{C}_{\alpha\beta}^\text{W} V]_{nm} + 
    [V^\dagger (\mathbb{B}_\alpha^\text{W})^\dagger V  J_\beta^\text{H} ]_{nm} \nonumber 
    \\ 
    & &
    +
    [J_\alpha^\text{H}  V^\dagger \mathbb{B}_\beta^\text{W} V ]_{nm} +
    [J_\alpha^\text{H} \mathbb{H}^\text{H} J_\beta^\text{H}]_{nm}.
\end{eqnarray}
\end{subequations}
Similarly, the quantum geometric tensor transforms as 
\begin{eqnarray}\label{eq:QGT_Hamiltonian_gauge}
    \mathbb{F}_{\alpha\beta,nm}^\text{H} &=& 
    \braket{\partial_\alpha u_n^\text{H} | \partial_\beta u_{m}^\text{H}} \nonumber
    \\
    &=&
    [V^\dagger \mathbb{F}_{\alpha\beta}^\text{W} V]_{nm} + 
    [V^\dagger \mathbb{A}_\alpha^\text{W} V J_\beta^\text{H} ]_{nm} \nonumber 
    \\
    & &
    +
    [J_\alpha^\text{H} V^\dagger \mathbb{A}_\beta^\text{W} V]_{nm} +
    [J_\alpha^\text{H} J_\beta^\text{H}]_{nm}.
\end{eqnarray}
Note the additional contributions due to $\mathbf{J}^\text{H}$.

A typical tight-binding method~\cite{Slater54PR} evaluates physical quantities using the eigenstates and eigenvalues of the reduced Hamiltonian $\mathbb{H}$ in $\bar{\mathcal{H}}$, obtained through the Wannierization process. It successfully reproduces the band structure of interest. However, this method fails to capture the full information associated with the $\mathbf{k}$-derivatives of the states. As shown in Eq.~\eqref{eq:Bloch_Hamiltonian_gauge}, the $\mathbf{k}$-derivatives of the states consists of the two contributions: the derivative of the coefficients $J_{mn}^{\text{H}}(\mathbf{k})$, and the derivative of the Wannier-gauge basis states $\ket{\partial_{\mathbf{k}}u_{m}^{\text{W}}}$. The former can be determined solely by the reduced Hamiltonian $\mathbb{H}$ within $\bar{\mathcal{H}}$. The latter, however, contains components outside $\bar{\mathcal{H}}$ and is encoded in the Wannier-gauge objects $\mathbb{A}_{\alpha}^{\text{W}}$, $\mathbb{B}_{\alpha}^{\text{W}}$, and $\mathbb{C}_{\alpha\beta}^{\text{W}}$. The standard tight-binding method uses only the states within $\bar{\mathcal{H}}$, thereby neglecting the latter contributions. Therefore, to recover this missing information, it is essential to compute the Wannier-gauge objects $\mathbb{A}_{\alpha}^\text{W}$, $\mathbb{B}_{\alpha}^\text{W}$, and $\mathbb{C}_{\alpha \beta}^\text{W}$ at the first-principles level and then introduced them via projection onto the reduced Hilbert space. The role of $\mathbb{A}_{\alpha}^\text{W}$ or the \textit{external Berry connection} has long been recognized in optical and Berry-curvature-related responses~\cite{Pedersen01PRB,Paul03PRB,Wang06PRB,Azpiroz22SPP}. When quantities related to orbital magnetism are evaluated, $\mathbb{B}_{\alpha}^\text{W}$ and $\mathbb{C}_{\alpha \beta}^\text{W}$ should also be included~\cite{Lopez12PRB}. This missing information is not only numerically significant but also depends on both the gauge and space selection. Consequently, the standard tight-binding method lacks invariance under these choices, and the independence of these choices can only be achieved when terms related to the objects are included. Finally, we emphasize that this gauge issue is applied not only to the orbital magnetism but also, in general, to physical quantities involving $\mathbf{k}$-derivatives of the state, such as the Berry curvature and quantum metric~\cite{Liu23arXiv}.

The practical importance of this missing information depends on the physical quantity and systems under consideration. For the anomalous Hall effect in single elements, the missing information is often relatively small. For example, in the case of the anomalous Hall conductivity of ferromagnetic Fe, 99\% of the contribution stems from the derivatives of the coefficients. However, for orbital magnetization, where both position and velocity jointly induce the state mixing, the basis-derivative contribution can be quantitatively important even in simple materials, as demonstrated by the results in Sec.~\ref{sec:first-principles_calculation}. Therefore, the inclusion of the $\mathbb{A}_{\alpha}^\text{W}$, $\mathbb{B}_{\alpha}^\text{W}$ and $\mathbb{C}_{\alpha \beta}^\text{W}$ is necessary for a gauge-invariant and quantitatively reliable evaluation.

We now turn to discussing the physical meaning of the Wannier-gauge matrices $\mathbb{A}$, $\mathbb{B}$, $\mathbb{C}$ as well as $J_{\alpha}$ introduced in Sec.~\ref{subsec:gauge_related_matrices}. When a Hamiltonian-gauge state is expressed in the basis of Wannier-gauge states, obtained by Fourier transforming the Wannier functions in real space to $\mathbf{k}$-space, it is written as a linear combination of those basis states according to Eq.~\eqref{eq:Bloch_Hamiltonian_gauge_derivative}. Differentiating this state with respect to $\mathbf{k}$ yields two contributions: the variation of the basis, captured by $\ket{\partial_{\alpha}u^{\text{W}}}$, and the variation of the coefficients, contained in the gauge correction $J_{\alpha}$.

$\mathbb{A}_{\alpha}^\text{W}$, $\mathbb{B}_{\alpha}^\text{W}$, and $\mathbb{C}_{\alpha \beta}^\text{W}$ are evaluated in {\it real space} as matrix elements between localized Wannier functions,
\begin{subequations}\label{eq:gauge_ABC_Wannier}
\begin{eqnarray}\label{eq:gauge_A_Wannier}
    \mathbb{A}_{\alpha , nm}^\text{W}  &=& \sum_{\mathbf{R}} e^{i\mathbf{k}\cdot\mathbf{R}} \braket{n\mathbf{0}|\hat{r}_{\alpha}|m\mathbf{R}},
\\[6pt]
\label{eq:gauge_B_Wannier}
    \mathbb{B}_{\alpha , nm}^\text{W}  &=& \sum_{\mathbf{R}} e^{i\mathbf{k}\cdot\mathbf{R}} \braket{n\mathbf{0}|\hat{H}(\hat{r}-R)_{\alpha}|m\mathbf{R}},
\\[6pt]
\label{eq:gauge_C_Wannier}
    \mathbb{C}_{\alpha \beta , nm}^\text{W}  &=& \sum_{\mathbf{R}} e^{i\mathbf{k}\cdot\mathbf{R}} \braket{n\mathbf{0}|\hat{r}_{\alpha}\hat{H}(\hat{r}-R)_{\beta}|m\mathbf{R}},
\end{eqnarray}
\end{subequations}
which contain the position and Hamiltonian operators. The physical meanings of these terms are illustrated with schematics in Fig.~\ref{fig:Wannier_ABC}. For simplicity of discussion, let us focus only on the intracell diagonal matrix elements. The Berry connection $\mathbb{A}_\alpha$ is related to the position matrix elements, describing the anomalous position of Bloch states~\cite{Go24PRB}. In the modern theory of electric polarization~\cite{Kingsmith93PRB, Vanderbilt93PRB}, the $\mathbf{k}$-integral of the Berry connection is related to the sum of the Wannier center positions. On the other hand, $\mathbb{B}_\alpha$ describes the velocity of Wannier functions. For instance, the velocity expectation value for the Wannier state $\ket{n\mathbf{0}}$ is given by $\braket{n\mathbf{0}|\hat{v}_{\alpha}|n\mathbf{0}} = -i(\mathbb{B}_{\alpha ,nn} - \mathbb{B}^{\dagger}_{\alpha ,nn})/\hbar$. Finally, $\mathbb{C}$ is responsible for the OAM of Wannier functions. It is $\mathbb{C}_{\alpha\beta}$ that the atomic OAM is captured within the modern theory. The OAM of $\ket{n\mathbf{0}}$ is given by $\braket{n\mathbf{0}|[\hat{\mathbf{r}} \times \hat{\mathbf{v}}]_\alpha |n\mathbf{0}} = i\varepsilon_{\alpha \beta \gamma}(\mathbb{C}_{\alpha \beta} - \{\mathbb{H}, \ \mathbb{F}_{\alpha \beta}\}/2)_{nn}/\hbar$.

While $\mathbb{A}$, $\mathbb{B}$, $\mathbb{C}$ encode the information of localized Wannier functions in {\it real space}, $J$ arises from the hybridization in {\it reciprocal space}. The physical meaning of $J$ is the position shift arising from the hybridization among Wannier-gauge states due to the Hamiltonian. This can be seen in the formula
\begin{align}\label{eq:J_Hamiltonian_gauge}
    \mathbf{J}^{\text{H}}_{nm}  &= 
    \begin{dcases}
        \frac{i[V^{\dagger} (\partial_\mathbf{k}\mathbb{H}^{\text{W}}) V]_{nm}}{\bar{\mathcal{E}}_{m} - \bar{\mathcal{E}}_{n} + i \eta}, \quad \text{for} \; n \neq m
        \\[2mm]
        0, \quad \text{for} \; n=m
    \end{dcases}
\end{align}
which is obtained by taking the derivative of Eq.~\eqref{eq:Hamiltonian_Hamiltonian_gauge}. Here, $\eta$ is an infinitesimally small positive real number, for which the limit $\eta \rightarrow 0+$ is taken in the last step of calculation. Note that Eq.~\eqref{eq:J_Hamiltonian_gauge} is similar to Eq.~\eqref{eq:connection-naive} but captures only the contribution due to the gauge transform within the reduced Hilbert space. Because $J$ is related to the position shift, $\sim \mathbb{H}J/\hbar$ may be interpreted as the anomalous velocity due to hybridizations induced by the Hamiltonian. This view is consistent with the behavior of $\mathbb{A}$, $\mathbb{B}$, and $\mathbb{C}$ due to the gauge transform from the Wannier gauge to the Hamiltonian gauge [Eq.~\eqref{eq:ABC_Hamiltonian_gauge}].

\subsection{Gauge-covariant expression}
\label{subsec:gauge_covariant_expression}

Now, we express the gauge-covariant objects [Eq.~\eqref{eq:modern_FGK}] in terms of the quantities mentioned above in order to compute the gauge-invariant orbital magnetization [Eq.~\eqref{eq:modern2}]. Recall that the above quantities include information about the derivative $\ket{\partial_{\alpha}u_{n}}$ that is lost when obtaining the reduced Hamiltonian through the band disentanglement procedure. To show this explicitly, we define the following gauge-covariant quantities for Eqs.~\eqref{eq:gauge_ABC} and \eqref{eq:QGT} by replacing the derivative $\ket{\partial_\alpha u_n^\text{W}}$ by $\widetilde{\partial}_\mathbf{k}$:
\begin{eqnarray}\label{eq:tilde_covariant_derivative}
    \ket{\widetilde{\partial}_\alpha u_{n}^\text{W} } &\equiv & \mathbb{Q}_{\mathbf{k}} \ket{\partial_\alpha u_n^\text{W}} \nonumber 
    \\
    &=&  \ket{\partial_\alpha u_n^\text{W}} + i \sum_{m\in \bar{\mathcal{H}}} \ket{u_m^\text{W}} \mathbb{A}_{\alpha,mn}^\text{W}.
\end{eqnarray}
$\ket{\partial_{\alpha}u_{n}^{\text{W}}}$ is clearly covariant under the transformation between the Hamiltonian gauge and the Wannier gauge. However, it depends on the space selection, since it quantifies the contributions of states outside the reduced Hilbert space to the derivative $\ket{\partial_{\alpha} u_{n}^{\text{W}}}$. Note that $\widetilde{\partial_{\alpha}}$ is defined such that the Berry connection vanishes,
\begin{eqnarray}\label{eq:tilde_A_Wannier}
    \widetilde{\mathbb{A}}_{\alpha,nm}^\text{W} 
    &=& 
    i \braket{u_{n}^\text{W} | \widetilde{\partial}_\alpha u_m^\text{W}}
    =
    0,
\end{eqnarray}
which also holds in the Hamiltonian gauge,
\begin{eqnarray}\label{eq:tilde_A_Hamiltonian}
    \widetilde{\mathbb{A}}_{\alpha,nm}^\text{H} = 
    i \braket{u_{n}^\text{H} | \widetilde{\partial}_\alpha u_m^\text{H}}
    =
    0.
\end{eqnarray}

The other geometry-related quantities defined with respect to $\widetilde{\partial_{\alpha}}$ are
\begin{subequations}\label{eq:gauge_ABC_covariant}
\begin{eqnarray}\label{eq:gauge_A_covariant}
    \widetilde{\mathbb{B}}_{\alpha,nm}^\text{W} &\equiv & i\bra{u_{n}^\text{W}}\hat{H}\ket{\widetilde{\partial}_{\alpha}u_{m}^\text{W}} \nonumber 
    \\
    &=& [\mathbb{B}_{\alpha}^\text{W} - \mathbb{H}^\text{W}\mathbb{A}_{\alpha}^\text{W}]_{nm},
    \\[6pt]
    \widetilde{\mathbb{C}}_{\alpha \beta,nm}^\text{W} &\equiv& 
    \bra{\widetilde{\partial}_{\alpha}u_{n}^\text{W}}\hat{H}\ket{\widetilde{\partial}_{\beta}u_{m}^\text{W}} \nonumber 
    \\
    &=&
    [\mathbb{C}_{\alpha \beta}^\text{W} - \mathbb{A}_{\alpha}^\text{W} {\mathbb{B}}_{\beta}^\text{W}  \nonumber 
    \\
    & & - (\mathbb{B}_{\alpha}^\text{W})^{\dagger}\mathbb{A}_{\beta}^\text{W}
    + \mathbb{A}_\alpha^\text{W} \mathbb{H}^\text{W} \mathbb{A}_\beta^\text{W} 
    ]_{nm},
    \\[6pt]
    \widetilde{\mathbb{F}}_{\alpha \beta, nm}^\text{W} 
    &\equiv& 
    \braket{\widetilde{\partial}_{\alpha}u_{n}^\text{W}| \widetilde{\partial}_{\beta}u_{m}^\text{W}} \nonumber 
    \\
    &=& 
    [\mathbb{F}_{\alpha \beta}^\text{W} - \mathbb{A}_{\alpha}^\text{W} \mathbb{A}_{\beta}^\text{W}]_{nm}.
\end{eqnarray}
\end{subequations}
The terms above in Eq.~\eqref{eq:gauge_ABC_covariant} keep the same form in different gauges. Transformation of these quantities onto the Hamiltonian gauge can be conveniently carried out by
\begin{subequations}
\begin{eqnarray}
    \widetilde{\mathbb{B}}_{\alpha,nm}^\text{H} 
    &\equiv&
    i\bra{u_{n}^\text{H}}\hat{H}\ket{\widetilde{\partial}_{\alpha}u_{m}^\text{H}}
    \nonumber 
    \\
    &=& [V^\dagger \widetilde{\mathbb{B}}_{\alpha}^\text{W} V]_{nm}
    \nonumber 
    \\
    &=&
    [V^\dagger \mathbb{B}_{\alpha}^\text{W} V]_{nm} - \bar{\mathcal{E}}_n [V^\dagger \mathbb{A}_{\alpha}^\text{W} V]_{nm},
    \\[6pt]
    \widetilde{\mathbb{C}}_{\alpha\beta,nm}^\text{H} 
    &\equiv& 
    \braket{\widetilde{\partial}_{\alpha}u_{n}^\text{H}|\hat{H}| \widetilde{\partial}_{\beta}u_{m}^\text{H}} \nonumber 
    \\
    &=&
    [V^\dagger \widetilde{\mathbb{C}}_{\alpha\beta}^\text{W} V]_{nm} \nonumber 
    \\
    &=& [V^\dagger \mathbb{C}_{\alpha\beta}^\text{W} V]_{nm} 
    - [V^\dagger \mathbb{A}_\alpha^\text{W} V \cdot V^\dagger \mathbb{B}_\beta^\text{W} V]_{nm} \nonumber 
    \\
    & & - [V^\dagger (\mathbb{B}_\alpha^\text{W})^\dagger V \cdot V^\dagger \mathbb{A}_\beta^\text{W} V ]_{nm} \nonumber
    \\
    & &
    + \sum_{l\in\bar{\mathcal{H}}} [V^\dagger \mathbb{A}_\alpha^\text{W} V]_{nl} \bar{\mathcal{E}}_l [V^\dagger \mathbb{A}_\beta^\text{W} V]_{lm},
    \\[6pt]
    \widetilde{\mathbb{F}}_{\alpha\beta,nm}^\text{H} &\equiv & 
    \braket{\widetilde{\partial}_{\alpha}u_{n}^\text{H}| \widetilde{\partial}_{\beta}u_{m}^\text{H}} \nonumber 
    \\
    &=&
    [V^\dagger \widetilde{\mathbb{F}}_{\alpha\beta}^\text{W} V]_{nm} \nonumber
    \\
    &=&
    [V^\dagger {\mathbb{F}}_{\alpha\beta}^\text{W} V]_{nm}
    - [V^\dagger \mathbb{A}_\alpha^\text{W} V \cdot V^\dagger \mathbb{A}_\beta^\text{W} V]_{nm}, \nonumber 
    \\
\end{eqnarray}
\end{subequations}
where we have used Eq.~\eqref{eq:Hamiltonian_Hamiltonian_gauge}. Thus, geometry-related quantities in this form behave as other ordinary operators which do not contain $\mathbf{k}$-derivatives. Therefore, the above forms of the geometric quantities ensure that their information in the Wannier gauge is covariantly transformed into the Hamiltonian gauge, which is crucial in dealing with effective theories with a finite number of Wannier basis.

Using the Wannier-gauge objects, we can express matrix elements of the three objects $\hat{F}_{\alpha \beta}$, $\hat{G}_{\alpha \beta}$, and $\hat{K}_{\alpha \beta}$ in \eqref{eq:modern_FGK} regardless of the choice of a gauge as follows:
\begin{subequations}\label{eq:covariant_FGK}
\begin{eqnarray}
    \label{eq:covariant_F}
    \hat{\mathcal{F}}_{\alpha\beta}
    &=&
    \sum_{nm\in \bar{\mathcal{H}} }
    \ket{u_n}
    \left[
    f \left( \widetilde{\mathbb{F}}_{\alpha\beta} +  A_\alpha g A_\beta \right) f
    \right]_{nm}
    \bra{u_m},
    \\[6pt]
    \label{eq:covariant_G}
    \hat{\mathcal{G}}_{\alpha\beta}
    &=&
    \sum_{nm\in \bar{\mathcal{H}} }
    \ket{u_n}
    \Big[f \Big(
    \widetilde{\mathbb{C}}_{\alpha\beta} 
    +
    \widetilde{\mathbb{B}}_\alpha^\dagger g A_\beta
    +
    A_\alpha g \widetilde{\mathbb{B}}_\beta
    \nonumber
    \\
    & & \ \ \ \ \ \ \ \ \ \ \ \ \ \ \ \ \ \ \ 
    +
    A_\alpha g \mathbb{H} g A_\beta
    \Big) f \Big]_{nm}
    \bra{u_m},
    \\[6pt]
    \label{eq:covariant_K}
    \hat{\mathcal{K}}_{\alpha\beta}
    &=&
    \frac{1}{2}
    \sum_{nm\in \bar{\mathcal{H}} }
    \ket{u_n}
    \Big[ f \Big(
    \{
    \mathbb{H},\ \widetilde{\mathbb{F}}_{\alpha\beta} + A_\alpha g A_\beta
    \}
    \Big ) f \Big]_{nm}
    \bra{u_m},
    \nonumber 
    \\
\end{eqnarray}
\end{subequations}
where the objects in $[\cdots]$ are matrices defined within the reduced Hilbert space $\bar{\mathcal{H}}$, and we have defined $A_\alpha$ such that it coincides with the Berry connection in the Hamiltonian gauge [Eq.~\eqref{eq:A_Hamiltonian_gauge}], that is
\begin{subequations}\label{eq:Berry_connection}
\begin{eqnarray}
    A_{\alpha}^\text{H} &\equiv & \mathbb{A}_\alpha^\text{H}  = V^\dagger \mathbb{A}_\alpha^\text{W} V + J_\alpha^\text{H}, 
    \\ [6pt]
    A_{\alpha}^\text{W} &\equiv & V \mathbb{A}_\alpha^\text{H} V^\dagger =
    \mathbb{A}_\alpha^\text{W} + V J_\alpha^\text{H} V^\dagger.
\end{eqnarray}    
\end{subequations}
We remark that we have dropped the superscript `W' or `H' in the above expressions because the above expressions are valid in both gauges. Moreover, $\hat{\mathcal{F}}$, $\hat{\mathcal{G}}$, and $\hat{\mathcal{K}}$ are written only in terms of Bloch states defined within the reduced Hilbert space $\bar{\mathcal{H}}$, for which we use calligraphic font.
Equation~\eqref{eq:covariant_FGK} may be written in terms of quantum mechanical operators (with `hat' symbols)
\begin{subequations}
\begin{eqnarray}
\label{eq:hat_F}
    \hat{\mathcal{F}}_{\alpha\beta}
    &=&
    \hat{P}\hat{\widetilde{\mathbb{F}}}_{\alpha\beta}\hat{P} + \hat{P}\hat{A}_\alpha \hat{Q}^{\text{I}} \hat{A}_\beta \hat{P}, 
    \\
    \label{eq:hat_G}
    \hat{\mathcal{G}}_{\alpha\beta}
    &=&
    \hat{P}\hat{\widetilde{\mathbb{C}}}_{\alpha \beta}\hat{P} + \hat{P}\hat{\widetilde{B}}_{\alpha}^{\dagger}\hat{Q}^{\text{I}} \hat{A}_{\beta} \hat{P} + \hat{P}\hat{A}_{\alpha}\hat{Q}^{\text{I}} \hat{\widetilde{B}}_{\beta} \hat{P}
    \nonumber
    \\
    & & \ \
    + \hat{P}\hat{A}_{\alpha}\hat{Q}^{\text{I}} \hat{\mathbb{H}} \hat{Q}^{\text{I}} \hat{A}_{\beta} \hat{P},
    \\
    \label{eq:hat_K}
    \hat{\mathcal{K}}_{\alpha\beta}
    &=&
    \frac{1}{2}\left( \hat{P}\{ \hat{\mathbb{H}}, \hat{\widetilde{\mathbb{F}}}_{\alpha\beta}\} \hat{P} + \hat{P}\{ \hat{\mathbb{H}}, \hat{A}_\alpha \hat{Q}^{\text{I}} \hat{A}_\beta \} \hat{P} \right).
\end{eqnarray}
\end{subequations}

Therefore, based on these objects, $M_{\gamma}^\mathrm{SR}(\mathbf{k})$ and $M_{\gamma}^\mathrm{CM}(\mathbf{k})$ [Eq.~\eqref{eq:modern_wavepacket}] can be calculated as 
\begin{subequations}\label{eq:modern3}
\begin{eqnarray}
    \label{eq:modern_SR_covariant}
    M_{\gamma}^\mathrm{SR} (\mathbf{k}) &=& \frac{e\varepsilon_{\alpha \beta \gamma}}{2\hbar} \text{Im} 
    \left[ \text{tr}\left( 
    \hat{\mathcal{G}}_{\mathbf{k}, \alpha \beta} - \hat{\mathcal{K}}_{\mathbf{k}, \alpha \beta}
    \right) \right] ,
    \\ [6pt]
    \label{eq:modern_CM_covariant}
    M_{\gamma}^\mathrm{CM} (\mathbf{k}) &=& \frac{e\varepsilon_{\alpha \beta \gamma}}{\hbar} \text{Im} 
    \left[ \text{tr}\left(
    \hat{\mathcal{K}}_{\mathbf{k}, \alpha \beta} - \mathcal{E}_{\text{F}}\hat{\mathcal{F}}_{\mathbf{k}, \alpha \beta} 
    \right) \right],
\end{eqnarray}
\end{subequations}
respectively. Here, `tr' denotes the trace over $\bar{\mathcal{H}}_{\mathbf{k}}$, which is in contrast to `Tr' which denotes the trace over the entire Hilbert space $\mathcal{H}_{\mathbf{k}}$. The LC and IC contributions can be constructed from the SR and CM contributions by Eq.~\eqref{eq:modern_wannier}. 
In our work, we employ the Hamiltonian gauge representation and numerically implement the following equations explicitly:
\begin{subequations}\label{eq:modern_covariant_Hamiltonian_gauge}
\begin{eqnarray}
    \label{eq:modern_covariant_Hamiltonian_gauge_SR}
    M_\gamma^\text{SR} (\mathbf{k})
    &=&
    \frac{e\varepsilon_{\alpha\beta\gamma}}{2\hbar}
    \sum_{n\in \bar{\mathcal{H}}}
    f_n^\text{H}
    \text{Im}
    \Big[ 
    \widetilde{\mathbb{C}}_{\alpha\beta,nn}^\text{H}
    -
    \mathcal{E}_n \widetilde{\mathbb{F}}_{\alpha\beta,nn}^\text{H}
    \nonumber 
    \\
    & & 
    + 
    \sum_{l\in \bar{\mathcal{H}}}
    \Big\{
    (\widetilde{\mathbb{B}}_{\alpha}^\text{H})^\dagger_{nl} g_l^\text{H} A_{\beta,ln}^\text{H}
    +
    A_{\alpha,nl}^\text{H} g_l^\text{H} \widetilde{\mathbb{B}}_{\beta,ln}^\text{H}
    \nonumber 
    \\
    & & \ \ \ \ \ \ \ 
    + 
    (\bar{\mathcal{E}}_l - \bar{\mathcal{E}}_n)
    A_{\alpha,nl}^\text{H} g_l^\text{H}  A_{\beta,ln}^\text{H}
    \Big\}
    \Big ],
    \\[6pt]
    \label{eq:modern_covariant_Hamiltonian_gauge_CM}
    M_\gamma^\text{CM} (\mathbf{k})
    &=&
    \frac{e\varepsilon_{\alpha\beta\gamma}}{\hbar}
    \sum_{n\in \bar{\mathcal{H}}}
    f_n^\text{H} (\bar{\mathcal{E}}_n - \mathcal{E}_\text{F}) 
    \nonumber 
    \\
    & & 
    \times 
    \text{Im}
    \Big[ 
    \widetilde{\mathbb{F}}_{\alpha\beta,nn}^\text{H}
    +
    \sum_{l\in \bar{\mathcal{H}}}
    A_{\alpha,nl}^\text{H} g_l^\text{H} A_{\beta,ln}^\text{H} 
    \Big].
\end{eqnarray}
\end{subequations}

This covariant form of the orbital magnetization was first derived by Lopez {\it et al.}~\cite{Lopez12PRB}. In Appendix~\ref{appendix:gauge_selection}, we explicitly show that the resulting orbital magnetization is independent of the choice of the Wannier functions, the gauge transformation $V(k)$ between the Hamiltonian gauge and the Wannier gauge, and any gauge transformation $U(\mathbf{k})$ between different Hamiltonian-gauge Bloch states, such that $[U,\mathbb{H}] = 0$ as long as the inner window is set above $\mathcal{E}_{\text{F}}$.

\subsection{Constructing orbital moment operator}
\label{subsec:orbital_moment_operator}

Equation~\eqref{eq:modern3} is a valid equation for describing the orbital moment in the {\it ground state}. However, quite often, one needs to analyze the orbital moment of excited states above the Fermi energy, or generally band-resolved orbital moment. With this motivation, we construct a quantum mechanical orbital moment operator $\hat{\mathbf{M}}^\text{SR/CM}$ (with `hat' symbol) such that Eq.~\eqref{eq:modern_covariant_Hamiltonian_gauge} is recovered by
\begin{equation}\label{eq:expectation_value}
    \mathbf{M}^\text{SR/CM} (\mathbf{k})
    =
    \text{tr}
    \left[ 
    \hat{P}_\mathbf{k} 
    \hat{\mathbf{M}}^\text{SR/CM} (\mathbf{k})
    \right].
\end{equation}

For this, we first define {\it occupation-weighted covariant derivative} for a Bloch state in the Hamiltonian gauge,
\begin{eqnarray}\label{eq:occupation-weighted}
    \ket{\bar{D}_\mu u_n^\text{H}}
    &\equiv &
    \hat{Q} \ket{\partial_\mu u_n^\text{H}} f_n^\text{H}
    +
    (\hat{\mathbb{1}}-\hat{Q}^\mathrm{I}) \ket{\partial_\mu u_n^\text{H}} g_n^\text{H}
    \nonumber
    \\
    &=&
    (\hat{\mathbb{Q}} + \hat{Q}^\mathrm{I}) \ket{\partial_\mu u_n^\text{H}} f_n^\text{H}
    +
    (\hat{\mathbb{Q}}+\hat{P}) \ket{\partial_\mu u_n^\text{H}} g_n^\text{H}.
    \nonumber 
    \\
    &=& \ket{\partial_\mu u_n^\text{H}}
    +
    i\sum_{m\in \bar{\mathcal{H}}} \ket{u_m^\text{H}}
    (
    f_m^\text{H} A_{mn}^\text{H} f_n^\text{H} +
    g_m^\text{H} A_{mn}^\text{H} g_n^\text{H}
    ).
    \nonumber 
    \\
\end{eqnarray}
Note that it is defined such that the occupation functions $f$ and $g$ are combined with their complementary projections, $\hat{Q}=\hat{\mathbb{Q}} + \hat{Q}^\mathrm{I}$ and $\hat{\mathbb{1}}-\hat{Q}^\mathrm{I}=\hat{\mathbb{Q}}+\hat{P}$, respectively. This enables us to define the orbital moment operators within the reduced Hilbert space $\bar{\mathcal{H}}$ as
\begin{widetext}
\begin{subequations}\label{eq:modern_operator_covariant_derivative}
\begin{eqnarray}
    \label{eq:modern_operator_covariant_derivative_SR}
    \hat{M}_\gamma^\text{SR}
    &=&
    \frac{ie\varepsilon_{\alpha\beta\gamma}}{2\hbar}
    \sum_{nm\in\bar{\mathcal{H}}}
    \ket{u_n^\text{H}}
    \bra{\bar{D}_\alpha u_n^\text{H}}
    \left( 
    \hat{H} -
    \frac{\bar{\mathcal{E}}_n + \bar{\mathcal{E}}_m}{2}
    \right)
    \ket{\bar{D}_\beta u_m^\text{H}}
    \bra{u_m^\text{H}},
    \\
    \label{eq:modern_operator_covariant_derivative_CM}
    \hat{M}_\gamma^\text{CM}
    &=&
    \frac{ie\varepsilon_{\alpha\beta\gamma}}{\hbar}
    \sum_{nm\in\bar{\mathcal{H}}}
    \ket{u_n^\text{H}}
    \bra{\bar{D}_\alpha u_n^\text{H}}
    \left( 
    \frac{\bar{\mathcal{E}}_n + \bar{\mathcal{E}}_m}{2}
    -
    \mathcal{E}_\text{F}
    \right)
    \ket{\bar{D}_\beta u_m^\text{H}}
    \bra{u_m^\text{H}}.
\end{eqnarray}
\end{subequations}
Explicit calculation from the definitions in the previous subsections results in
\begin{subequations}\label{eq:modern_operator_braket}
\begin{eqnarray}
    \label{eq:modern_operator_braket_SR}
    \hat{M}_{\gamma}^\text{SR}
    &=&
    \frac{ie\varepsilon_{\alpha\beta\gamma}}{2\hbar}
    \sum_{n,m\in\bar{\mathcal{H}}}
    \ket{u_n}
    \Bigg[
    \widetilde{\mathbb{C}}_{\alpha\beta} 
    -
    \frac{1}{2}
    \{
    \mathbb{H},\ \widetilde{\mathbb{F}}_{\alpha\beta}
    \}
    +
    f \Big(
    \widetilde{\mathbb{B}}_\alpha^\dagger g A_\beta
    +
    A_\alpha g \widetilde{\mathbb{B}}_\beta + A_\alpha g \mathbb{H} A_\beta  
    \Big) f 
    -
    \frac{1}{2}
    f 
    \{
    \mathbb{H},\  A_\alpha g A_\beta
    \}
    f \nonumber 
    \\
    & & \ \ \ \ \ \ \ \ \ \ \ \ \ \ \ \ \ \ \ \ \ \ \ 
    +
    g \Big(
    \widetilde{\mathbb{B}}_\alpha^\dagger f A_\beta
    +
    A_\alpha g \widetilde{\mathbb{B}}_\beta + A_\alpha f \mathbb{H} A_\beta  
    \Big) g
    -
    \frac{1}{2}
    g 
    \{
    \mathbb{H},\ A_\alpha f A_\beta
    \}
    g 
    \Bigg]_{nm} \bra{u_m},
    \\[6pt]
    \label{eq:modern_operator_braket_CM}
    \hat{M}_{\gamma}^\text{CM}
    &=&
    \frac{ie\varepsilon_{\alpha\beta\gamma}}{\hbar}
    \sum_{n,m\in\bar{\mathcal{H}}}
    \ket{u_n}
    \Bigg[ 
    \frac{1}{2}
    \{
    \mathbb{H} - \mathcal{E}_\text{F} \mathbb{P},\ \widetilde{\mathbb{F}}_{\alpha\beta}
    \}
    +
    \frac{1}{2}
    f
    \{
    \mathbb{H} - \mathcal{E}_\text{F} \mathbb{P},\ A_\alpha g A_\beta
    \}
     f 
    +
    \frac{1}{2}
    g
    \{
    \mathbb{H}-\mathcal{E}_\text{F} \mathbb{P},\ A_\alpha f A_\beta
    \}
     g 
    \Bigg]_{nm}
    \bra{u_m}, \nonumber 
    \\
\end{eqnarray}
\end{subequations}
or alternatively,
\begin{subequations}\label{eq:modern_operator_abstract}
\begin{eqnarray}
    \label{eq:modern_operator_abstract_SR}
    \hat{M}_{\gamma}^\text{SR}
    &=&
    \frac{ie\varepsilon_{\alpha\beta\gamma}}{2\hbar}
    \Bigg[
    \hat{\widetilde{\mathbb{C}}}_{\alpha \beta} 
    -
    \{ \hat{\mathbb{H}}, \hat{\widetilde{\mathbb{F}}}_{\alpha\beta}\}
    +
    \hat{P}\hat{\widetilde{B}}_{\alpha}^{\dagger}\hat{Q}^{\text{I}} \hat{A}_{\beta} \hat{P} 
    + \hat{P}\hat{A}_{\alpha}\hat{Q}^{\text{I}} \hat{\widetilde{B}}_{\beta} \hat{P}
    + \hat{P}\hat{A}_{\alpha}\hat{Q}^{\text{I}} \hat{\mathbb{H}} \hat{Q}^{\text{I}} \hat{A}_{\beta} \hat{P}
    -
    \frac{1}{2} \hat{P}\{ \hat{\mathbb{H}},\ \hat{A}_\alpha \hat{Q}^{\text{I}} \hat{A}_\beta \} \hat{P}
    \nonumber 
    \\
    & & \ \ \ \ \ \ \ \ \ 
    +
    \hat{Q}^\text{I}\hat{\widetilde{B}}_{\alpha}^{\dagger}\hat{P} \hat{A}_{\beta} \hat{Q}^\text{I} 
    + \hat{Q}^\text{I}\hat{A}_{\alpha}\hat{P} \hat{\widetilde{B}}_{\beta} \hat{Q}^\text{I}
    + \hat{Q}^\text{I}\hat{A}_{\alpha}\hat{P} \hat{\mathbb{H}} \hat{Q}^{\text{I}} \hat{A}_{\beta} \hat{Q}^\text{I}
    -
    \frac{1}{2} \hat{Q}^\text{I}\{ \hat{\mathbb{H}},\ \hat{A}_\alpha \hat{P} \hat{A}_\beta \} \hat{Q}^\text{I}
    \Bigg],
    \\[6pt]
    \label{eq:modern_operator_abstract_CM}
    \hat{M}_{\gamma}^\text{CM}
    &=&
    \frac{ie\varepsilon_{\alpha\beta\gamma}}{\hbar}
    \Bigg[ 
    \frac{1}{2} \{ \hat{\mathbb{H}}-\mathcal{E}_\text{F} \hat{\mathbb{P}},\ \hat{\widetilde{\mathbb{F}}}_{\alpha\beta}\}
    +
    \frac{1}{2} \hat{P}\{ \hat{\mathbb{H}} - \mathcal{E}_\text{F} \hat{\mathbb{P}}, \hat{A}_\alpha \hat{Q}^{\text{I}} \hat{A}_\beta \} \hat{P}
    +\frac{1}{2}
    \hat{Q}^\text{I}\{ \hat{\mathbb{H}} - \mathcal{E}_\text{F} \hat{\mathbb{P}}, \ \hat{\widetilde{\mathbb{F}}}_{\alpha\beta}\} \hat{Q}^\text{I}
    \Bigg].
\end{eqnarray}
\end{subequations}
\end{widetext}
in operator form. We emphasize that the definition above is independent of the choice of a gauge, so we have dropped the superscript `H' or `W' in Eq.~\eqref{eq:modern_operator_braket}. In the Hamiltonian gauge, nonetheless, Eq.~\eqref{eq:modern_operator_braket} recovers the result of the ground state orbital moment, Eq.~\eqref{eq:modern_covariant_Hamiltonian_gauge} by taking the trace with the density matrix [Eq.~\eqref{eq:expectation_value}]. The above Eqs.~\eqref{eq:modern_operator_covariant_derivative}-\eqref{eq:modern_operator_abstract} are one of the main results of this paper.

The band-resolved orbital moment in the Hamiltonian gauge is thus given by
\begin{widetext}
\begin{subequations}\label{eq:modern_band_resolved}
\begin{eqnarray}
    \label{eq:modern_band_resolved_SR}
    \bra{u_n^\text{H}} \hat{M}_\gamma^\text{SR} \ket{u_n^\text{H}}
    &=&
    \frac{e\varepsilon_{\alpha\beta\gamma}}{2\hbar}
    \text{Im}
    \Bigg[
    \widetilde{\mathbb{C}}_{\alpha\beta,nn}^\text{H}
    -
    \bar{\mathcal{E}}_n
    \widetilde{\mathbb{F}}_{\alpha\beta,nn}^\text{H} \nonumber 
    \\
    & & 
    +
    \sum_{l\in\bar{\mathcal{H}}}
    \Big\{
    f_n^\text{H} \Big(
    (\widetilde{\mathbb{B}}_{\alpha,nl}^\text{H})^\dagger g_l^\text{H} A_{\beta,ln}^\text{H}
    +
    A_{\alpha,nl}^\text{H} g_l^\text{H} \widetilde{\mathbb{B}}_{\beta,ln}^\text{H} + A_{\alpha,nl}^\text{H} g_l^\text{H} \mathcal{E}_l A_{\beta,ln}^\text{H}  
    \Big) 
    -
    f_n^\text{H} \bar{\mathcal{E}}_n
    A_{\alpha,nl}^\text{H} g_l A_{\beta,ln}^\text{H} \nonumber 
    \\
    & & \ \ \ \ \ \ \ \ 
    +
    g_n^\text{H} \Big(
    (\widetilde{\mathbb{B}}_{\alpha,nl}^\text{H})^\dagger f_l^\text{H} A_{\beta,ln}^\text{H}
    +
    A_{\alpha,nl}^\text{H} f_l^\text{H} \widetilde{\mathbb{B}}_{\beta,ln}^\text{H} + A_{\alpha,nl}^\text{H} f_l^\text{H} \bar{\mathcal{E}}_l A_{\beta,ln}^\text{H}  
    \Big) 
    -
    g_n^\text{H} \bar{\mathcal{E}}_n
    A_{\alpha,nl}^\text{H} f_l^\text{H} A_{\beta,ln}^\text{H}
    \Big\}
    \Bigg],
    \\[6pt]
    \label{eq:modern_band_resolved_CM}
    \bra{u_n^\text{H}} \hat{M}_\gamma^\text{CM} \ket{u_n^\text{H}}
    &=&
    \frac{e\varepsilon_{\alpha\beta\gamma}}{\hbar}
    (\bar{\mathcal{E}}_n - \mathcal{E}_\text{F})
    \text{Im}
    \Bigg[
    \widetilde{\mathbb{F}}_{\alpha\beta,nn}^\text{H}
    +
    \sum_{l\in\bar{\mathcal{H}}}
    \left( 
    f_n^\text{H} A_{\alpha,nl}^\text{H} g_l^\text{H} A_{\beta,ln}^\text{H} + g_n^\text{H} A_{\alpha,nl}^\text{H} f_l^\text{H} A_{\beta,ln}^\text{H}
    \right)
    \Bigg].
\end{eqnarray}
\end{subequations}
\end{widetext}
Note that these expressions agree with the total orbital moment [Eq.~\eqref{eq:modern_covariant_Hamiltonian_gauge}] such that
\begin{equation}\label{eq:band_sum}
    M_\gamma^\text{SR/CM}
    =
    \sum_n f_n \bra{u_n^\text{H}} \hat{M}_\gamma^\text{SR/CM} \ket{u_n^\text{H}}.
\end{equation}

We emphasize that the occupation functions $f$ and $g$ are necessary for the consistency of the band summation of each SR/CM contributions, Eq.~\eqref{eq:band_sum}. Otherwise, the result obtained by summing over individual bands does not agree with the total. Interestingly, for the total orbital moment $M_\gamma = M_\gamma^\text{SR} + M_\gamma^\text{CM}$, the corresponding operator does not need to include occupation functions. That is, the total orbital moment operator can be defined simply as
\begin{align}\label{eq:orbital_moment_operator_total}
    \hat{M}_{\gamma}(\mathbf{k}) = \frac{ie\varepsilon_{\alpha \beta \gamma}}{2\hbar} &\left[ \widetilde{\hat{\mathbb{C}}}_{\alpha \beta} + \frac{1}{2} \{\hat{\mathbb{H}}, \hat{\widetilde{\mathbb{F}}}_{\alpha \beta} \} - 2\mathcal{E}_{\text{F}} \hat{\widetilde{\mathbb{F}}}_{\alpha \beta} 
    \right. \notag
    \\
    &\left. 
    + \hat{\widetilde{\mathbb{B}}}_{\alpha}^{\dagger} \hat{A}_{\beta} 
    + \hat{A}_{\alpha} \hat{\widetilde{\mathbb{B}}}_{\beta}
    + \hat{A}_{\alpha} \hat{\mathbb{H}} \hat{A}_{\beta}  \right. \notag
    \\
    &\left. 
    + \frac{1}{2} \{ \hat{\mathbb{H}},\ \hat{A}_{\alpha} \hat{A}_{\beta} \}
    -2 \mathcal{E}_{\text{F}} \hat{A}_{\alpha} \hat{A}_{\beta} \right],
\end{align}
which still satisfy the band summation consistency,
\begin{equation}\label{eq:summation_total}
    M_\gamma
    =
    \sum_n f_n \bra{u_n^\text{H}} \hat{M}_\gamma \ket{u_n^\text{H}}.
\end{equation}
A derivation of Eq.~\eqref{eq:orbital_moment_operator_total} is provided in Appendix~\ref{appendix:linear_forms}. This expression enables more efficient calculations of orbital magnetization by reducing the computational cost. We emphasize, however, that it guarantees only that the trace remains unchanged, while the off-diagonal components vary. Therefore, in our work, we do not employ Eq.~\eqref{eq:orbital_moment_operator_total} but Eqs.~\eqref{eq:modern_operator_covariant_derivative}-\eqref{eq:modern_band_resolved}.

Finally, we comment on the validity of the operators introduced in Eqs.~\eqref{eq:modern_operator_covariant_derivative}-\eqref{eq:modern_band_resolved}. These operators--obtained by promoting Eq.~\eqref{eq:modern_wavepacket} to operator form while preserving gauge covariance--reproduce the orbital magnetization exactly and produce a gauge-independent, consistent band-resolved orbital moment. When off-diagonal matrix elements are essential, as for the orbital current, there remains an ongoing debate over what constitutes an appropriate operator for OAM~\cite{Pezo22PRB,Busch23PRR,Culcer25PRL,Cysne26arXiv}. We address this issue deeply in Appendix.~\ref{appendix:operator_discussion}. We expect that this discussion will provide useful guidance for treating quantities such as the orbital Hall conductivity and orbital-related susceptibilities.

\subsection{Term-by-term analysis: $J$-decomposition}
\label{subsec:term-by-term_analysis}

The results in Eqs.~\eqref{eq:modern_covariant_Hamiltonian_gauge}, \eqref{eq:modern_operator_braket}, and \eqref{eq:modern_band_resolved} contain terms in different powers of $J_\alpha$, which is in the definition of $A_\alpha$ [Eq.~\eqref{eq:Berry_connection}] and captures the contribution due to coherent band hybridization, as seen from Eq.~\eqref{eq:J_Hamiltonian_gauge}. Thus, we expand $M_{\gamma}$ in powers of $J_{\alpha}$ as
\begin{align}\label{eq:J-decomp}
    M_{\gamma} &= M_{\gamma}^{(0)} + M_{\gamma}^{(1)} + M_{\gamma}^{(2)},
\end{align}
where 
\begin{equation}\label{eq:J-decomp2}
    M_{\gamma}^{(i)} \propto (J_{\alpha})^{i}, \qquad i =0,1,2.
\end{equation}
Similarly, both $M^{\text{SR}}_{\gamma}$ and $M^{\text{CM}}_{\gamma}$ can also be expanded in powers of $J_{\alpha}$. This decomposition was first introduced in Ref.~\cite{Lopez12PRB} to analyze the numerical stability of each term in computing the anomalous Hall conductivity. However, since this decomposition depends on the choice of the Wannier basis, no attempt was made to extract physical meaning from it beyond the numerical perspective. Nevertheless, the decomposition can be employed to analyze how the ACA and the standard tight-binding method are related to the total orbital magnetization, thereby deepening our understanding of the modern theory of orbital magnetism.

If the electronic wave functions are well localized, then they are already smooth in $\mathbf{k}$-space. This implies that the basis variation is dominant, whereas $J_{\alpha}$ is relatively small. Therefore, $M_{\gamma}$ can be approximately described by $M_{\gamma}^{(0)}$. In contrast, if the electronic wave functions are delocalized, then the mixing is required to construct smooth Wannier-gauge states. This implies that the variation of the coefficients becomes dominant and $M_{\gamma}^{(2)}$ provides a large portion of $M_{\gamma}$.

Typical tight-binding calculations compute physical quantities solely on the basis of an effective Hamiltonian, e.g. obtained by Wannierization. In practice, the derivative of the Bloch state is commonly handled through perturbation theory and thus captures only $J$ term in Eq.~\eqref{eq:Bloch_Hamiltonian_gauge_derivative}. This approach works only by expanding the entire Hilbert space, so in the effective model within the reduced Hilbert space $\bar{\mathcal{H}}$, this suffers from the truncation problem. Furthermore, it uses the interpolated energy $\bar{\mathcal{E}}_{n}$ rather than the ab initio energy ${\mathcal{E}}_{n}$, which usually deviate for high-energy states. Importantly, this naive approach neglects the $\mathbf{k}$-dependence of the basis, the effect of the anomalous position~\cite{Gobel24PRL}. These are captured by the gauge-objects $\mathbb{A}$, $\mathbb{B}$, and $\mathbb{C}$, as we have introduced in Eq.~\eqref{eq:gauge_ABC}. This discards the information on $\mathbb{A}_{\alpha}$, $\mathbb{B}_{\alpha}$, and $\mathbb{C}_{\alpha \beta}$, while capturing only that on $J_{\alpha}$. Consequently, the results obtained from naively employing the $\mathbf{k}\cdot \mathbf{p}$ perturbation theory or the standard tight-binding calculations are equal to the result obtained by calculating only $M^{(2)}_{\gamma}$. This also implies that calculations performed on effective models capture only part of the contribution to the total orbital magnetization, $M_\gamma^\text{(2)}$, which completely misses the ACA contribution.

The $J$-decomposition also enables us to clarify the relationship between $M_{\gamma}$ and the orbital magnetization evaluated by the ACA, $M_{\gamma}^{\text{ACA}}$. Assume that the Wannier functions are chosen to closely resemble atomic orbitals and that the intercell contribution is negligible. Under this assumption, the orbital magnetization calculated solely from $\mathbb{A}_{\alpha}$, $\mathbb{B}_{\alpha}$, and $\mathbb{C}_{\alpha \beta}$ reflects the contribution of the atomic-orbital basis to $M_{\gamma}$. To extract only the self-rotation part from this contribution, we define $M^{\text{SR(on)}}_{\gamma}$ by computing $M^{\text{SR}(0)}_{\gamma}$ (note that $M^{(0)}_{\gamma} = M^{\text{SR}(0)}_{\gamma} + M^{\text{CM}(0)}_{\gamma}$) from Eq.~\eqref{eq:gauge_ABC_Wannier} while including only the intracell contribution $\braket{\mathbf{0}|[\cdot]|\mathbf{0}}$. The ACA, which captures the self-rotation contribution of atomic orbitals within the MT sphere, corresponds to $M^{\text{SR(on)}}_{\gamma}$ within the MT sphere. Therefore, if the electronic wave functions are well localized around the atomic nuclei and MT is sufficiently large, $M^{\text{ACA}}_{\gamma}$ provides a value close to $M^{\text{SR(on)}}_{\gamma}$. In contrast, if the wave functions are delocalized in the interstitial regions or MT is set to be too small, then $M^{\text{ACA}}_{\gamma}$ captures only a fraction of $M^{\text{SR(on)}}_{\gamma}$. A comparison between $M_{z}^{\text{ACA}}$ and $M_{\gamma}^{\text{SR(on)}}$ is presented in Appendix~\ref{appendix:SR_onsite}.

In the following section, we apply the term-by-term analysis to various classes of real materials. To examine the contributions of the ACA and the standard tight-binding method through the decomposition of Eq.~\eqref{eq:J-decomp}, we set the Wannier basis close to atomic orbitals. The computational details are provided in Appendix~\ref{appendix:computational_details}.

\section{First-principles calculation}\label{sec:first-principles_calculation}

\subsection{Computational methods}\label{subsec:computation_method_outline}

\begin{figure*}[t]
\includegraphics[width=500pt]{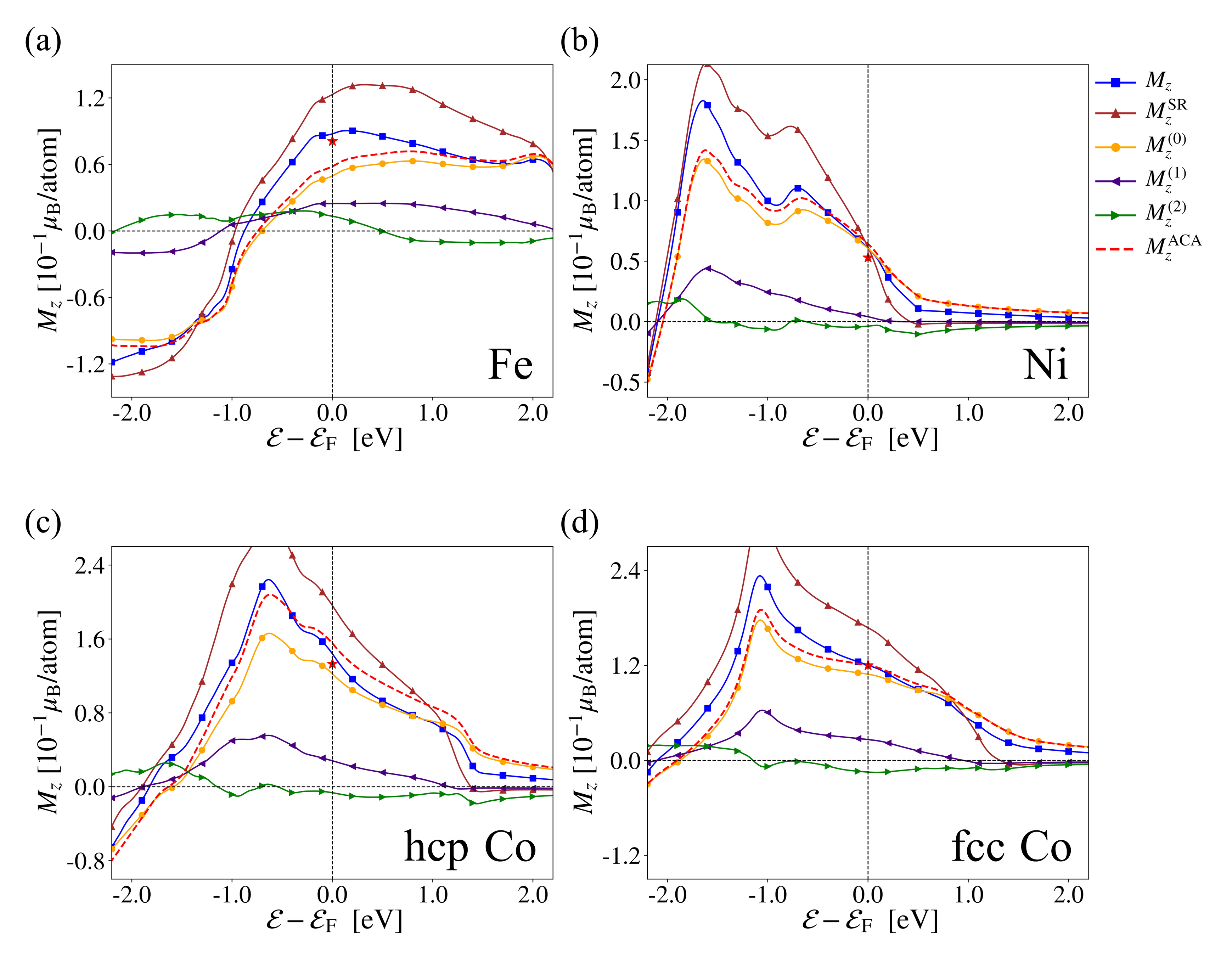}
\caption{\label{fig4}\textbf{Modern theory and ACA of orbital magnetization in ferromagnetic $d$-transition metals with DFT+U.} Comparison between the modern theory and ACA for (a) bcc Fe with spin-quantization axis [001], (b) fcc Ni [111], (c) hcp Co [0001], and (d) fcc Co [111]. The energy dependence of the modern theory orbital magnetization $M_{z}$ (blue), $M^{(0)}_{z}$ (yellow), $M^{(1)}_{z}$ (purple), $M^{(2)}_{z}$ (green), self-rotation contribution $M^{\text{SR}}_{z}$ (brown line), and ACA $M^{\text{ACA}}_{z}$ (red dashed line) are shown. Experimental values of the orbital magnetization at $\mathcal{E}_{\text{F}}$~\cite{Ceresoli10PRB} are presented by red stars.}
\end{figure*}

\begin{figure*}[t]
\includegraphics[width=500pt]{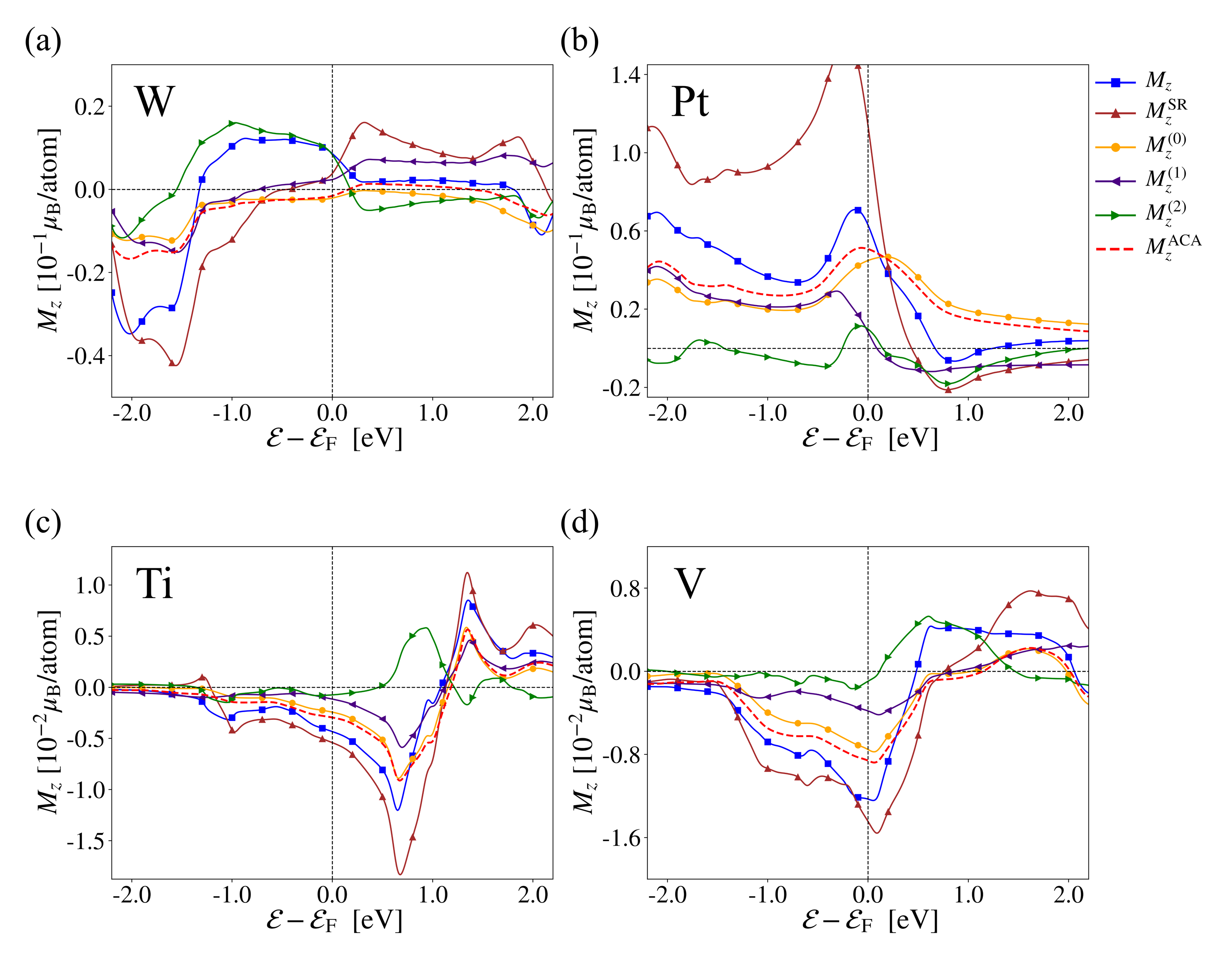}
\caption{\label{fig5}\textbf{Modern theory and ACA for orbital magnetization in various $d$-transition metals.} Comparison between the modern theory and ACA for (a) magnetic bcc W with spin-quantization axis [001], (b) fcc Pt [111], (c) fcc Ti [111], and (d) bcc V [001]. The energy dependence of $M_{z}$ (blue), $M^{(0)}_{z}$ (yellow), $M^{(1)}_{z}$ (purple), $M^{(2)}_{z}$ (green), $M^{\text{SR}}_{z}$ (brown), and $M^{\text{ACA}}_{z}$ (red dashed line) are shown.}
\end{figure*}

The electronic structures of real materials are calculated using the \textsc{Fleur} package~\cite{FLEUR} that implements the full-potential linearized augmented plane wave (FLAPW) method~\cite{Wimmer81PRB} of density functional theory (DFT). We use the Perdew-Burke-Ernzerhof exchange-correlation functional within the generalized gradient approximation~\cite{Perdew96PRL}. The BZ is sampled using the Monkhorst-Pack $\mathbf{k}$-mesh~\cite{Monkhorst76PRB}. The SOC is included through the second variation scheme. For the case of Fe, Ni, Co, and $T_{d}$-WTe$_{2}$ monolayer, we applied the DFT+U method within the self-consistent DFT cycle~\cite{Shick99PRB}. The ACA OAM is calculated by Eq.~\eqref{eq:ACA}. We transform the Bloch states obtained from the DFT calculation into Wannier states using the \textsc{Wannier90} package~\cite{Pizzi20PCM}. Then, the reduced Hamiltonian and the other operators are transformed to the basis of Wannier functions to obtain the tight-binding model~\cite{Freimuth08PRB}. The Wannier-gauge objects $\mathbb{A}_{\alpha}$, $\mathbb{B}_{\alpha}$, $\mathbb{C}_{\alpha\beta}$ and $\mathbb{F}_{\alpha\beta}$ are calculated following the approach described in Refs.~\cite{Lopez12PRB} and \cite{Urru25PRB}. Using the Wannier-gauge objects, reduced Hamiltonian, and ACA OAM operator, both the ACA [Eq.~\eqref{eq:ACA-sum}] and modern theory orbital magnetization [Eq.~\eqref{eq:band_sum}] is calculated by employing the \textsc{Orbitrans} code~\cite{Orbitrans}. The BZ integration is carried out on the Monkhorst-Pack $\mathbf{k}$-mesh~\cite{Monkhorst76PRB}. The computational details, including the parameters used in the material calculations, are provided in the Appendix~\ref{appendix:computational_details}. The calculation results are summarized in Table~\ref{table:1} and illustrated in Figs.~\ref{fig4}-\ref{fig10}.

\subsection{\textit{d}-transition metals}\label{subsec:d_transition_metals}

In $d$-transition metals, orbital magnetism mainly originates from electrons partially filling the $d$ orbitals. Compared to electrons in the $s$ or $p$ orbitals, $d$ electrons tend to be spatially strongly localized around the nucleus and exhibit weaker overlap with the orbital of neighboring atoms. As a result, in these elements the self-rotation contribution is generally more important than the CM contribution. In the case of the \textit{J}-decomposition, $M^{(0)}_{\gamma}$ accounts for most of $M_{\gamma}$, while the contributions of $M^{(1)}$ and $M^{(2)}$ are insignificant. These imply that the ACA largely reproduces the total orbital magnetization, whereas the standard tight-binding method captures only a small fraction of it.

We begin with the single-element bulk ferromagnets Fe, Co, and Ni. The results for bcc Fe, fcc Co, hcp Co, and fcc Ni are presented in Fig. \ref{fig4}. We obtain $M_{z} = 0.06653 \, \mu_{B}/\text{atom}$ and $M_{z}^{\text{ACA}} = 0.04482 \, \mu_{B}/\text{atom}$ for bcc Fe [001]; $M_{z} = 0.07500 \, \mu_{B}/\text{atom}$ and $M_{z}^{\text{ACA}} = 0.07913 \, \mu_{B}/\text{atom}$ for hcp Co [0001]; $M_{z} = 0.07329 \, \mu_{B}/\text{atom}$ and $M_{z}^{\text{ACA}} = 0.06884 \, \mu_{B}/\text{atom}$ for fcc Co [111]; and $M_{z} = 0.04196 \, \mu_{B}/\text{atom}$ and $M_{z}^{\text{ACA}} = 0.04262 \, \mu_{B}/\text{atom}$ for Ni [111] at $\mathcal{E}_{\text{F}}$. Here, the braket $[\cdots]$ indicates the orientation of the spin-quantization axis. These results are consistent with previous first-principles calculation~\cite{Ceresoli10PRB} and Wannier interpolation~\cite{Lopez12PRB,Hanke16PRB} studies.

We include the Hubbard $U$ and $J$ parameters in the DFT calculation to account for electron-electron correlations. With the $U$ and $J$ corrections, we obtain values in closer agreement with the experimental data [red stars in Figs.~\ref{fig4}(a)-\ref{fig4}(c)]: the calculated values are $M_{z} = 0.08760 \, \mu_{B}/\text{atom}$ and $M_{z}^{\text{ACA}} = 0.06817 \, \mu_{B}/\text{atom}$ for bcc Fe [001]; $M_{z} = 0.1440 \, \mu_{B}/\text{atom}$ and $M_{z}^{\text{ACA}} = 0.1552 \, \mu_{B}/\text{atom}$ for hcp Co [0001]; $M_{z} = 0.1203 \, \mu_{B}/\text{atom}$ and $M_{z}^{\text{ACA}} = 0.120 \, \mu_{B}/\text{atom}$ for fcc Co [111]; and $M_{z} = 0.06081 \, \mu_{B}/\text{atom}$ and $M_{z}^{\text{ACA}} = 0.06435 \, \mu_{B}/\text{atom}$ for Ni [111] at $\mathcal{E}_{\text{F}}$; the experimental values are $M_{z} = 0.081 \, \mu_{B}/\text{atom}$ for bcc Fe [001]; $M_{z} = 0.133 \, \mu_{B}/\text{atom}$ for hcp Co [0001]; $M_{z} = 0.120 \, \mu_{B}/\text{atom}$ for fcc Co [111]; and $M_{z} = 0.053 \, \mu_{B}/\text{atom}$ for Ni [111]~\cite{Meyer61JAP}. 

For all three materials, modern theory and ACA show good agreement: $M_{z}^{\text{ACA}}$ reproduces both the value of $M_{z}$ at $\mathcal{E}_{\text{F}}$ and its overall trend with respect to $\mathcal{E}$. In particular, the discrepancies between $M_{z}$ and $M_{z}^{\text{ACA}}$ at $\mathcal{E}_{\text{F}}$--defined as $\vert (M_{z} - M_{z}^{\text{ACA}})/M_{z} \vert$--are 8\% in hcp Co (0.2\% in fcc Co) and 6\% in Ni. The discrepancy is largest in Fe and smallest in Ni, reflecting the fact that the wave functions of the electron are the least localized in Fe and the most localized in Ni (see Fig.~\ref{fig7} for detailed discussion). Among the terms decomposed in powers of $J$, $M_{z}^{(0)}$ shows a magnitude and $\mathcal{E}$ dependence similar to those of $M_{z}^{\text{ACA}}$. On the other hand, $M^{(2)}_{z}$ differs from $M_{z}$ not only in its value at $\mathcal{E}_{\text{F}}$ but also in its overall trend with respect to $\mathcal{E}$. It is found to have a smaller value than both $M_{z}$ and $M_{z}^{\text{ACA}}$, and in the case of Co, it even exhibits the opposite sign. This behavior implies that the `naive' consideration of only $J$ terms in tight-binding methods fails to describe the modern theory orbital magnetization in these materials. For all three materials, $M^{\text{SR}}_{z}$ shows a similar $\mathcal{E}$-dependence to $M_{z}$ but with larger values.

To explore this trend not only in $3d$ but also in $4d$ and $5d$ compounds, which do not have spontaneous magnetization, we apply an external exchange field on the spin and obtain the self-consistent electronic structure in full-relativistic manner such that the orbital magnetism is naturally accounted for. The exchange field is applied at the level of first-principles by adding a spin-Zeeman term to the Hamiltonian inside the MT region, with its magnitude fixed at 0.544 eV. Our calculation for nonmagnetic metals with magnetization induced by a spin-Zeeman term is not merely a theoretical toy model for analyzing the modern theory of orbital magnetization, but also relates to real experimental situations in thin films. Transition metals such as Pt, Rh, and Pd are paramagnetic materials in bulk and thus do not possess permanent magnetic moments. However, owing to their high magnetic susceptibilities, they can acquire magnetic moments by overcoming the Stoner criterion through the magnetic proximity effect or size reduction. For example, Pt can exhibit magnetization when forming atomic chains~\cite{Strigl15NatCom} or Pt-ferromagnet alloys~\cite{Albert71AdvChem,Simopoulos96PRB,Meier11PRB}. The induced magnetic moment of Pt has been reported to reach up to 0.29 $\mu_{\text{B}}$ per atom in Ni/Pt multilayers~\cite{Wilhelm00PRL} and up to 0.5 $\mu_{\text{B}}$ per atom in Fe/Pt multilayers~\cite{Antel99PRB}. These values are comparable to the induced spin magnetic moment of 0.315 $\mu_{\text{B}}$ per atom obtained in our calculations with the spin-Zeeman term. We note that the present calculation is not intended to quantitatively predict the orbital magnetization of a specific thin-film structure, but rather to provide a qualitative evaluation of the trends expected in spin-polarized metals. In this sense, our calculations may serve as a qualitative theoretical description of the orbital magnetization in such magnetized nonmagnetic metals.

The results are presented in Fig.~\ref{fig5}. Here, we discuss the results on only four materials: bcc W, fcc Pt, fcc Ti, and bcc V (the results on exhaustive list of materials are shown in Table~\ref{table:1}). In Pt, Ti, and V, $M_{z}$ and $M_{z}^{\text{ACA}}$ exhibit similar $\mathcal{E}$ dependences, while the discrepancies between $M_{z}$ and $M_{z}^{\text{ACA}}$ are slightly greater than those of Fe, Co, and Ni --19\% in Pt, 32\% in Ti and 30\% in V. Nevertheless, the ACA still successfully captures about 70\% of the modern theory results in these materials. In all three materials, $M_{z}^{(0)}$ shows close values at $\mathcal{E}_{\text{F}}$ and $\mathcal{E}$-dependence on $M_{z}^{\text{ACA}}$, which is similar to the trend observed in the ferromagnets. $M_{z}^{(1)}$ and $M_{z}^{(2)}$ provide slightly larger contributions than in ferromagnets, but they remain smaller than $M_{z}^{(0)}$, and their $\mathcal{E}$-dependence is distinctly different from that of $M_{z}$. The general trend of $M_{z}^{\text{SR}}$ resembles that of $M_{z}$, but its magnitude is larger in these materials. It is similar to the behavior found in ferromagnets.

In contrast, W exhibits a markedly different trend. In W, the ACA no longer reproduces modern theory: $M_{z}$ deviates not only from its general trend, but also exhibits the opposite sign at $\mathcal{E}_{\text{F}}$. $M_{z}^{(0)}$ is comparable to $M_{z}^{\text{ACA}}$ at $\mathcal{E}_{\text{F}}$, but does not provide a dominant contribution to $M_{z}$. Meanwhile, $M_{z}^{(1)}$ and $M_{z}^{(2)}$ contribute significantly to $M_{z}$. In addition, the difference between $M_{z}^{\text{SR}}$ and $M_{z}$ also becomes pronounced. The difference in the term-by-term decompositions between W and the other materials (Fe, Co, Ni, Pt, Ti, and V) reveals that the microscopic nature of the orbital magnetism in W is different from that of other compounds, which are distinguished by comparing the ACA and the modern theory results.

\begin{figure}[t]
\includegraphics[width=245pt]{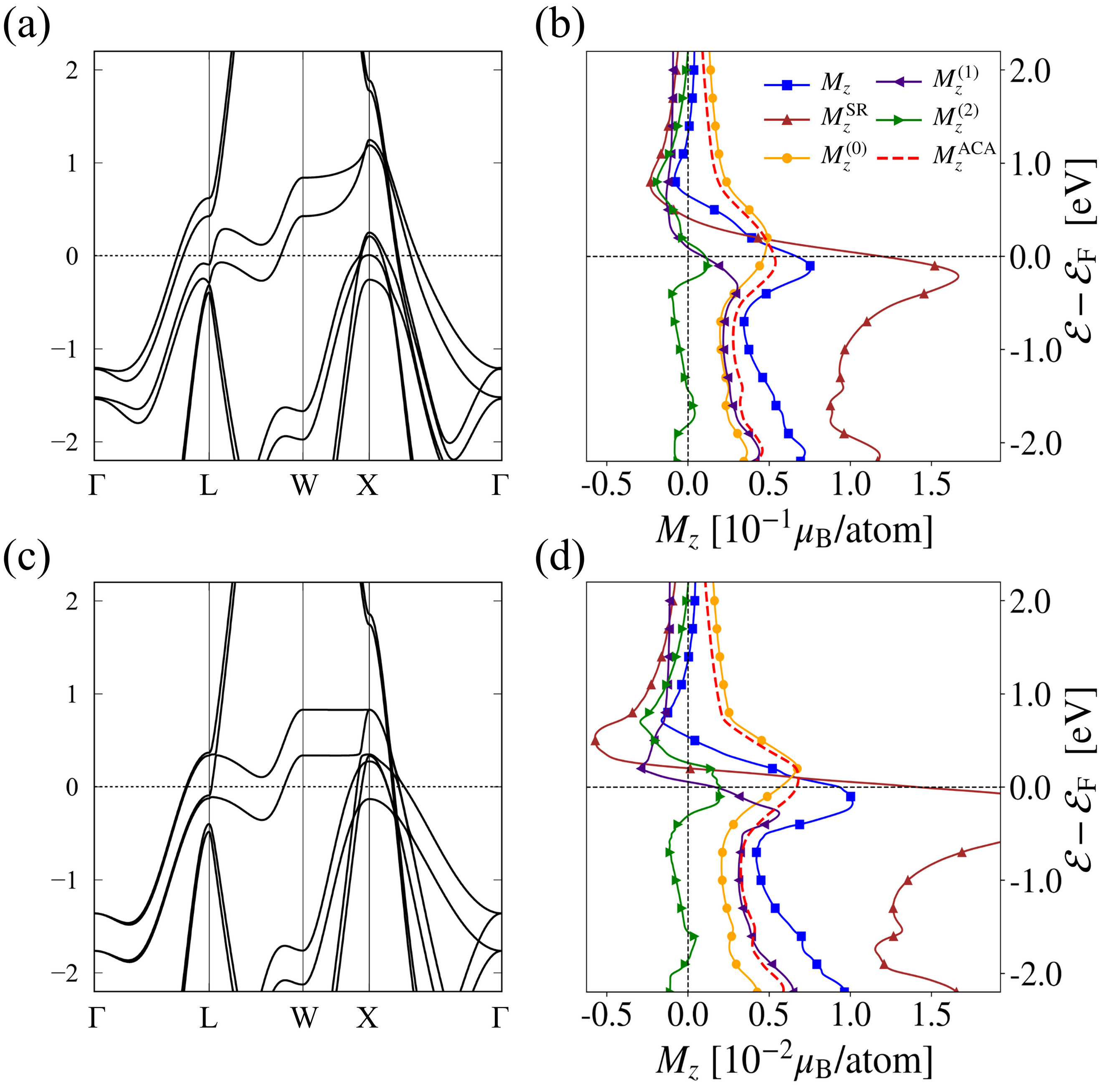}
\caption{\label{fig6}\textbf{Spin-orbit coupling dependence of orbital magnetization.} (a) Band structure of magnetic fcc Pt [001]. (b) Orbital magnetization (blue) of magnetic Pt with $M^{(0)}_{z}$ (yellow), $M^{(1)}_{z}$ (purple), $M^{(2)}_{z}$ (green), $M^{\text{SR}}_{z}$ (brown), and $M^{\text{ACA}}_{z}$ (red dashed line). (c) Band structure of modified magnetic Pt with the spin-orbit coupling strength reduced to 10\%. (d) Corresponding orbital magnetization of modified magnetic Pt is shown.}
\end{figure}

\begin{figure}[t]
\includegraphics[width=245pt]{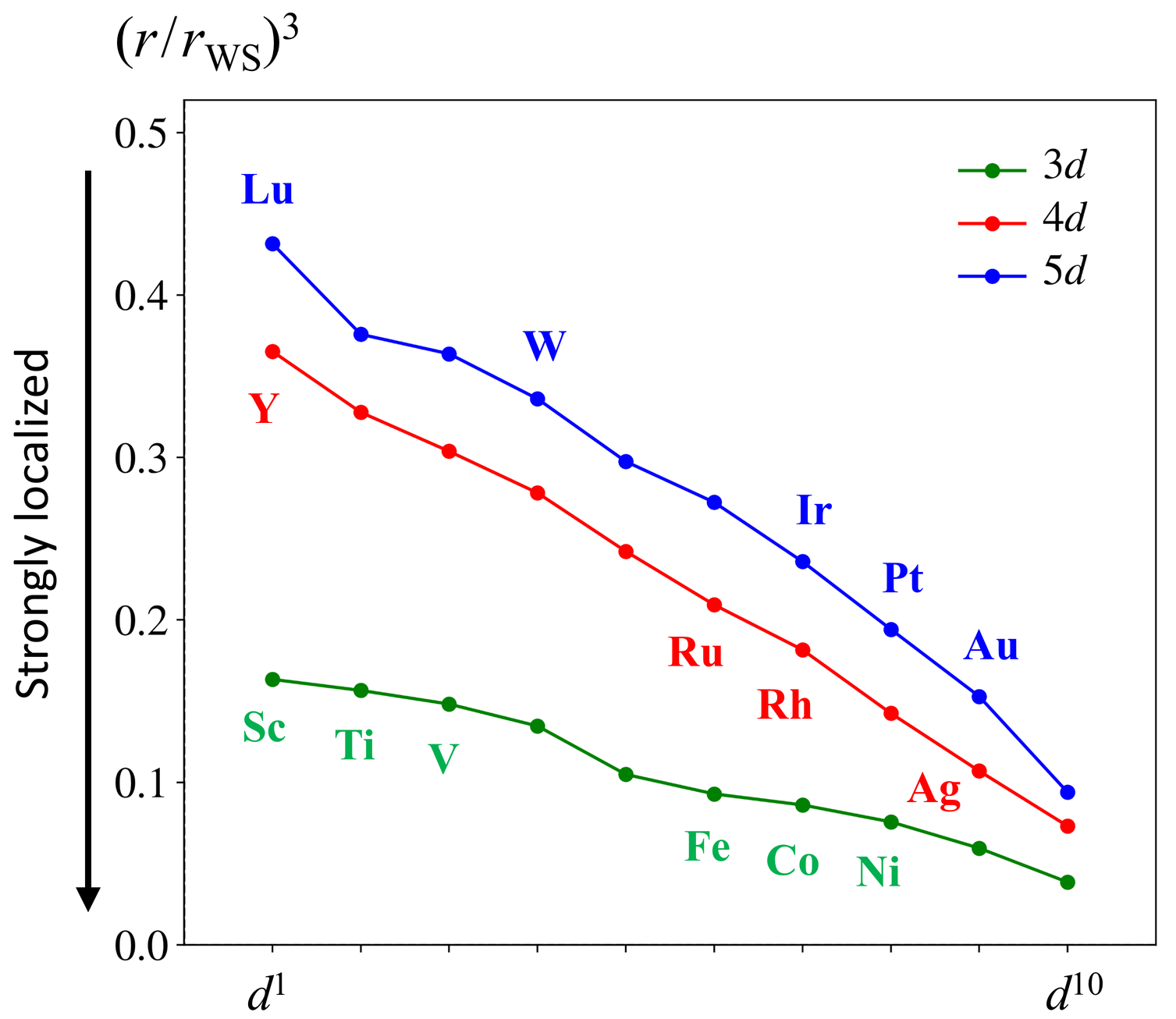}
\caption{\label{fig7} \textbf{Degree of spatial localization of the wave functions in $d$-transition metals.} The ratio of $d$-shell volume to Wigner-Seitz volume, $(r/r_{\text{WS}})^{3}$, for 3$d$ (green), 4$d$ (red), and 5$d$ (blue) transition elements are presented. The data are replotted from Ref.~\protect\cite{Marel88PRB}; see Fig. 1 therein.}
\end{figure}

\begin{figure*}[t]
\includegraphics[width=500pt]{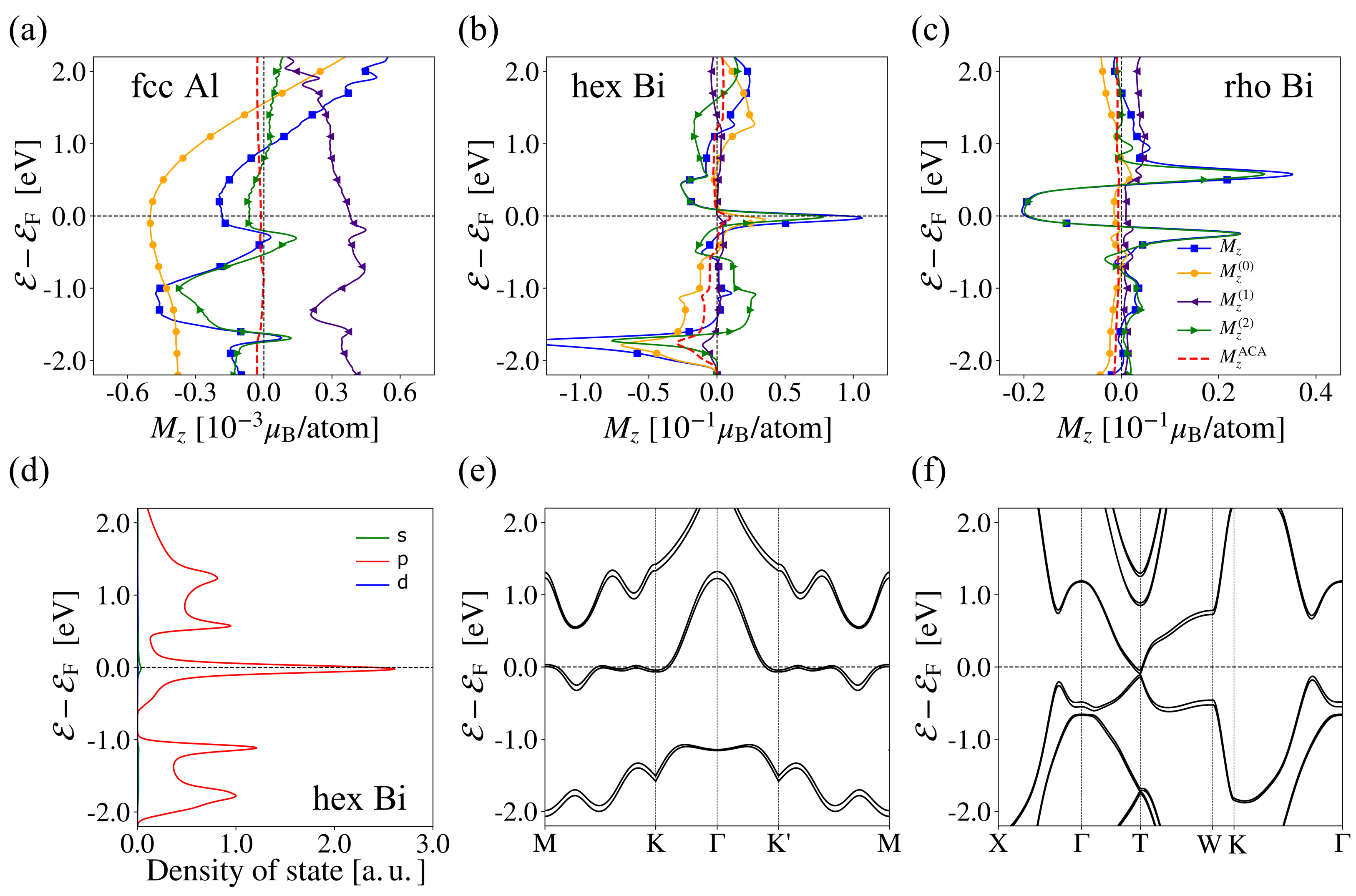}
\caption{\label{fig8}\textbf{Failure of the ACA in $sp$ metals.} (a)--(c) Orbital magnetization of (a) fcc Al with spin-quantization axis [111], (b) two-dimensional magnetic hex Bi with spin-quantization axis oriented out-of-plane, and (c) rho Bi [001]. The energy dependence of $M_{z}$ (blue), $M^{(0)}_{z}$ (yellow), $M^{(1)}_{z}$ (purple), $M^{(2)}_{z}$ (green), and $M^{\text{ACA}}_{z}$ (red dashed line) are shown. (d) Density of state and (e) band structure of the magnetic hex Bi monolayer [001]. (f) Band structure of the magnetic rho Bi [001].}
\end{figure*}

We first examine the effect of the spin-orbit coupling (SOC). While the crystal field causes the orbital quenching that restricts the OAM in equilibrium, the SOC restores a finite orbital magnetization by entangling the spin and OAM. Accordingly, strong orbital magnetization is observed in materials with strong SOC (e.g., Pt, W), whereas only weak orbital magnetization appears in those with weak SOC (e.g., Ti, V). To evaluate the effect of SOC in detail, we compare the orbital magnetization of fcc Pt [001] with that of modified Pt in which the SOC strength is artificially reduced to 10\%. The orbital magnetization of fcc Pt [001] and the modified Pt, together with their electronic band structures, are shown in Fig.~\ref{fig6}. In the modified Pt, the magnitude of the orbital magnetization decreases by about an order of magnitude, but the relative weights of each term and their energy dependence remain almost unchanged. This implies that although changes in the SOC strength alter the band structure near avoided crossings and thereby cause slight variations in the relative contributions of each term, such changes are not the dominant factor in determining the overall trends. In addition, the difference in the term-by-term decompositions between W and Pt infers that the SOC is not the dominant factor in determining the relative contributions of each term.

Next, we focus on spatial localization of the wave functions of electrons. The degree of spatial localization of the wave functions of electrons is quantified by the ratio of the $d$-shell volume to the Wigner-Seitz cell volume, $(r/r_{\text{WS}})^{3}$. Figure.~\ref{fig7} shows the values of $(r/r_{\text{WS}})^{3}$ for 3$d$-, 4$d$-, and 5$d$-transition metals reported in Ref.~\cite{Marel88PRB}. In general, the wave functions of electrons are more localized around the nucleus for elements with shorter periods, and within the same period for those with larger atomic numbers. Figure 7 shows that the $d$ electrons in W are more delocalized than those in the other elements. In contrast, elements such as Co and Ni have electrons that are strongly localized near the atomic nuclei. In materials with strongly localized electrons, the hopping between sites is suppressed, leading to broad and smooth Bloch states in $\text{k}$-space. Moreover, a large portion of the wave functions of electrons reside within the MT spheres. These explain why the difference between the ACA and modern theory becomes significant and why $M_{z}^{(1)}$ and $M_{z}^{(2)}$ provide pronounced contributions to W, whereas the ACA reproduces the results of modern theory well and $M_{z}^{(0)}$ contributes dominantly to $M_{z}$ in other materials such as Co and Ni. The detailed analysis is presented in Appendix~\ref{appendix:localization} and Fig.~\ref{figA2}. We note that the degree of spatial localization of the wave functions of electrons provides a general trend for the relative contributions of each term, but the precise values are influenced by details such as the crystal structure, the number of electrons filling the shell, and the strength of the SOC.

\subsection{\textit{sp} metals}\label{subsec:sp_metals}

In the previous subsection, we analyzed $d$-transition metals and confirmed that the degree of electron localization plays a crucial role in determining the ACA and each term of the modern theory of orbital magnetization. To make this more evident, we now investigate the orbital magnetization in $sp$ metals. In $sp$ metals, orbital magnetization mainly originates from electrons partially filling the $p$ orbitals. Compared with $d$ electrons, the $s$ and $p$ electrons are strongly delocalized from the nucleus and more easily hop to neighboring atoms with large kinetic energy. As a result, in these elements, the CM contribution becomes relatively larger than in $d$-transition metals. For example, intersite hoppings of $s$ electrons can make the CM contribution, although the ACA is zero for $s$ orbitals by definition. In addition, the $M_{z}^{(2)}$ term becomes highly significant, while $M_{z}^{(0)}$ captures only a fraction of $M_{z}$. This indicates that the ACA completely fails to describe total orbital magnetization, whereas the standard tight-binding model provides a more favorable description than the ACA. Furthermore, since the intercell contribution becomes non-negligible, $M_{z}^{(0)}$ and $M_{z}^{\text{ACA}}$ exhibit noticeable differences.

The results are presented in Fig.~\ref{fig8}. As in the case of nonmagnetic $d$-transition metals, we introduce a spin-Zeeman term of 0.544 eV, as most $sp$ metals do not spontaneously exhibit nonzero magnetization. As examples of $sp$ metals, we show the results on Al in fcc structure and Bi in hexagonal (hex) and rhombohedral (rho) structures. The orbital magnetization of Al as a function of the energy is shown in Fig.~\ref{fig8}(a). We note that, in Al with weak SOC, the positive spin magnetic moment induced by an external magnetic field is much larger than the orbital moment and thus drives the paramagnetism. Unlike $d$-transition metals, the orbital magntization computed within the ACA differs substantially from that obtained from the modern theory. Moreover, $M_{z}^{(0)}$ fails to reproduce $M_{z}$. This is an example showing that, in materials dominated by delocalized $p$ character, the ACA can yield orbital magnetization that are distinctly different from those of the modern theory.

In hexagonal Bi, we find dominant $p$ character near the Fermi energy, shown in Fig.~\ref{fig8}(d). This arises from the two almost flat bands with narrow gap [Fig.~\ref{fig8}(e)], thus giving rise to the van Hove singularity of the density of states. The orbital magnetization of hexagonal Bi is shown in Fig.~\ref{fig8}(b). A remarkable peak of $M_{z}$ appears near $\mathcal{E}_{\text{F}}$, which is about 12 times larger than $M_{z}^{\text{ACA}}$. This result indicates that the presence of the van Hove singularity alone does not make strong orbital magnetization in the ACA, but interband hybridization is required, which leads to strong orbital magnetization within the modern theory. We identify that the strong peak in $M_{z}$ originates from the interband effect between two nearly degenerate $p$ bands located near $\mathcal{E}_{\text{F}}$. In contrast, $M_{z}^{\text{ACA}}$, determined by the atomic OAM of each band, does not exhibit such a pronounced peak. This result clearly demonstrates the fundamental difference in origin between the modern theory orbital magnetism, which is strongly affected by interband hybridization, and the ACA, which is governed by intraband orbital characters. We find that $M_{z}^{(0)}$ and $M_{z}^{(2)}$ contribute 19.7\% and 79.8\% to $M_{z}$, respectively. Moreover, $M_{z}^{\text{ACA}}$ accounts for only 42.4\% of $M_{z}^{(0)}$. On the other hand, $M_{z}^{\text{SR}}$ only amounts to 54.2\% of $M_{z}$ and $M_{z}^{\text{CM}}$ is non-negligible (Table~\ref{table:1}).

\begin{figure*}[t]
\includegraphics[width=500pt]{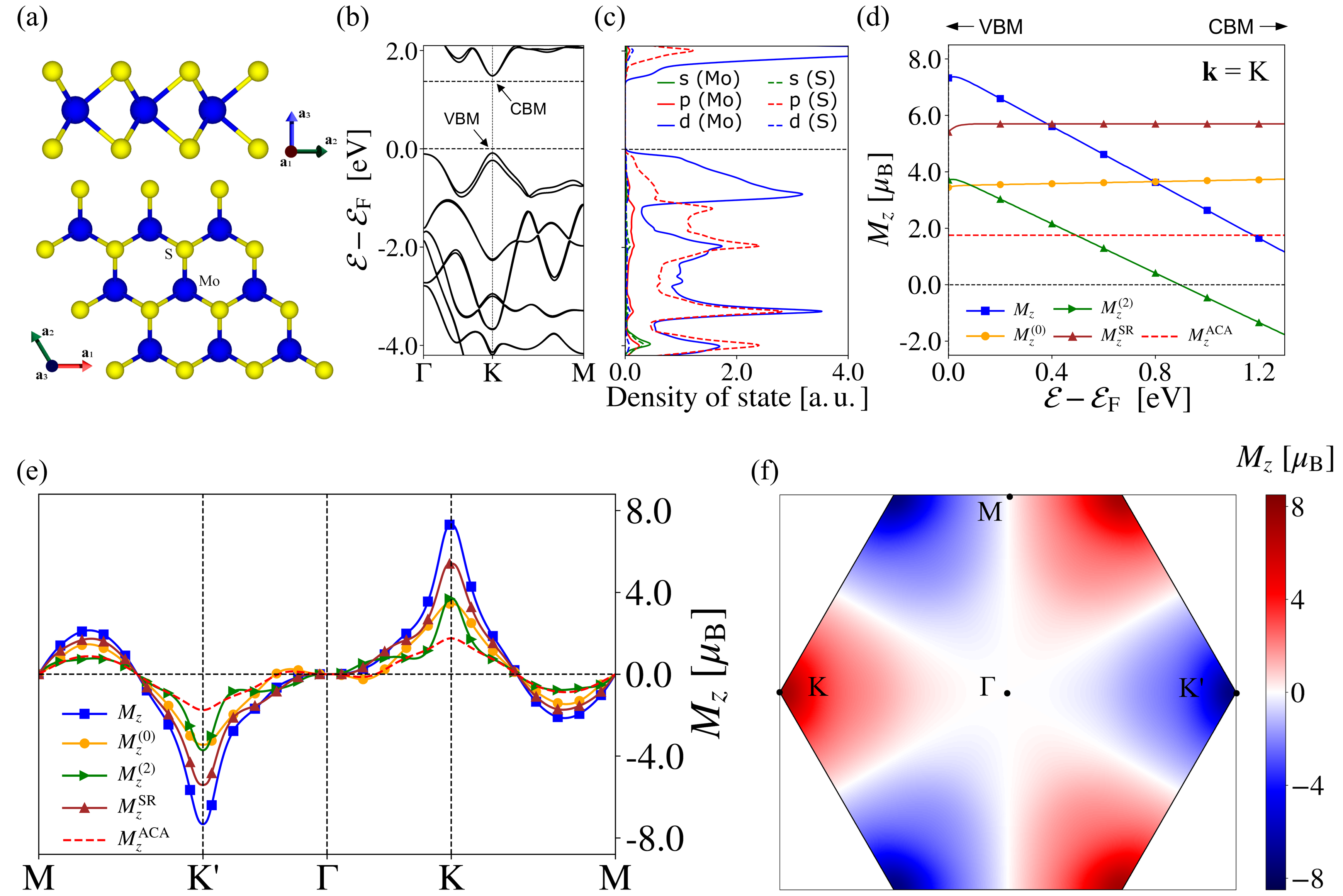}
\caption{\label{fig9}\textbf{Modern theory and ACA orbital magnetization in 1H-Mo$\text{S}_{2}$ monolayer.} (a) Lattice structure of 1H-Mo$\text{S}_{2}$ monolayer. (b) Band structure of the 1H-MoS$_{2}$ monolayer. The insulating energy gap is presented between the valence band maximum (VBM) and conduction band minimum (CBM) at $\text{K}$ point. (c) Density of states of 1H-Mo$\text{S}_{2}$. (d),(e) Comparison between the modern theory and ACA local orbital magnetization for 1H-Mo$\text{S}_2$ monolayer. The $\mathbf{k}$-resolved $M_{z}(\textbf{k})$ (blue), $M_{z}^{(0)}(\textbf{k})$ (yellow), $M_{z}^{(2)}(\textbf{k})$ (green), $M_{z}^{\text{SR}}(\textbf{k})$ (brown), and $M_{z}^{\text{ACA}}(\textbf{k})$ (red dashed line) are plotted along $\text{M}$--$\text{K}^{\prime}$--$\Gamma$--$\text{K}$--$\text{M}$ line in (e). The energy dependence of the orbital magnetization at $\text{K}$ is presented in (d). (f) $\mathbf{k}$-resolved $M_{z}$ plotted over the two-dimensional BZ.}
\end{figure*}

An interesting feature of the orbital magnetism in $sp$ metals is its sensitive dependence on the crystal structure. It is because $sp$ orbitals are widely spread beyond the MT region, thereby strongly influenced by the potential and orbital hybridization induced by neighboring sites. This is in contrast to $d$-transition metals, where localized $d$ states are only weakly influenced by the crystal structure, while $s$ orbitals do not {\it see} the discrete symmetry of neighboring sites. To demonstrate this point, we investigate Bi in rhombohedral structure. The result of the orbital magnetization in rhombohedral Bi is shown in Fig.~\ref{fig8}(c). We find that $M_{z}$ is negative at $\mathcal{E}_{\text{F}}$, in contrast to hexagonal Bi. Also, $M_{z}$ exhibits opposite(positive)-sign peaks at about 0.3 eV below and 0.5 eV above $\mathcal{E}_{\text{F}}$. Such a strong $\mathcal{E}$-dependence arises from interband coupling between adjacent bands, similar to hexagonal Bi. In contrast, $M_{z}^{\text{ACA}}$, which is governed by the intrinsic orbital moments of individual bands, shows only a weak $\mathcal{E}$-dependence.

\subsection{Transition metal dichalcogenides}
\label{subsec:TMDs}

The TMDs are important material examples worth investigating orbital magnetism for several reasons. First, two-dimensionality makes the contributions from individual bands pronounced, which makes the TMDs ideal for studying band hybridization effects. Second, many of them have distinct broken symmetry. For example, in 1H-MoS$_{2}$, the lack of inversion symmetry allows for the emergence of a finite orbital moment around the two valleys, whereas time-reversal symmetry enforces opposite orbital moments at $\text{K}$ and $\text{K}^{\prime}$. A polar vector in the plane makes the orbital moment aligned out-of-plane. The symmetry-breaking effect is even more dramatic in T$_d$-WTe$_2$, where the distorted and buckled crystal structure makes the band structure highly anisotropic. Finally, all these properties make the Berry curvature effect crucial. The Berry curvature vanishes if both time-reversal and spatial inversion symmetries coexist. In TMDs, however, broken inversion symmetry generates large Berry curvature locally in each $\mathbf{k}$, although the integral of the Berry curvature over the Brillouin zone vanishes due to time-reversal symmetry.

The analysis of $\mathbf{k}$-resolved orbital moment of 1H-MoS$_{2}$ is presented in Fig.~\ref{fig9}. In 1H-MoS$_{2}$, both the valence band maximum (VBM) and the conduction band minimum (CBM) are located at the two valleys $K$ and $K^{\prime}$ [Fig.~\ref{fig9}(b)], and hence these points are important to interpret optical transitions (e.g. MCD, photoluminescence) and quantum geometric properties (e.g. Berry curvature, the modern theory of orbital magnetism, the spin Hall effect, and the orbital Hall effect)~\cite{Cao12NatCom,Zeng12NatNano,Mak12NatNano,Lee17NatMat,Bhowal20PRB,Wu13NatPhys,Tahir14PRB,Canonico20PRB,Bhowal20PRB2,Xue20PRB,Sallen12PRB,Cysne22PRB,Arakawa25PRB}. The distinct $C_{3v}$ symmetry of 1H-MoS$_{2}$ leads to valley-contrasting orbital moment. This orbital-momentum locking is confirmed in our calculation, Figs.~\ref{fig9}(e) and \ref{fig9}(f).

The CBM at the valleys originates from the spin-degenerate Mo $d_{z^{2}}$ state, while the VBM is described mainly as the $d_{x^{2}-y^{2}} \pm id_{xy}$ state splitted by the SOC at $K(K^{\prime})$, respectively. This yields a positive (negative) orbital moment of valence band electrons at $K(K^{\prime})$, respectively. This picture is consistent with our ACA calculation in Fig.~\ref{fig9}(e) (red dashed line). $M_{z}(\mathbf{k})$ at each valley shows the same sign as $M_{z}^{\text{ACA}}(\mathbf{k})$, but is about four times larger, indicating that the ACA captures only a small fraction of modern theory. This difference arises from the interorbital motion of electrons, reinforced by the $p-d$ hopping between the $p$ orbitals of S and the $d$ orbitals of Mo. The density of state of 1H-MoS$_{2}$ shown in Fig.~\ref{fig9}(c), together with an atomic-orbital-based analysis~\cite{Bhowal20PRB,Bhowal20PRB2}, supports this interpretation. $M_{z}^{\text{SR}}(\mathbf{k})$ [brown line in Fig.~\ref{fig9}(e)] is similar to $M_{z}(\mathbf{k})$ but with a smaller magnitude.

\begin{figure*}[t]
\includegraphics[width=500pt]{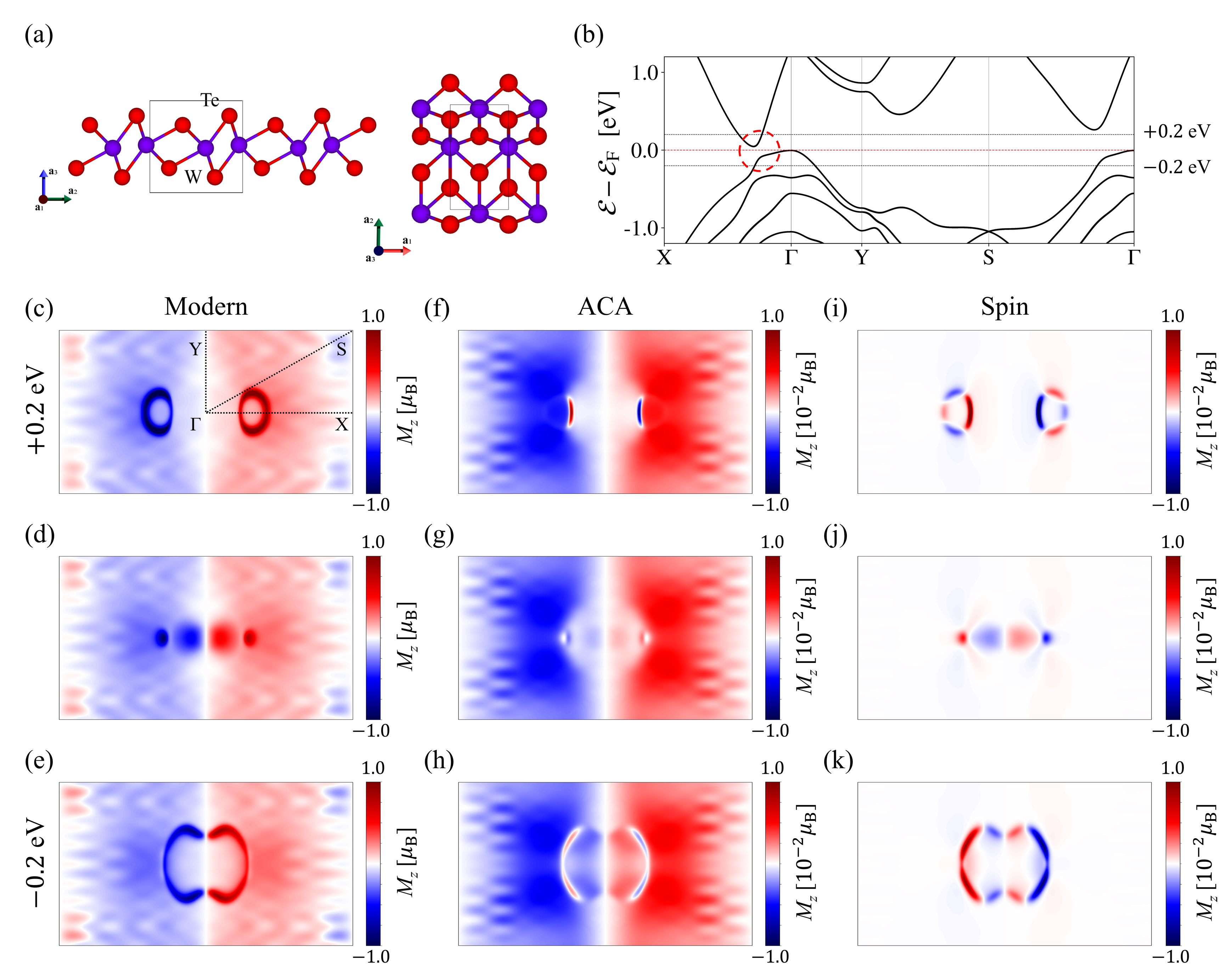}
\caption{\label{fig10}\textbf{Comparison of the modern theory and ACA orbital magnetization in T$_{d}$-WTe$_{2}$ monolayer.} (a) Lattice structure of T$_d$-WTe$_{2}$ monolayer. (b) Calculated band structure. (c)--(e) $\mathbf{k}$-resolved modern theory orbital magnetization $M_{z}$ plotted in the two-dimentional BZ, at energy levels of (c) $\mathcal{E}=\mathcal{E}_{\text{F}}+0.2$ eV, (d) $\mathcal{E}=\mathcal{E}_{\text{F}}$, and (e) $\mathcal{E}=\mathcal{E}_{\text{F}}-0.2$ eV. (f)--(h) Corresponding $\mathbf{k}$-resolved ACA orbital magnetization $M_{z}^{\text{ACA}}$ at (f) $\mathcal{E}=\mathcal{E}_{\text{F}}+0.2$ eV, (g) $\mathcal{E}=\mathcal{E}_{\text{F}}$, and (h) $\mathcal{E}=\mathcal{E}_{\text{F}}-0.2$ eV. (i)--(k) $\mathbf{k}$-resolved spin moments at (i) $\mathcal{E}=\mathcal{E}_{\text{F}}+0.2$ eV, (j) $\mathcal{E}=\mathcal{E}_{\text{F}}$, and (k) $\mathcal{E}=\mathcal{E}_{\text{F}}-0.2$ eV.}
\end{figure*}

Now, we consider the decomposition in the order of $J_{\alpha}$. $M_{z}^{(2)}(\mathbf{k})$ exhibits almost the same trend as $M_{z}(\mathbf{k})$, although its magnitude is about half as large. The previous study~\cite{Bhowal20PRB2}, which performed both first-principles calculation with Wannier interpolation and the effective model calculation based on a four-band tight-binding model, reported good agreement between the two approaches near the valleys. Our result for $M_{z}^{(2)}(\mathbf{k})$ is consistent with both the trend and the magnitude of these two calculations. This verifies that $M_{z}^{(2)}(\mathbf{k})$ corresponds to the results obtained only from the effective Hamiltonian constructed by the Wannier interpolation or tight-binding model. $M_{z}^{(0)}(\mathbf{k})$ exhibits a magnitude peak similar to $M_{z}^{(2)}(\mathbf{k})$ at both valleys, but with a wider width.

We also examine the energy dependence of the local orbital magnetization. At zero temperature, orbital magnetization is related to the intrinsic anomalous Hall conductivity $\sigma_{xy}$ through the St$\check{\textrm{r}}$eda formula~\cite{Streda82PCM,Xiao10RMP}
\begin{align}\label{eq:streda}
    \sigma_{xy} &= -e\left( \frac{\partial \textbf{M}}{\partial \mu_{\text{c}}} \right)_{\textbf{B},T},
\end{align}
where $\mu_{\text{c}}$ is the chemical potential. In 1H-MoS$_{2}$, time-reversal symmetry constrains the BZ integral of the Berry curvature to vanish, resulting in $\sigma_{xy} = 0$, but breaking the inversion symmetry allows both Berry curvature and orbital magnetization to have finite values at each $\mathbf{k}$. Although Eq.~\eqref{eq:streda} was first derived for BZ-integrated quantities, it can also be obtained by directly differentiating the modern theory formula of orbital magnetization [Eq.~\eqref{eq:modern}], which implies that the same relation applies to orbital magnetization and Berry curvature at each $\mathbf{k}$. Then, the presence of a Berry-curvature texture indicates that the $\mathbf{k}$-local orbital magnetization can exhibit an energy dependence within the insulating energy gap. The results presented in Figs.~\ref{fig9}(e) and~\ref{fig9}(f) are obtained by choosing $\mathcal{E}_{\text{F}}$ near the VBM, as in previous studies~\cite{Bhowal20PRB,Bhowal20PRB2}. Figure~\ref{fig9}(d) shows the dependence of the $\mathbf{k}$-local orbital magnetization on $\mathcal{E}$ at the K valley within the insulating gap. As $\mathcal{E}$ increases, $M_{z}$ decreases linearly with a constant slope $\vert \Delta M_{z} (\mathbf{k} = \text{K}) /\Delta \mathcal{E} \vert = 4.96$ $\mu_{\text{B}}/\text{eV}$. This value corresponds to the calculated value of the Berry curvature at K, $\Omega (\mathbf{k} = \text{K}) = 18.9$ $\text{\AA}^{2}$, and agrees with previous study~\cite{Feng16CMS}. The slight nonlinear variation near $\mathcal{E} \approx \mathcal{E}_{\text{F}}$ originates from the influence of VBM due to the smearing corresponding to $T = 300$ K introduced in the numerical calculation. On the other hand, $M_{z}^{\text{SR}}$ and $M_{z}^{\text{ACA}}$ do not have chemical potential dependence, so they remain constant within the gap. As $\mathcal{E}$ increases, $M_{z}^{(0)}$ slightly increases with a slope $\vert \Delta M_{z}^{(0)} (\mathbf{k} = \text{K}) /\Delta \mathcal{E} \vert = 0.175$ $\mu_{\text{B}}/\text{eV}$, while $M_{z}^{(2)}$ decreases with a slope $\vert \Delta M_{z}^{(2)} (\mathbf{k} = \text{K}) /\Delta \mathcal{E} \vert = 4.37$ $\mu_{\text{B}}/\text{eV}$, which is close to that of $M_{z}$. This shows that the gauge-correction term is sensitive to changes in $\mathcal{E}$ in the gap.

Figure~\ref{fig10} presents the $\mathbf{k}$-resolved orbital moment of T$_{d}$-WTe$_{2}$ monolayer, defined as a monolayer directly isolated from the three-dimensional T$_{d}$ bulk lattice. The T$_{d}$ phase of the WTe$_{2}$ monolayer is derived from a slight structural distortion of the 1T$^{\prime}$ phase. 1T$^{\prime}$-WTe$_{2}$ has two important symmetries: mirror symmetry $\mathcal{M}_{a_{1}}$ and twofold screw rotational symmetry $\mathcal{C}_{2a_{1}}$, the combination of which gives rise to the inversion symmetry~\cite{Xu18NatPhys}. In T$_{d}$ phase, the $\mathcal{C}_{2a_{1}}$ symmetry is weakly broken, resulting loss of the inversion symmetry. This broken symmetry permits to emergence of a non-vanishing Berry curvature texture and orbital moment texture in $\mathbf{k}$-space, making T$_{d}$-WTe$_{2}$ as a candidate for observing exotic transport phenomena, such as the quantum nonlinear Hall effect and the orbital Rashba-Edelstein effect.

1T$^{\prime}$-WTe$_{2}$ was theoretically predicted to have a non-trivial $\mathbb{Z}_{2}$ topological phase driven by a band inversion near the $\Gamma$ point~\cite{Qian14SCI}. Subsequent works~\cite{Tang17NatPhys,Wu18SCI} confirmed it to be a quantum spin Hall insulator with a bulk band gap of tens of meV. Previous theoretical and experimental studies~\cite{Xu18NatPhys,You18PRB,Torbatian20PRAppl,Vila21PRR} reported that the T$_{d}$-WTe$_{2}$ monolayer also exhibits a small indirect band gap near the $\Gamma$ point upon the inclusion of SOC. Here, we investigate the orbital moments texture within the $T_{d}$-WTe$_{2}$ monolayer structure as proposed by these prior works.

In our calculation, the indirect insulating energy gap ($\approx$ 50 meV) appears on the $\text{X}$--$\Gamma$ line [Fig.~\ref{fig10}(b)], which is consistent with the previous studies~\cite{Tang17NatPhys,You18PRB,Xu18NatPhys}. The texture of $M_{z}$ is nearly two orders of magnitude stronger than $M_{z}^{\text{ACA}}$ [Figs.~\ref{fig10}(d) and \ref{fig10}(g)]. We emphasize the pronounced peak of $M_{z}$ that appears near the avoided band crossing located on the $\text{X}$--$\Gamma$ line [red dotted circle in Fig.~\ref{fig10}(b)]. At this point, the narrow energy gap induced by the SOC is not only responsible for the nontrivial topological phase, but also generates a strong orbital magnetization driven by the spiky Berry curvature, while it affects the ACA only weakly. At $\mathcal{E} = \mathcal{E}_{\text{F}} + 0.2$ eV, electron doping fills states near the CBM, forming an electron pocket and thereby modifying the shape of the $M_{z}$ peak [Fig.~\ref{fig10}(c)].

These results show that, similar to the Berry curvature, the modern theory orbital magnetization is strongly influenced by the electronic band structure and the geometric properties of the eigenstate. This suggests that topological materials are promising platforms for exploring orbitronic phenomena originating in the nonlocal nature of circulating currents. This also indicates that the modern-theory orbital magnetism may dominate in the TMD family with similar band structures. Indeed, a recent theoretical study~\cite{Canonico25arXiv} reported that, in a MoTe$_{2}$/Ferromagnet heterostructure, an itinerant OAM generated by the orbital Rashba-Edelstein effect can be much larger than both the ACA and spin contributions.

\section{Concluding remarks}\label{sec:concluding_remarks}

In this paper, we have presented an anatomy of the modern theory of orbital magnetism based on the term-by-term analysis. Using the Wannier-based gauge-invariant formalism of the orbital magnetization, we have introduced two decompositions schemes and evaluated contributions in each scheme for $d$-transition metals, $sp$ metals, and TMDs by combining first-principles calculation with Wannier interpolation, These analyses reveal the intrinsic properties and provide physical intuitions of the modern theory of orbital magnetism.

In the first decomposition, the orbital magnetization is separated into $M_{\gamma}^{\text{SR}}$ and $M_{\gamma}^{\text{CM}}$, distinguishing the local and nonlocal orbital angular motions of electrons in real space. Our results for $d$-transition and $sp$ metals confirm that self-rotation dominates when electrons are strongly localized around the nuclei, whereas itinerant effects become important when the electronic states are delocalized. Since $M_{\gamma}^{\text{SR}}$ is directly related to MCD measurement, this finding provides an important clue to understanding magneto-optical phenomena. Our results also provide quantitative confirmation of the suggestion proposed in Ref.~\cite{Souza08PRB} that $sp$-metals may exhibit a large ratio $\vert M_{\gamma}^{\text{CM}}/M_{\gamma}^{\text{SR}}\vert$.

The $J$-decomposition is the second decomposition we consider. The $J$-decomposition, introduced in the Wannier representation, reveals how the orbital angular motion of the Wannier basis contributes to the orbital magnetization. By choosing a Wannier basis close to atomic orbitals and evaluating the intracell self-rotation term of $M_{\gamma}^{(0)}$, we establish a direct connection between the ACA and the modern theory. For $d$-transition metals, where electrons are well-localized near the nuclei (e.g., Ni, Co, and V), the ACA reproduces more than 70\% of the total orbital magnetization. In contrast, for $sp$ metals and TMDs, the ACA completely fails to reproduce the modern theory. In Al, hexagonal and rhombohedral Bi, the modern theory provides an orbital magnetization significantly larger than the ACA. In particular, the result for hexagonal Bi highlights the origin of the difference between the ACA and modern theory orbital magnetization: the interband effects between two nearly degenerate $p$ bands at $\mathcal{E}_{\text{F}}$ produce a strong peak in the modern theory, whereas the ACA, determined by the intrinsic orbital moments of individual bands, cannot capture such features. This distinction emphasizes that the modern theory orbital magnetization is strongly related to the geometric properties of the Bloch states. In 1H-MoS$_{2}$, where the $p$ orbitals of chalcogen atoms mediate hopping between $d$ orbitals of transition metals, the ACA only captures a small portion of the orbital moment in the $K$/$K^{\prime}$ valleys. Notably, within the gap between the VBM and the CBM, the ACA remains unchanged, whereas the modern theory orbital magnetization varies linearly with energy. In monolayer T$_{d}$-WTe2, the $\mathbf{k}$-resolved orbital moment obtained from the modern theory is nearly two orders of magnitude larger than the ACA and, in particular, exhibits a pronounced peak near the avoided band crossing, originating from band hybridization. This suggests that the TMDs family provides a promising platform for investigating phenomena associated with modern theory orbital magnetism.

The $J$-decomposition also clarifies the extent to which effective models can describe the orbital magnetization. Although effective models successfully reproduce the band structure within a target energy window, they are inherently limited in evaluating quantities that depend on state derivatives---such as the orbital magnetization and Berry curvature---because of the truncation problem. In addition, $\mathbf{k}$-fixed approaches like the $\mathbf{k} \cdot \mathbf{p}$ perturbation theory or the naive calculation of the standard tight-binding method miss the contributions from the Wannier basis, capturing only the $M_{\gamma}^{(2)}$ term. In $d$-transition metals, however, $M_{\gamma}^{(2)}$ accounts for only a small fraction of the total orbital magnetization, so that the ACA is a more suitable approximation.

Finally, we have also provided new insights into the gauge-covariant formalism of orbital magnetization. The formalism naturally includes both the effects captured by the ACA and effective models, so that it properly describes the magnetic moment arising from both local and nonlocal orbital motions of electrons. We have explicitly proved that the gauge-covariant formula for metals, first introduced in Ref.~\cite{Lopez12PRB}, is independent of the gauge selection as long as the inner window includes all occupied states with an appropriate initial projection of orbital characters. Recently, several studies~\cite{Bhowal21PRB,Cysne22PRB,Busch23PRR,Gobel24PRL,Pezo22PRB,Go24NanoLett,Lee25PRB} have employed OAM operators derived from the modern theory of orbital magnetization to compute the orbital-related quantities such as the orbital Hall conductivity. However, it remains unclear whether such formulations are suitable for quantitative calculations in real materials, especially because of the gauge problem. In our work, we have constructed a quantum-mechanical operator for the orbital magnetization based on the gauge-covariant formalism, which behaves like `ordinary' operators. However, we have also discussed its implications and limitations. While this operator is appropriate for calculating the orbital magnetization, its applicability to other quantities such as the orbital current operator requires further studies, in particular, how they correspond to physically measurable effects. Nevertheless, we expect that our formulation provides a foundation for establishing practical and physically transparent approaches to a wide range of orbital-related phenomena, such as the orbital Hall effect, orbital Rashba-Edelstein effect, and magnetoelectric effects, in real materials. As orbitronics continues to emerge as a promising paradigm for information processing based on orbital degrees of freedom, establishing such rigorous and gauge-consistent theoretical tools will be essential for identifying materials, interpreting experiments, and designing next-generation orbital-based electronic and spintronic devices.

\begin{acknowledgments}
We thank Ivo Souza, Garima Ahuja, Felix Lüpke, Lukasz Plucinski, Jongjun M. Lee, Byeonghyeon Choi, and Suik Cheon for fruitful discussions. D.G. acknowledges fruitful discussion with Aurélien Manchon, Börge Göbel, Jagoda Sławińska, and Tatiana G. Rappoport. H.-W. Lee, H. Lee, and I. Baek were financially supported by the National Research Foundation of Korea (NRF) grant funded by the Korean government (MSIT) (No. RS-2024-00356270, RS-2024-00410027). This work was supported by the Jülich Supercomputing Centre (jiff40) and the National Supercomputing Center (KSC-2025-CRE-0068). We also acknowledge support by the Deutsche Forschungsgemeinschaft (DFG) in the framework of TRR 288 $-$ 422213477 (Project B06), and by the EIC Pathfinder OPEN grant 101129641 ``OBELIX''. D.G. was supported by the POSCO Science Fellowship of POSCO TJ Park Foundation and a Korea University Grant.
\end{acknowledgments}

\begin{table*}[ht!]
\begin{center}
\small
\begin{tabular}{@{\hspace{8pt}} c @{\hspace{4pt}} c @{\hspace{11pt}} c @{\hspace{11pt}} c @{\hspace{11pt}} c @{\hspace{11pt}} c @{\hspace{11pt}} c @{\hspace{11pt}} c c @{\hspace{10pt}} c @{\hspace{4pt}}}
\hline
\hline
\multirow{2}{*}{\rule{0pt}{3.0ex}Materials} & \multirow{2}{*}{\rule{0pt}{3.0ex}SQA} & \multicolumn{6}{@{}c@{}}{\makebox[2.5cm][c]{\hspace{-10pt}\rule{0pt}{2.5ex}Modern theory}} & & \multirow{2}{*}{\rule{0pt}{3.0ex}$M_{z}^{\text{ACA}}$} \\
\cmidrule(r){3-8}
 & & $M_{z}$ & $M^{\text{SR}}_{z}$ & $M^{\text{CM}}_{z}$ & $M^{(0)}_{z}$ & $M^{(2)}_{z}$ & $M^{\text{SR}(\text{on})}_{z}$ & & \\
\hline
\rule{0pt}{3.2ex}Fe (bcc) & 001 & 6.653$\times$$10^{-2}$ & 1.023$\times$$10^{-1}$ & -3.578$\times$$10^{-2}$ & 3.933$\times$$10^{-2}$ & 6.968$\times$$10^{-3}$ & 4.989$\times$$10^{-2}$ & & 4.482$\times$$10^{-2}$ \\[2mm]
\rule{0pt}{3.0ex}Fe$^{\star}$ (bcc) & 001 & 8.760$\times$$10^{-2}$ & 1.231$\times$$10^{-1}$ & -3.549$\times$$10^{-2}$ & 4.962$\times$$10^{-2}$ & 1.327$\times$$10^{-3}$ & 6.413$\times$$10^{-2}$ & & 6.817$\times$$10^{-2}$ \\[2mm]
\rule{0pt}{3.0ex}Co (hcp) & 0001 & 7.500$\times$$10^{-2}$ & 1.107$\times$$10^{-1}$ & -3.572$\times$$10^{-2}$ & 6.211$\times$$10^{-2}$ & -2.787$\times$$10^{-3}$ & 7.260$\times$$10^{-2}$ & & 7.913$\times$$10^{-2}$ \\[2mm]
\rule{0pt}{3.0ex}Co$^{\star}$ (hcp) & 0001 & 1.440$\times$$10^{-1}$ & 1.963$\times$$10^{-1}$ & -5.224$\times$$10^{-2}$ & 1.224$\times$$10^{-1}$ & -6.453$\times$$10^{-3}$ & 1.416$\times$$10^{-1}$ & & 1.552$\times$$10^{-1}$ \\[2mm]
\multirow{2}{*}{Co (fcc)} & 111 & 7.329$\times$$10^{-2}$ & 1.080$\times$$10^{-1}$ & -3.470$\times$$10^{-2}$ & 6.200$\times$$10^{-2}$ & -5.019$\times$$10^{-3}$ & 7.261$\times$$10^{-2}$ & & 6.884$\times$$10^{-2}$ \\[1.5mm]
& 001 & 7.774$\times$$10^{-2}$ & 1.145$\times$$10^{-1}$ & -3.675$\times$$10^{-2}$ & 6.576$\times$$10^{-2}$ & -5.325$\times$$10^{-3}$ & 7.701$\times$$10^{-2}$ & & 7.302$\times$$10^{-2}$ \\[2mm]
\multirow{2}{*}{Co$^{\star}$ (fcc)} & 111 & 1.203$\times$$10^{-1}$ & 1.675$\times$$10^{-1}$ & -4.728$\times$$10^{-2}$ & 1.088$\times$$10^{-1}$ & -1.480$\times$$10^{-2}$ & 1.259$\times$$10^{-1}$ & & 1.200$\times$$10^{-1}$ \\[2mm]
& 001 & 1.195$\times$$10^{-1}$ & 1.665$\times$$10^{-1}$ & -4.703$\times$$10^{-2}$ & 1.080$\times$$10^{-1}$ & -1.466$\times$$10^{-2}$ & 1.250$\times$$10^{-1}$ & & 1.192$\times$$10^{-1}$ \\[2mm]
\multirow{2}{*}{Ni (fcc)} & 111 & 4.196$\times$$10^{-2}$ & 4.160$\times$$10^{-2}$ & 3.526$\times$$10^{-4}$ & 4.010$\times$$10^{-2}$ & 6.230$\times$$10^{-4}$ & 4.379$\times$$10^{-2}$ & & 4.262$\times$$10^{-2}$ \\[1.5mm]
& 001 & 4.574$\times$$10^{-2}$ & 4.479$\times$$10^{-2}$ & 9.524$\times$$10^{-4}$ & 4.319$\times$$10^{-2}$ & -4.911$\times$$10^{-4}$ & 4.730$\times$$10^{-2}$ & & 4.594$\times$$10^{-2}$ \\[2mm]
\multirow{2}{*}{Ni$^{\star}$ (fcc)} & 111 & 6.081$\times$$10^{-2}$ & 6.147$\times$$10^{-2}$ & -6.608$\times$$10^{-4}$ & 6.049$\times$$10^{-2}$ & -3.737$\times$$10^{-3}$ & 6.654$\times$$10^{-2}$ & & 6.435$\times$$10^{-2}$ \\[1.5mm]
& 001 & 6.028$\times$$10^{-2}$ & 6.083$\times$$10^{-2}$ & -5.495$\times$$10^{-4}$ & 5.985$\times$$10^{-2}$ & -3.566$\times$$10^{-3}$ & 6.587$\times$$10^{-2}$ & & 6.369$\times$$10^{-2}$ \\[2mm]
\multirow{2}{*}{Ti (fcc)} & 111 & -4.331$\times$$10^{-3}$ & -5.411$\times$$10^{-3}$ & 1.080$\times$$10^{-3}$ & -2.431$\times$$10^{-3}$ & -7.451$\times$$10^{-4}$ & -3.149$\times$$10^{-3}$ & & -2.954$\times$$10^{-3}$ \\[1.5mm]
& 001 & -4.622$\times$$10^{-3}$ & -5.781$\times$$10^{-3}$ & 1.159$\times$$10^{-3}$ & -2.586$\times$$10^{-3}$ & -7.929$\times$$10^{-4}$ & -3.391$\times$$10^{-3}$ & & -3.158$\times$$10^{-3}$ \\[2mm]
\multirow{2}{*}{Pt (fcc)} & 111 & 6.304$\times$$10^{-2}$ & 1.139$\times$$10^{-1}$ & -5.082$\times$$10^{-2}$ & 4.487$\times$$10^{-2}$ & 9.858$\times$$10^{-3}$ & 5.310$\times$$10^{-2}$ & & 5.075$\times$$10^{-2}$ \\[1.5mm]
& 001 & 6.673$\times$$10^{-2}$ & 1.192$\times$$10^{-1}$ & -5.243$\times$$10^{-2}$ & 4.681$\times$$10^{-2}$ & 1.070$\times$$10^{-2}$ & 5.369$\times$$10^{-2}$ & & 5.327$\times$$10^{-2}$ \\[2mm]
\multirow{2}{*}{Ir (fcc)} & 111 & 1.439$\times$$10^{-2}$ & 4.654$\times$$10^{-2}$ & -3.215$\times$$10^{-2}$ & 3.819$\times$$10^{-3}$ & -3.293$\times$$10^{-3}$ & 1.206$\times$$10^{-2}$ & & 9.956$\times$$10^{-3}$ \\[1.5mm]
& 001 & 1.431$\times$$10^{-2}$ & 4.752$\times$$10^{-2}$ & -3.321$\times$$10^{-2}$ &  3.701$\times$$10^{-3}$ & -3.262$\times$$10^{-3}$ & 1.212$\times$$10^{-2}$ & & 9.927$\times$$10^{-3}$ \\[2mm]
\multirow{2}{*}{Ag (fcc)} & 111 & -5.549$\times$$10^{-5}$ & -9.260$\times$$10^{-4}$ & 8.705$\times$$10^{-4}$ & 1.309$\times$$10^{-3}$ & -5.375$\times$$10^{-4}$ & 8.428$\times$$10^{-4}$ & &8.489$\times$$10^{-4}$ \\[1.5mm]
& 001 & 3.525$\times$$10^{-5}$ & -9.260$\times$$10^{-4}$ & -8.705$\times$$10^{-4}$ & 1.309$\times$$10^{-3}$ & -5.375$\times$$10^{-4}$ & 8.428$\times$$10^{-4}$ & & 8.489$\times$$10^{-4}$ \\[2mm]
\multirow{2}{*}{Au (fcc)} & 111 & 2.627$\times$$10^{-3}$ & -1.922$\times$$10^{-3}$ & 4.548$\times$$10^{-3}$ & 6.568$\times$$10^{-3}$ & -1.217$\times$$10^{-3}$ & 3.977$\times$$10^{-3}$ & & 4.426$\times$$10^{-3}$ \\[1.5mm]
& 001 & 2.267$\times$$10^{-3}$ & -2.502$\times$$10^{-3}$ & 4.768$\times$$10^{-3}$ & 5.552$\times$$10^{-3}$ & -1.790$\times$$10^{-3}$ & 4.732$\times$$10^{-2}$ & & 4.688$\times$$10^{-3}$ \\[2mm]
\multirow{2}{*}{Rh (fcc)} & 111 & 1.017$\times$$10^{-2}$ & 3.665$\times$$10^{-2}$ & -2.648$\times$$10^{-2}$ & 5.929$\times$$10^{-3}$ & -2.033$\times$$10^{-3}$ & 1.065$\times$$10^{-2}$ & & 9.390$\times$$10^{-3}$ \\[1.5mm]
& 001 & 1.133$\times$$10^{-2}$ & 4.464$\times$$10^{-2}$ & -3.331$\times$$10^{-2}$ & 6.452$\times$$10^{-3}$ & -2.391$\times$$10^{-3}$ & 1.092$\times$$10^{-2}$ & & 9.859$\times$$10^{-3}$ \\[2mm]
\multirow{1}{*}{Al (fcc)} & 111 & -1.833$\times$$10^{-4}$ & -1.454$\times$$10^{-4}$ & -3.789$\times$$10^{-5}$ & -4.997$\times$$10^{-4}$ & -6.364$\times$$10^{-5}$ & -7.682$\times$$10^{-4}$ & & -1.315$\times$$10^{-5}$ \\[1.5mm]
\rule{0pt}{3.0ex}V (bcc) & 001 & -1.230$\times$$10^{-2}$ & -1.444$\times$$10^{-2}$ & 2.144$\times$$10^{-3}$ & -7.529$\times$$10^{-3}$ & -9.578$\times$$10^{-4}$ & -9.002$\times$$10^{-3}$ & & -8.596$\times$$10^{-3}$ \\[2mm]
\rule{0pt}{3.0ex}W (bcc) & 001 & 8.595$\times$$10^{-3}$ & 3.827$\times$$10^{-3}$ & 4.768$\times$$10^{-3}$ & -2.101$\times$$10^{-3}$ & 8.332$\times$$10^{-3}$ & -1.794$\times$$10^{-3}$ & & -1.562$\times$$10^{-3}$ \\[2mm]
\rule{0pt}{3.0ex}Bi (hex) & 001 & 9.406$\times$$10^{-2}$ & 5.097$\times$$10^{-2}$ & 4.310$\times$$10^{-2}$ & 1.857$\times$$10^{-2}$ & 7.504$\times$$10^{-2}$ & 2.980$\times$$10^{-2}$ & & 7.880$\times$$10^{-3}$ \\[2mm]
\rule{0pt}{3.0ex}Bi (rho) & 001 & -1.960$\times$$10^{-2}$ & 3.808$\times$$10^{-3}$ & -2.341$\times$$10^{-2}$ & -1.431$\times$$10^{-3}$ & -1.912$\times$$10^{-2}$ & -3.428$\times$$10^{-5}$ & & -6.132$\times$$10^{-4}$ \\[2mm]
\rule{0pt}{3.0ex}Ru (hcp) & 0001 & 4.142$\times$$10^{-3}$ & 1.126$\times$$10^{-2}$ & -7.116$\times$$10^{-3}$ & 1.477$\times$$10^{-3}$ & -1.240$\times$$10^{-3}$ & 4.633$\times$$10^{-3}$ & & 3.595$\times$$10^{-3}$ \\[2mm]
\hline
\hline
\end{tabular}
\end{center}
\caption{\textbf{Orbital magnetization of various materials at $\mathcal{E}_{\text{F}}$.} The results of $M_{z}$, $M^{\text{SR}}_{z}$, $M^{\text{CM}}_{z}$, $M^{(0)}_{z}$, $M^{(2)}_{z}$, $M^{\text{SR}(\text{on})}_{z}$, and $M^{\text{ACA}}_{z}$ are shown. Here, SQA indicates the orientation of the spin-quantization axis. The asterisks attached to Fe, Ni, and Co indicate that the results are obtained with DFT+U. For all materials except the $d$-transition ferromagnets (Fe, Co, Ni), the artificial spin-Zeeman interaction with a magnitude of 0.544 eV is applied. All quantities are given in units of $\mu_{\text{B}}$/atom.}
\label{table:1}
\end{table*}
\clearpage

\appendix


\section{Construction of gauge-covariant objects for orbital magnetization}
\label{appendix:gauge-invariant-objects}

\subsection{Introduction to the gauge-covariant derivative}

In this section, we revisit the derivation of the gauge-invariant formula in Eq.~\eqref{eq:modern2}. Since the partial derivative of a state does not, in general, transform consistently (i.e., in a gauge-covariant manner) under a gauge transformation, the covariant derivative must be introduced when evaluating physical quantities that depend on state derivatives, such as the Berry curvature, the quantum metric, and the modern theory orbital magnetization. We start with the simplest case without degeneracy to illustrate the spirit of constructing gauge-covariant derivative~\cite{Nakahara90Book,Vanderbilt18Book}. The partial derivative of $\ket{u_{n\mathbf{k}}}$, $\ket{\partial_{\alpha} u_{n\mathbf{k}}}$, does not transform covariantly under the $U(1)$ gauge transformation, $\ket{u_{n\mathbf{k}}} \rightarrow \ket{u^{\prime}_{n\mathbf{k}}} = e^{i\phi_{n}(\mathbf{k})}\ket{u_{n\mathbf{k}}}$:
\begin{align}\label{eq:U1_gauge_transform}
    \ket{\partial_{\alpha} u^{\prime}_{n\mathbf{k}}} = i\partial_{\alpha}\phi_{n}(\mathbf{k}) \ket{u^{\prime}_{n\mathbf{k}}} + e^{i\phi_{n}(\mathbf{k})}\ket{\partial_{\alpha} u_{n\mathbf{k}}},
\end{align}
where $\phi_{n} (\mathbf{k})$ is a real scalar function. Introducing the Berry connections $A_{\alpha , n} = i\braket{u_{n\mathbf{k}}|\partial_{\alpha}u_{n\mathbf{k}}}$ and $A^{\prime}_{\alpha , n} = i\braket{u^{\prime}_{n\mathbf{k}}|\partial_{\alpha}u^{\prime}_{n\mathbf{k}}}$, Eq.~\eqref{eq:U1_gauge_transform} can be expressed by
\begin{align}\label{eq:U1_gauge_form}
    \ket{\partial_{\alpha} u^{\prime}_{n\mathbf{k}}} +i\ket{u^{\prime}_{n\mathbf{k}}}A^{\prime}_{\alpha , n} = e^{i\phi_{n}(\mathbf{k})} \left[ \ket{\partial_{\alpha} u_{n\mathbf{k}}} + i\ket{u_{n\mathbf{k}}}A_{\alpha , n}\right].
\end{align}
Now, we define the gauge-covariant derivative of the state $\ket{u_{n\mathbf{k}}}$ of $U(1)$ gauge transformation, $\ket{D_{\alpha}u_{n\mathbf{k}}}$, by
\begin{align}\label{eq:cov_derivative}
    \ket{D_{\alpha}u_{n\mathbf{k}}} = D_{\alpha}\ket{u_{n\mathbf{k}}} = \left( \partial_{\alpha} + iA_{\alpha , n}\right) \ket{u_{n\mathbf{k}}}.
\end{align}
Then, we obtain $\ket{D_{\alpha}u^{\prime}_{n\mathbf{k}}} = \ket{D_{\alpha}u_{n\mathbf{k}}}$. This covariant derivative can also be expressed as
\begin{align}\label{eq:cov_derivative2}
    \ket{D_{\alpha}u_{n\mathbf{k}}} = \hat{Q}_{n\mathbf{k}} \ket{\partial_{\alpha}u_{n\mathbf{k}}} = (\hat{\mathds{1}} - \hat{P}_{n\mathbf{k}}) \ket{\partial_{\alpha}u_{n\mathbf{k}}},
\end{align}
where $\hat{P}_{n\mathbf{k}} = \ket{u_{n\mathbf{k}}}\bra{u_{n\mathbf{k}}}$ is the projection operator. Note that $\hat{Q}_{n\mathbf{k}}$ eliminates the additional term, which breaks the covariance, by removing the components of $\ket{\partial_{\alpha}u_{n\mathbf{k}}}$ that are parallel to $\ket{u_{n\mathbf{k}}}$.

Next, we extend this to the $U(N)$ gauge problem that arises from the $N$-fold energy degeneracy. In this situation, $U(N)$ gauge transformation, represented by the $N$-dimensional unitary matrix $U(\mathbf{k})$, such that
\begin{align}\label{eq:UN_gauge_transform}
    \ket{u_{n\mathbf{k}}} \rightarrow \ket{u_{n\mathbf{k}}^{\prime}} = \sum_{m \in \mathcal{D}} \ket{u_{m\mathbf{k}}}U_{mn}(\mathbf{k}),
\end{align}
mixes the states in the $N$-dimensional degenerate eigenspace $\mathcal{D} = \mathcal{D}(\mathbf{k})$, a subspace of the full Hilbert space $\mathcal{H}$. Since the gauge transformation acts on $\mathcal{D}$, we refer to this subspace ``$\mathcal{D}$'' as the ``\textit{active subspace}'' of $U$, and to its complement subspace $\mathcal{H} - \mathcal{D}$ as the ``\textit{inactive subspace}'' of $U$~\cite{Liu23arXiv}. Under the gauge transformation $U$, the partial derivative of state transforms by
\begin{align}\label{eq:UN_derivative}
    \ket{\partial_{\alpha} u^{\prime}_{n\mathbf{k}}} = \sum_{m \in \mathcal{D}} \left[ \ket{u_{m\mathbf{k}}} \partial_{\alpha}U_{mn}(\mathbf{k}) + \ket{\partial_{\alpha}u_{m\mathbf{k}}}U_{mn}(\mathbf{k}) \right].
\end{align}
Introducing the non-Abelian Berry connections $A_{mn} = i\braket{u_{m\mathbf{k}}|\partial_{\alpha}u_{n\mathbf{k}}}$ and $A^{\prime}_{mn} = i\braket{u^{\prime}_{m\mathbf{k}}|\partial_{\alpha}u^{\prime}_{n\mathbf{k}}}$, Eq.~\eqref{eq:UN_derivative} can be expressed by
\begin{align}\label{eq:UN_gauge_form}
    \ket{\partial_{\alpha} u^{\prime}_{n\mathbf{k}}} + i\sum_{l \in \mathcal{D}}\ket{u^{\prime}_{l\mathbf{k}}}A^{\prime}_{ln} = \sum_{m \in \mathcal{D}} \left[ \ket{\partial_{\alpha}u_{m\mathbf{k}}} + i\sum_{l \in \mathcal{D}} \ket{u_{l\mathbf{k}}} A_{lm} \right].
\end{align}
Here, we define the gauge-covariant derivative of $u_{n\mathbf{k}}$ of the $U(N)$ gauge transformation by
\begin{align}\label{eq:UN_cov_derivative}
    \ket{D_{\alpha}u_{n\mathbf{k}}} = \ket{\partial_{\alpha}u_{n\mathbf{k}}} + i\sum_{m \in \mathcal{D}} \ket{u_{m\mathbf{k}}} A_{\alpha ,mn}.
\end{align}
As in the case of the $U(1)$ gauge, this covariant derivative can also be expressed as
\begin{align}\label{eq:UN_cov_derivative2}
    \ket{D_{\alpha}u_{n\mathbf{k}}} = \hat{Q}_{\mathcal{D}\mathbf{k}} \ket{\partial_{\alpha}u_{n\mathbf{k}}} = (\hat{\mathds{1}} - \hat{P}_{\mathcal{D}\mathbf{k}}) \ket{\partial_{\alpha}u_{n\mathbf{k}}},
\end{align}
where $\hat{P}_{\mathcal{D}\mathbf{k}} = \sum_{n \in \mathcal{D}}\hat{P}_{n\mathbf{k}}$ is the projection operator in the active subspace $\mathcal{D}$. Here, $\hat{Q}_{\mathcal{D}\mathbf{k}}$ throws the derivative of the state into the inactive subspace, thereby eliminating the additional term that breaks the covariance. In a similar manner, we can define the covariant derivative not only for states but also for operators~\cite{Nakahara90Book}.

This formulation, which ensures the covariance of the state derivative under unitary transformations, can be naturally extended beyond the $U(N)$ gauge problem to arbitrary unitary transformations. For a given $U$, one can define an active subspace $\mathcal{D}$ and introduce the corresponding projection operators, from which a covariant operator with respect to $U$ can be constructed. It should be noted, however, that this covariance is achieved by discarding part of the information contained in the derivative of the states. The validity of the introduced covariant derivative is determined by how much physically meaningful information is lost in this reduction. In the case of the $U(N)$ gauge problem, the covariant derivative defined in Eq.~\eqref{eq:UN_cov_derivative} removes only the dummy information arising from the mixing between degenerate energy eigenstates and thus does not eliminate any physically meaningful quantity. On the other hand, if the introduced covariant derivative removes physically meaningful information that contributes significantly to the quantity of interest, then--even though it transforms covariantly under the target unitary transformations--the resulting quantity becomes unreliable, as the relevant information is lost. Therefore, this provides a criterion for assessing the validity of a given covariant derivative.

\subsection{Gauge-covariant objects for orbital magnetization}

Now, we construct the gauge-covariant objects for orbital magnetization. First, we focus on the quantum geometric tensor $\hat{F}_{\mathbf{k},\alpha\beta} = \ket{u_{n\mathbf{k}}}\braket{\partial_{\alpha}u_{n\mathbf{k}}|\partial_{\beta}u_{m\mathbf{k}}}\bra{u_{m\mathbf{k}}}$, one of the building blocks of the orbital magnetization and Berry curvature. This is not covariant under the gauge transformation and thus generates additional terms. To eliminate these additional terms, we introduce the covariant derivative $D_{\alpha}$, which yields $\ket{u_{n\mathbf{k}}}\braket{D_{\alpha}u_{n\mathbf{k}}|D_{\beta}u_{m\mathbf{k}}}\bra{u_{m\mathbf{k}}}$. Other building blocks can also be obtained as gauge-covariant forms in a similar way. They can be expressed by products of the projection operators~\cite{Ceresoli06PRB}:
\begin{subequations}\label{eq:building_blocks}
\begin{align}\label{eq:building_blocks_F}
    \hat{F}_{\mathcal{D}\mathbf{k}, \alpha \beta} = (\partial_{\alpha} \hat{P}_{\mathcal{D}\mathbf{k}}) \hat{Q}_{\mathcal{D}\mathbf{k}} (\partial_{\beta} \hat{P}_{\mathcal{D}\mathbf{k}}),
\end{align}
\begin{align}\label{eq:building_blocks_G}
    \hat{G}_{\mathcal{D}\mathbf{k}, \alpha \beta} = (\partial_{\alpha} \hat{P}_{\mathcal{D}\mathbf{k}}) \hat{Q}_{\mathcal{D}\mathbf{k}} \hat{H}_{\mathbf{k}} \hat{Q}_{\mathcal{D}\mathbf{k}} (\partial_{\beta} \hat{P}_{\mathcal{D}\mathbf{k}}),
\end{align}
\begin{align}\label{eq:building_blocks_K}
    \hat{K}_{\mathcal{D}\mathbf{k}, \alpha \beta} = \hat{H}_{\mathbf{k}} (\partial_{\alpha} \hat{P}_{\mathcal{D}\mathbf{k}}) \hat{Q}_{\mathcal{D}\mathbf{k}} (\partial_{\beta} \hat{P}_{\mathcal{D}\mathbf{k}}).
\end{align}
\end{subequations}
These objects transform covariantly under the gauge transformation $U$ whose active subspace is $\mathcal{D}$. Let us divide the Hilbert space into several proper subspaces such as $\mathcal{H} = \bigoplus_{l}\mathcal{D}_{l}$ and consider gauge transformations $U_{l}$ whose active subspaces are $\mathcal{D}_{l}$. Then, $M_{\gamma}(\mathbf{k})$ can be expressed as a trace of a linear combination of these objects:
\begin{align}\label{eq:D_modern}
    M_{\gamma}(\mathbf{k}) = \frac{\varepsilon_{\alpha \beta \gamma}e}{2\hbar}\sum_{l} \biggr\{ f_{\mathcal{D}_{l}\mathbf{k}}\text{Im}\biggr[ &\hat{G}_{\mathcal{D}_{l}\mathbf{k}, \alpha \beta} + \hat{K}_{\mathcal{D}_{l}\mathbf{k}, \alpha \beta} \notag
    \\[2mm]
    &- 2\mathcal{E}_{\text{F}}\hat{F}_{\mathcal{D}_{l}\mathbf{k}, \alpha \beta} \biggr] \biggr\} ,
\end{align}
and this expression is manifestly gauge-invariant under $U_{l}$.

Now, we evaluate $M_{\gamma}(\mathbf{k})$ from a given electronic structure. If the band structure exhibits a $N$-fold degeneracy at $\mathbf{k}$, it is necessary to account for a $U(N)$ gauge transformation, and the active subspaces $\mathcal{D}_{l}$ must be selected such that $M_{\gamma}(\mathbf{k})$ remains invariant under this transformation. If there exists a band crossing, $\mathcal{D}_{l}$ varies with $\mathbf{k}$ near that point. It makes the derivative $\partial_{\alpha}\hat{P}_{D\mathbf{k}}$ singular, and this yields unpredictable calculation results~\cite{Thonhauser12IJMPB}. To eliminate this singularity, it must be assumed that the projection operator is smooth with respect to $\mathbf{k}$. This can be achieved by grouping the entangled bands over the BZ into a single active subspace~\cite{Liu23arXiv}. For example, consider a situation in which the Hilbert space $\mathcal{H}$ can be decomposed into $\mathcal{H}_{A}$ and $\mathcal{H}_{B}$, separated by a finite energy gap. In this case, each of $\mathcal{H}_{A}$ and $\mathcal{H}_{B}$ can be treated as an active subspace, and the corresponding projection operators $\hat{P}_{A}$ and $\hat{P}_{B}$ can be chosen to be smooth functions of $\mathbf{k}$. However, note that such a global extension of active subspaces may cause $\hat{Q}_{A(B)}$ to excessively remove geometric information from the states. Therefore, to achieve smoothness while avoiding such over-reduction of information, the active subspace should be expanded only as much as needed to ensure smoothness but kept as small as possible. In most band structures, the occupied states are entangled. In such cases, to ensure smoothness, one can select an active subspace that encompasses all occupied states~\cite{Ceresoli06PRB,Lopez12PRB,Liu23arXiv}.

In the case of the orbital magnetization, Eqs.~\eqref{eq:building_blocks} and \eqref{eq:D_modern} are replaced by Eqs.~\eqref{eq:modern_FGK} and \eqref{eq:modern2}, respectively, where the projection operators $\hat{P}_{\mathcal{D}\mathbf{k}}$ and $\hat{Q}_{\mathcal{D}\mathbf{k}}$ are replaced by the ground state projectors $\hat{P}_{\mathbf{k}}$ and $\hat{Q}_{\mathbf{k}} = \hat{\mathds{1}} - \hat{P}_{\mathbf{k}}$. To evaluate the orbital magnetization, we sum over all the occupied states. The covariant derivative defined using the projection operator $\hat{Q}_{\mathbf{k}}$ throws the derivative of an occupied state onto the unoccupied subspace, thereby discarding the information between the occupied states. However, when the summation is performed over all the occupied states, these missing contributions cancel each other, and consequently, no physically meaningful information is lost. Therefore, Eq.~\eqref{eq:modern2} is a valid gauge-invariant formula for the orbital magnetization of modern theory that is also suitable for numerical calculations. We note that, in metals, band degeneracies located at the Fermi surface can lead to numerically unstable points. However, the number of such point is usually limited and does not significantly influence the $\mathbf{k}$-summed results.

\section{Derivation of the gauge-invariant form of orbital magnetization}\label{appendix:derivation_orbital_magnetization}

In this section, we derive Eq.~\eqref{eq:modern2} from the definition of $\hat{F}_{\alpha \beta}$, $\hat{G}_{\alpha \beta}$, and $\hat{K}_{\alpha \beta}$. We begin by expressing the derivation of the ground state projector in the Wannier gauge.
\begin{align}\label{eq:partial_P}
    \partial_{\alpha}\hat{P} = \sum_{nn^{\prime}}^{J} &\left( \ket{\partial_{\alpha} u_{n}^{\text{W}}}f_{nn^{\prime}}^{\text{W}}\bra{u_{n^{\prime}}^{\text{W}}} + \ket{u_{n}^{\text{W}}}f_{nn^{\prime}}^{\text{W},\alpha}\bra{u_{n^{\prime}}^{\text{W}}} \right. \nonumber
    \\
    &\left. + \ket{u_{n}^{\text{W}}}f_{nn^{\prime}}^{\text{W}}\bra{\partial_{\alpha} u_{n^{\prime}}^{\text{W}}} \right) ,
\end{align}
where $f_{nm}^{\text{W},\alpha} = \partial_{\alpha} f_{nm}^{\text{W}}$. Here, we use the reduced notation introduced in Sec.~\ref{subsec:Wannier_reresentation}.

We focus on the occupation matrix $f^{\text{W}} = [f^{\text{W}}_{nm}]_{L \times L}$ and its complement $g^{\text{W}} = [g^{\text{W}}_{nm}]_{L \times L}$ in the Wannier gauge and reveal their properties. Here, $L = \text{dim}\bar{\mathcal{H}}$. Recall that the occupation matrices are non-diagonal in Wannier gauge. At zero temperature, $f^{\text{W}}$ and $g^{\text{W}}$ are idempotent, as well as projection operators. They commute with the Hamiltonian $\mathbb{H}$ and satisfy $[f^{\text{W}},g^{\text{W}}] = \{f^{\text{W}},g^{\text{W}}\} = 0$. The relation $f^{\text{W}} = Vf^{\text{H}}V^{\dagger}$ ($f^{\text{H}} = [f^{\text{H}}_{n}]_{L \times L}$) entails
\begin{align}\label{eq:f_W_derivative}
    f_{nm}^{\text{W},\alpha} &= i[f^{\text{W}},J_{\alpha}]_{nm},
\end{align}
where $J_{\alpha} = VJ^{\text{H}}_{\alpha}V^{\dagger}$ and $J^{\text{H}}_{\alpha}$ is the connection defined in Eq.~\eqref{eq:J_H}.

Using these properties, Eq.~\eqref{eq:partial_P} can be written by
\begin{align}\label{eq:partial_P2}
    \partial_{\alpha}\hat{P} = \sum_{nn^{\prime}}^{L} &\Big\{ \ket{\partial_{\alpha} u_{n}^{\text{W}}}f_{nn^{\prime}}^{\text{W}}\bra{u_{n^{\prime}}^{\text{W}}} + i\ket{u_{n}^{\text{W}}}[f^{\text{W}},J_{\alpha}]_{nm}\bra{u_{n^{\prime}}^{\text{W}}} \nonumber
    \\
    & + \ket{u_{n}^{\text{W}}}f_{nn^{\prime}}^{\text{W}}\bra{\partial_{\alpha} u_{n^{\prime}}^{\text{W}}} \Big\} .
\end{align}
Since
\begin{align}\label{eq:partial_P_bbQ}
    ( \partial_{\alpha}\hat{P} ) \hat{\mathbb{Q}} = \sum_{nn^{\prime}}^{L} \ket{u_{n}^{\text{W}}}f_{nn^{\prime}}^{\text{W}}\bra{\partial_{\alpha}u_{n^{\prime}}^{\text{W}}}\hat{\mathbb{Q}},
\end{align}
and
\begin{align}\label{eq:partial_P_QI}
    ( \partial_{\alpha}\hat{P} ) \hat{Q}^{\text{I}} = i\sum_{nn^{\prime}}^{L} \ket{u_{n}^{\text{W}}}(f^{\text{W}}A_{\alpha}g^{\text{W}})_{nn^{\prime}}\bra{u_{n^{\prime}}^{\text{W}}} ,
\end{align}
we have
\begin{align}\label{eq:partial_P_Q}
    ( \partial_{\alpha}\hat{P} ) \hat{Q} &= ( \partial_{\alpha}\hat{P} ) (\hat{\mathbb{Q}} + \hat{Q}^{\text{I}}) \notag
    \\[2mm]
    &= \sum_{nn^{\prime}}^{L}\Big\{ \ket{u^{\text{W}}_{n}}f_{nn^{\prime}}^{\text{W}}\bra{\partial_{\alpha}u^{\text{W}}_{n^{\prime}}}\hat{\mathbb{Q}} ] 
    \nonumber
    \\
    & \quad \quad \quad \; \; + i\ket{u^{\text{W}}_{n}}(\hat{f}^{\text{W}}A_{\alpha}\hat{g})_{nn^{\prime}}^{\text{W}}\bra{u^{\text{W}}_{n^{\prime}}} \Big\} .
\end{align}
Similarly, we obtain
\begin{align}\label{eq:Q_partial_P}
    \hat{Q}( \partial_{\alpha}\hat{P} ) = \sum_{nn^{\prime}}^{L} \Big\{& \hat{\mathbb{Q}}\ket{\partial_{\alpha}u^{\text{W}}_{n}}f^{\text{W}}_{nn^{\prime}}\bra{u^{\text{W}}_{n^{\prime}}} \nonumber
    \\
    &- i\ket{u^{\text{W}}_{n}}(g^{\text{W}}A_{\alpha}f^{\text{W}})_{nn^{\prime}}\bra{u^{\text{W}}_{n^{\prime}}} \Big\} .
\end{align}
Then, Eq.~\eqref{eq:covariant_F} is obtained by straightforward calculations. $\hat{\mathcal{G}}_{\alpha\beta}$ [Eq.~\eqref{eq:covariant_G}] and $\hat{\mathcal{K}}_{\alpha\beta}$ [Eq.~\eqref{eq:covariant_K}] are derived in the same way.

The imaginary parts of Eq.~\eqref{eq:covariant_FGK} are given by
\begin{subequations}\label{eq:imaginary_covariant_FGK}
\begin{align}\label{eq:imaginary_covariant_F}
    \text{Im} \, \hat{\mathcal{F}}_{\alpha \beta} &= -\frac{1}{2} \text{Re} \left[ \hat{P}\widetilde{\mathbb{\Omega}}_{\alpha \beta}\hat{P} \right] + \text{Im} \left[ \hat{P} A_{\alpha} \hat{Q}^{\text{I}} A_{\beta} \hat{P} \right] ,
\end{align}
\begin{align}\label{eq:imaginary_covariant_G}
    \text{Im} \, \hat{\mathcal{K}}_{\alpha \beta} &= -\frac{1}{4} \text{Re} \left[ \hat{P}\{ \mathbb{H}, \widetilde{\mathbb{\Omega}}_{\alpha \beta}\} \hat{P} \right] + \frac{1}{2}\text{Im} \left[ \hat{P} \{ \mathbb{H}, A_{\alpha} \hat{Q}^{\text{I}} A_{\beta} \} \hat{P} \right] ,
\end{align}
\begin{align}\label{eq:imaginary_covariant_K}
    \text{Im} \, \hat{\mathcal{G}}_{\alpha \beta} = &-\frac{1}{2} \text{Re} \left[ \hat{P}\widetilde{\mathbb{\Lambda}}_{\alpha \beta}\hat{P} \right] + \text{Im} \left[ \hat{P}\widetilde{\mathbb{B}}^{\dagger}_{\alpha}\hat{Q}^{\text{I}}A_{\beta}\hat{P} \right. \notag
    \\[2mm]
    &\left. + \hat{P}A_{\alpha}\hat{Q}^{\text{I}}\widetilde{\mathbb{B}}_{\beta}\hat{P} + \hat{P} A_{\alpha} \hat{Q}^{\text{I}} \mathbb{H} \hat{Q}^{\text{I}} A_{\beta} \hat{P} \right] ,
\end{align}
\end{subequations}
where
\begin{align}\label{eq:Lambda}
    \widetilde{\mathbb{\Lambda}}_{\alpha \beta} = i\widetilde{\mathbb{C}}_{\alpha \beta} - i\widetilde{\mathbb{C}}_{\alpha \beta}^{\dagger}.
\end{align}
and
\begin{align}\label{eq:Omega}
    \widetilde{\mathbb{\Omega}}_{\alpha \beta} = i\widetilde{\mathbb{F}}_{\alpha \beta} - i\widetilde{\mathbb{F}}_{\alpha \beta}^{\dagger}.
\end{align}

\section{Gauge selection and orbital magnetization}\label{appendix:gauge_selection}

In this section, we investigate how the orbital magnetization [Eq.~\eqref{eq:modern3}] and the orbital moment operators [Eq.~\eqref{eq:modern_operator_covariant_derivative}] transform between two different Wannier gauges. Here, we assume that both two Wannier gauges are selected within the same reduced Hilbert space. Then, the reduced Hamiltonian $\mathbb{H}_{\mathbf{k}} = \hat{\mathbb{P}}_{\mathbf{k}}\hat{H}_{\mathbf{k}}\hat{\mathbb{P}}_{\mathbf{k}}$ is identical in the two gauge.  Let us denote the reduced Hilbert space as $\widetilde{\mathcal{H}}_{\mathbf{k}}$, the occupied space as $\widetilde{\mathcal{H}}^{\textrm{o}}_{\mathbf{k}}$, and the unoccupied space as $\widetilde{\mathcal{H}}^{\textrm{u}}_{\mathbf{k}}$ ($\widetilde{\mathcal{H}}^{\textrm{o}}_{\mathbf{k}} \oplus \widetilde{\mathcal{H}}^{\textrm{u}}_{\mathbf{k}} = \widetilde{\mathcal{H}}_{\mathbf{k}}$). The projection operators on $\widetilde{\mathcal{H}}_{\mathbf{k}}$, $\widetilde{\mathcal{H}}^{\textrm{o}}_{\mathbf{k}}$, $\widetilde{\mathcal{H}}^{\textrm{u}}_{\mathbf{k}}$, and $\mathcal{H}_{\mathbf{k}} - \widetilde{\mathcal{H}}_{\mathbf{k}}$ are $\hat{\mathbb{P}}_{\mathbf{k}}$, $\hat{P}_{\mathbf{k}}$, $\hat{Q}^{\text{I}}_{\mathbf{k}}$, and $\hat{\mathbb{Q}}_{\mathbf{k}}$, respectively. Recall that Eqs.~\eqref{eq:modern3} and \eqref{eq:modern_operator_covariant_derivative} are constructed using the covariant derivative $\hat{Q}_{\mathbf{k}} \ket{u_{n\mathbf{k}}}$. As shown in Appendix~\ref{appendix:gauge-invariant-objects}, this covariant derivative guaranties that Eqs.~\eqref{eq:modern3} and \eqref{eq:modern_operator_covariant_derivative} are invariant and covariant, respectively, under any unitary transformation whose active subspace lies within $\widetilde{H}^{\textrm{o}}_{\mathbf{k}}$. However, it does not ensure the covariance of the state derivative under a unitary transformation that mixes the occupied and unoccupied states. Since the Wannierization process for metals necessarily involves such mixing, it is not obvious whether Eqs.~\eqref{eq:modern3} and \eqref{eq:modern_operator_covariant_derivative} are gauge invariant and covariant under transformations between different Wannier gauges.

Although unitary transformations between Wannier gauges mix the occupied and unoccupied subspaces, we show that if the space selection is fixed, then Eqs.~\eqref{eq:modern3} and \eqref{eq:modern_operator_covariant_derivative} remain gauge invariant and covariant, respectively, under any transformation between two Wannier gauges.

To prove this, we take two arbitrary Wannier gauges $\text{W}_{1}$ and $\text{W}_{2}$, and introduce the corresponding Hamiltonian gauges $\text{H}_{1}$ and $\text{H}_{2}$, associated with each Wannier gauge. Unitary transformations $W$, $V_{1}$, $V_{2}$, and $G$ are defined as follows:
\begin{subequations}\label{eq:transform_def}
\begin{align}\label{eq:AB1a}
    W: \ket{u_{n\mathbf{k}}^{\text{W}_{2}}} = \sum_{m}^{L} \ket{u_{m\mathbf{k}}^{\text{W}_{1}}}W_{mn}(\mathbf{k}),
\end{align}
\begin{align}\label{eq:AB1b}
    V_{1}: \ket{u_{n\mathbf{k}}^{\text{H}_{1}}} = \sum_{m}^{L} \ket{u_{m\mathbf{k}}^{\text{W}_{1}}}(V_{1})_{mn}(\mathbf{k}),
\end{align}
\begin{align}\label{eq:AB1c}
    V_{2}: \ket{u_{n\mathbf{k}}^{\text{H}_{2}}} = \sum_{m}^{L} \ket{u_{m\mathbf{k}}^{\text{W}_{2}}}(V_{2})_{mn}(\mathbf{k}),
\end{align}
\begin{align}\label{eq:AB1d}
    G: \ket{u_{n\mathbf{k}}^{\text{H}_{2}}} = \sum_{m}^{L} \ket{u_{m\mathbf{k}}^{\text{H}_{1}}}G_{mn}(\mathbf{k}).
\end{align}
\end{subequations}
Note that even if $\text{W}_{1}$ and $\text{W}_{2}$ were fixed, $V_{1}$ and $V_{2}$ are not uniquely determined. Here, since we assumed that both the Wannier gauges $\text{W}_{1}$ and $\text{W}_{2}$ are defined on the same space $\bar{\mathcal{H}}$, all of these are transformations on $\bar{\mathcal{H}}$.

The objects constructed using the covariant derivative with respect to $\bar{\mathcal{H}}$, such as $\widetilde{\mathbb{B}}_{\alpha}$, $\widetilde{\mathbb{C}}_{\alpha \beta}$, $\widetilde{\mathbb{F}}_{\alpha \beta}$, $\widetilde{\mathbb{\Omega}}_{\alpha \beta}$, and $\widetilde{\mathbb{\Lambda}}_{\alpha \beta}$, are covariant under the transformation $W$, $V_{1}$, $V_{2}$, and $G$. For example, $\widetilde{\mathbb{F}}_{\alpha \beta}^{\text{W}_{1}} = \sum_{n,m}\ket{u_{n}^{\text{W}_{1}}}\braket{\partial_{\alpha}u_{n}^{\text{W}_{1}}|\hat{\mathbb{Q}}|\partial_{\beta}u_{m}^{\text{W}_{1}}}\bra{u_{m}^{\text{W}_{1}}}$ on the gauge $W_{1}$ transforms as follows:
\begin{align}\label{eq:F_transform}
    \widetilde{\mathbb{F}}_{\alpha \beta}^{\text{W}_{1}} &= W\widetilde{\mathbb{F}}_{\alpha \beta}^{\text{W}_{2}}W^{\dagger} = V_{1}\widetilde{\mathbb{F}}_{\alpha \beta}^{\text{H}_{1}}V_{1}^{\dagger} = V_{1}G\widetilde{\mathbb{F}}_{\alpha \beta}^{\text{H}_{2}}G^{\dagger}V_{1}^{\dagger}.
\end{align}
Here, $\widetilde{\mathbb{F}}_{\alpha \beta}^{\text{H}_{i}}$ denotes $\sum_{n,m}\ket{u_{n}^{\text{H}_{i}}}\braket{\partial_{\alpha}u_{n}^{\text{H}_{i}}|\hat{\mathbb{Q}}|\partial_{\beta}u_{m}^{\text{H}_{i}}}\bra{u_{m}^{\text{H}_{i}}}$, where $\textrm{H}_{i}$ indicates the gauge $\textrm{H}_{1}$ or $\textrm{H}_{2}$.

In contrast, the Berry connection $\mathbb{A}_{\alpha}$ and $\mathbb{B}_{\alpha}$ are not covariant under these transformations:
\begin{align}\label{eq:A_rule}
    U^{\dagger}\mathbb{A}_{\alpha}U &= \mathbb{A}_{\alpha}^{\prime} - J_{\alpha}^{\prime},
\end{align}
\begin{align}\label{eq:B_rule}
    U^{\dagger}\mathbb{B}_{\alpha}U &= \mathbb{B}_{\alpha}^{\prime} - \mathbb{H}^{\prime}J_{\alpha}^{\prime},
\end{align}
where $J^{\prime}_{\alpha} = U^{\dagger}(\partial_{\alpha}U)$, $J_{\alpha} = (\partial_{\alpha}U)U^{\dagger}$ and $U: \ket{u} \mapsto \ket{u^{\prime}}$ indicate one of $W$, $V_{1}$, $V_{2}$, and $G$. To address this, we define the gauge-corrected Berry connections $A_{\alpha}^{1}$ and $A_{\alpha}^{2}$ that incorporate the gauge corrections $J^{1}_{\alpha} = (\partial_{\alpha}V_{1})V_{1}^{\dagger}$ and $J^{2}_{\alpha} = (\partial_{\alpha}V_{2})V_{2}^{\dagger}$ corresponding to transformations $V_{1}$ and $V_{2}$, respectively:
\begin{align}\label{eq:corrected_A}
    A^{1}_{\alpha} &= \mathbb{A}^{\text{W}_{1}}_{\alpha} + J^{1}_{\alpha}, \quad A^{2}_{\alpha} = \mathbb{A}^{\text{W}_{2}}_{\alpha} + J^{2}_{\alpha}.
\end{align}
Then, $A^{1}_{\alpha}$ and $A^{2}_{\alpha}$ transform consistently under $V_{1}$ and $V_{2}$, respectively:
\begin{align}\label{eq:transformation_A}
    V_{1}^{\dagger}A^{1}_{\alpha}V_{1} &= \mathbb{A}^{\text{H}_{1}}_{\alpha}, \quad V_{2}^{\dagger}A^{2}_{\alpha}V_{2} = \mathbb{A}^{\text{H}_{2}}_{\alpha}.
\end{align}
Under the transformation $W$, $A^{1}_{\alpha}$ and $A^{2}_{\alpha}$ follow the rule:
\begin{align}\label{eq:AB7}
    A^{1}_{\alpha} &= WA^{2}_{\alpha}W^{\dagger} + i WV_{2}(\partial_{\alpha}G^{\dagger})GV_{2}^{\dagger}W^{\dagger}.
\end{align}
Note that $G$ commutes with the Hamiltonian $\mathbb{H}$ for every $\mathbf{k}$.

Assume that the energy spectrum $\{ \bar{\mathcal{E}}_{n}^{1} \}$ of Hamiltonian gauge $\textrm{H}_{1}$ is the same as the energy spectrum $\{ \bar{\mathcal{E}}_{n}^{2} \}$ of Hamiltonian gauge $\textrm{H}_{2}$. In the absence of degeneracy, $G$ is a diagonal matrix for every $\mathbf{k}$, and this implies that $(\partial_{\alpha}G^{\dagger})G$ is also diagonal. Then, the second term in Eq.~\eqref{eq:AB7} vanishes between $\hat{P}$ and $\hat{Q}^{\text{I}}$:
\begin{subequations}\label{eq:PAQ_QAP}
\begin{align}\label{eq:PAQ}
    \hat{P}^{\text{W}_{1}}A^{1}_{\alpha}\hat{Q}^{\text{IW}_{1}} &= W(\hat{P}^{\text{W}_{2}}A^{2}_{\alpha}\hat{Q}^{\text{IW}_{2}})W^{\dagger},
\end{align}
\begin{align}\label{eq:QAP}
    \hat{Q}^{\text{IW}_{1}}A^{1}_{\alpha}\hat{P}^{\text{W}_{1}} &= W(\hat{Q}^{\text{IW}_{2}}A^{2}_{\alpha}\hat{P}^{\text{W}_{2}})W^{\dagger}.
\end{align}
\end{subequations}
When degeneracy is present, $(\partial_{\alpha}G^{\dagger})G$ becomes block-diagonal, leading to the same conclusion. Since the reduced Hamiltonian is the same in the two gauge, $\{ \bar{\mathcal{E}}^{1}_{n} \} = \{ \bar{\mathcal{E}}^{2}_{n} \}$. Therefore, Eqs.~\eqref{eq:modern3} and \eqref{eq:modern_operator_covariant_derivative} remain gauge invariant and covariant, respectively, under any transformation between two Wannier gauges. In fact, Eq.~\eqref{eq:imaginary_covariant_FGK} holds as long as $G$ does not mix the occupied and unoccupied states. This implies that if the energy spectrum of the occupied or the unoccupied states is identical, then the additional term appearing in Eq.~\eqref{eq:AB7} vanishes.

The above argument is valid if there exists no degeneracy on the Fermi surface. If degeneracy occurs exactly on the Fermi surface, the additional term may not vanish. However, since the number of such $\mathbf{k}$ points is usually limited, their contribution can be neglected when integrating over the entire BZ. Consequently, $\hat{P}A_{\alpha}\hat{Q}^{\text{I}}$ and $\hat{Q}^{\text{I}}A_{\alpha}\hat{P}$ are almost gauge covariant under the transformation between the Wannier gauges.

\section{Space selection and orbital magnetization}\label{appendix:space_selection}

In the previous section, we considered the case in which the same space is selected but only the gauge choice is different. We now turn to the situation in which the two Wannier gauges are defined over different subspaces. We assume that the Wannier gauge $\text{W}_{1}$ is defined in the subspace $\bar{\mathcal{H}}^{\text{in}} \oplus \Delta \bar{\mathcal{H}}^{1}$ and $\text{W}_{2}$ is defined in $\bar{\mathcal{H}}^{\text{in}} \oplus \Delta \bar{\mathcal{H}}^{2}$. Here, $\bar{\mathcal{H}}^{\text{in}}$ denotes the subspace spanned by the states within the inner window. The projection operators onto $\bar{\mathcal{H}}^{\text{in}}$, $\Delta \bar{\mathcal{H}}^{1}$, and $\Delta \bar{\mathcal{H}}^{2}$ are denoted by $\hat{\mathbb{P}}^{\text{in}}$, $\Delta \hat{\mathbb{P}}^{1}$, and $\Delta \hat{\mathbb{P}}^{2}$, respectively. The dimensions of each space are $\text{dim}\,\bar{\mathcal{H}}^{\text{inner}} = L_{\text{c}}$, $\text{dim}\,\Delta \bar{\mathcal{H}}^{1} = L_{1}$, and $\text{dim}\,\Delta \bar{\mathcal{H}}^{2} = L_{2}$. We first assume that $\Delta\bar{\mathcal{H}}^{1} \ge \Delta\bar{\mathcal{H}}^{2}$. We introduce the corresponding Hamiltonian gauges $\text{H}_{1}$ and $\text{H}_{2}$ associated with $\text{W}_{1}$ and $\text{W}_{2}$. Unitary transformations $V_{1}$ and $V_{2}$ are defined by
\begin{subequations}\label{eq:space_transform}
\begin{align}\label{eq:space_transform_1}
    V_{1}: \ket{u_{n\mathbf{k}}^{\text{H}_{1}}} = \sum_{m}^{L_{\text{c}}+L_{1}} \ket{u_{m\mathbf{k}}^{\text{W}_{1}}}(V_{1})_{mn}(\mathbf{k}),
\end{align}
\begin{align}\label{eq:space_transform_2}
    V_{2}: \ket{u_{n\mathbf{k}}^{\text{H}_{2}}} = \sum_{m}^{L_{\text{c}}+L_{2}} \ket{u_{m\mathbf{k}}^{\text{W}_{2}}}(V_{2})_{mn}(\mathbf{k}).
\end{align}
\end{subequations}
Transformations $W: W_{1} \rightarrow W_{2}$ and $G: H_{1} \rightarrow H_{2}$ are not unitary, represented by $(L_{\text{c}}+L_{1}) \times (L_{\text{c}}+L_{2})$ matrices. The reduced Hamiltonian of each gauge is given by
\begin{align}\label{eq:space_Hamiltonian}
    \mathbb{H}^{i} = (\hat{\mathbb{P}}^{\text{in}} + \Delta \hat{\mathbb{P}}^{i})\hat{H}(\hat{\mathbb{P}}^{\text{in}} + \Delta \hat{\mathbb{P}}^{i}),
\end{align}
where $i = 1$ or $2$. Using the commutation relation $[\hat{\mathbb{P}}^{\text{in}}, \hat{H}] = 0$ and the orthogonality $\hat{\mathbb{P}}^{\text{in}} \Delta \hat{\mathbb{P}}^{i} = 0$, we obtain that
\begin{align}\label{eq:space_PHP}
    \hat{\mathbb{P}}^{\text{in}}\hat{H}\Delta \hat{\mathbb{P}}^{i} = \Delta \hat{\mathbb{P}}^{i}\hat{H}\hat{\mathbb{P}}^{\text{in}} = 0.
\end{align}
Thus, $\mathbb{H}^{1}$ and $\mathbb{H}^{2}$ share the common block $\hat{\mathbb{P}}^{\text{in}}\hat{H}\hat{\mathbb{P}}^{\text{in}}$.

Since the reduced Hilbert spaces of $\text{W}_{1}$ and $\text{W}_{2}$ differ, covariant objects such as $\widetilde{\mathbb{B}}_{\alpha}$, $\widetilde{\mathbb{C}}_{\alpha \beta}$, $\widetilde{\mathbb{F}}_{\alpha \beta}$, $\widetilde{\mathbb{\Omega}}_{\alpha \beta}$, and $\widetilde{\mathbb{\Lambda}}_{\alpha \beta}$ are no longer covariant under the transformations between the two gauges. For example, we consider
\begin{align}\label{eq:space_F}
    \widetilde{\mathbb{F}}_{\alpha \beta, nm}^{\text{H}_{i}} = \braket{\partial_{\alpha}u_{n}^{\text{H}_{i}}|(\mathbb{1} - \hat{\mathbb{P}}^{\text{in}} - \Delta \hat{\mathbb{P}}^{i})|\partial_{\beta}u_{m}^{\text{H}_{i}}}.
\end{align}
Assume that $\bar{\mathcal{H}}^{\text{in}}$ contains all the occupied states (i.e., the inner window is chosen above $\mathcal{E}_{\text{F}}$). Then, we have
\begin{align}\label{eq:space_PFP}
    [\hat{P}^{\text{in}}(\widetilde{\mathbb{F}}_{\alpha \beta}^{\text{H}_{2}} - G^{\dagger}\widetilde{\mathbb{F}}_{\alpha \beta}^{\text{H}_{1}}G)\hat{P}^{\text{in}}]_{nm} = f_{n}\bra{\partial_{\alpha}u_{n}^{\text{H}_{2}}}& \notag
    \\[2mm]
    \cdot (\Delta \hat{\mathbb{P}}^{1} - \Delta \hat{\mathbb{P}}^{2})\ket{\partial_{\beta}u_{m}^{\text{H}_{2}}}f_{m}&,
\end{align}
which does not vanish. Here, $\hat{P}^{\text{in}} = \hat{\mathbb{P}}^{\text{in}}\hat{P}\hat{\mathbb{P}}^{\text{in}}$ and $\hat{Q}^{\text{in}} = \hat{\mathbb{P}}^{\text{in}}\hat{Q}\hat{\mathbb{P}}^{\text{in}}$. On the other hand, we also have
\begin{align}\label{eq:space_AQA}
    (\mathbb{A}^{\text{H}_{i}}_{\alpha}\hat{Q}^{\text{H}_{i}}\mathbb{A}^{\text{H}_{i}}_{\beta})_{nm} = \braket{\partial_{\alpha}u_{n}^{\text{H}_{i}}|\hat{Q}^{\text{I}_{i}}|\partial_{\beta}u_{m}^{\text{H}_{i}}}.
\end{align}
We use $[\hat{P}^{\text{H}_{2}},G]=0$ and Eq.~\eqref{eq:AB7} to obtain the following:
\begin{align}\label{eq:space_AQA_rule}
    [\hat{P}^{\text{in}}(\mathbb{A}^{\text{H}_{2}}_{\alpha}\hat{Q}^{\text{H}_{2}}\mathbb{A}^{\text{H}_{2}}_{\beta} - G^{\dagger}\mathbb{A}^{\text{H}_{1}}_{\alpha}\hat{Q}^{\text{H}_{1}}\mathbb{A}^{\text{H}_{1}}_{\beta}G)\hat{P}^{\text{in}}]_{nm}& \notag
    \\[2mm]
    = f_{n}\bra{\partial_{\alpha}u_{n}^{\text{H}}}(\hat{Q}^{\text{I}_{2}}-\hat{Q}^{\text{I}_{1}})\ket{\partial_{\beta}u_{m}^{\text{H}}}f_{m}&.
\end{align}
Note that the additional terms proportional to $(\partial_{\alpha}G^{\dagger})G$ in Eq.~\eqref{eq:space_AQA_rule} vanish as long as the inner window is chosen above $\mathcal{E}_{\text{F}}$. From the relation $\hat{Q}^{\text{I}_{i}} = \Delta \hat{\mathbb{P}}^{i} + \hat{\mathbb{P}}^{\text{in}} - \hat{P}$, we arrive at
\begin{align}\label{eq:space_QGT}
    \hat{\mathbb{P}}^{\text{in}}\hat{\mathcal{F}}_{\alpha \beta}^{\text{H}_{2}}\hat{\mathbb{P}}^{\text{in}} = \hat{\mathbb{P}}^{\text{in}}G^{\dagger}\hat{\mathcal{F}}_{\alpha \beta}^{\text{H}_{1}}G\hat{\mathbb{P}}^{\text{in}}.
\end{align}
which shows that $\text{tr}\{\hat{\mathcal{F}}_{\alpha \beta}\}$ yields the same values for an arbitrary space and its subspace. This implies that $\text{tr}\{\hat{\mathcal{F}}_{\alpha \beta}\}$ is invariant with respect to the choice of the reduced Hilbert space as long as the inner window is chosen above $\mathcal{E}_{\text{F}}$. Note that we have shown that $\hat{\mathcal{F}}_{\alpha \beta}$ is covariant under the transformation between the Wannier and Hamiltonian gauge in the previous section (Appendix.~\ref{appendix:gauge_selection}). Similarly, $\text{tr}\{\hat{\mathcal{K}}_{\alpha \beta}\}$ and $\text{tr}\{\hat{\mathcal{G}}_{\alpha \beta}\}$ are also invariant under the choice of the reduced Hilbert space. Therefore, Eq.~\eqref{eq:modern3} is independent of space selection.

Now, we turn to Eq.~\eqref{eq:modern_operator_covariant_derivative}, which contains contributions from both the occupied and unoccupied states. Since even the energy spectra of $\hat{H}$ and $\mathbb{H}$ cannot be guaranteed to be identical outside the inner window, we consider only the unoccupied states within the inner window. For these unoccupied states, Eq.~\eqref{eq:space_PFP} is replaced by
\begin{align}\label{eq:space_PFP_2}
    [\hat{Q}^{\text{in}}(\widetilde{\mathbb{F}}_{\alpha \beta}^{\text{H}_{2}} - G^{\dagger}\widetilde{\mathbb{F}}_{\alpha \beta}^{\text{H}_{1}}G)\hat{Q}^{\text{in}}]_{nm} = g_{n}\bra{\partial_{\alpha}u_{n}^{\text{H}_{2}}}& \notag
    \\[2mm]
    \cdot (\Delta \hat{\mathbb{P}}^{1} - \Delta \hat{\mathbb{P}}^{2})\ket{\partial_{\beta}u_{m}^{\text{H}_{2}}}g_{m}&,
\end{align}
and Eq.~\eqref{eq:space_AQA_rule} is replaced by
\begin{align}\label{eq:space_AQA_rule_2}
    [\hat{Q}^{\text{in}}(\mathbb{A}^{\text{H}_{2}}_{\alpha}\hat{P}^{\text{H}_{2}}\mathbb{A}^{\text{H}_{2}}_{\beta} - G^{\dagger}\mathbb{A}^{\text{H}_{1}}_{\alpha}\hat{P}^{\text{H}_{1}}\mathbb{A}^{\text{H}_{1}}_{\beta}G)\hat{Q}^{\text{in}}]_{nm} = 0.
\end{align}
Thus, an additional term proportional to $\Delta \hat{\mathbb{P}}^{1} - \Delta \hat{\mathbb{P}}^{2}$ emerges, indicating that the unoccupied-state part of Eq.~\eqref{eq:modern_operator_covariant_derivative} depends on space selection, and this dependence can be mitigated by choosing the inner window. Therefore, to obtain reasonable results from the unoccupied-state contribution of Eq.~\eqref{eq:modern_operator_covariant_derivative}, the inner window should be chosen sufficiently high.
 
\begin{figure*}[t!]
\includegraphics[width=500pt]{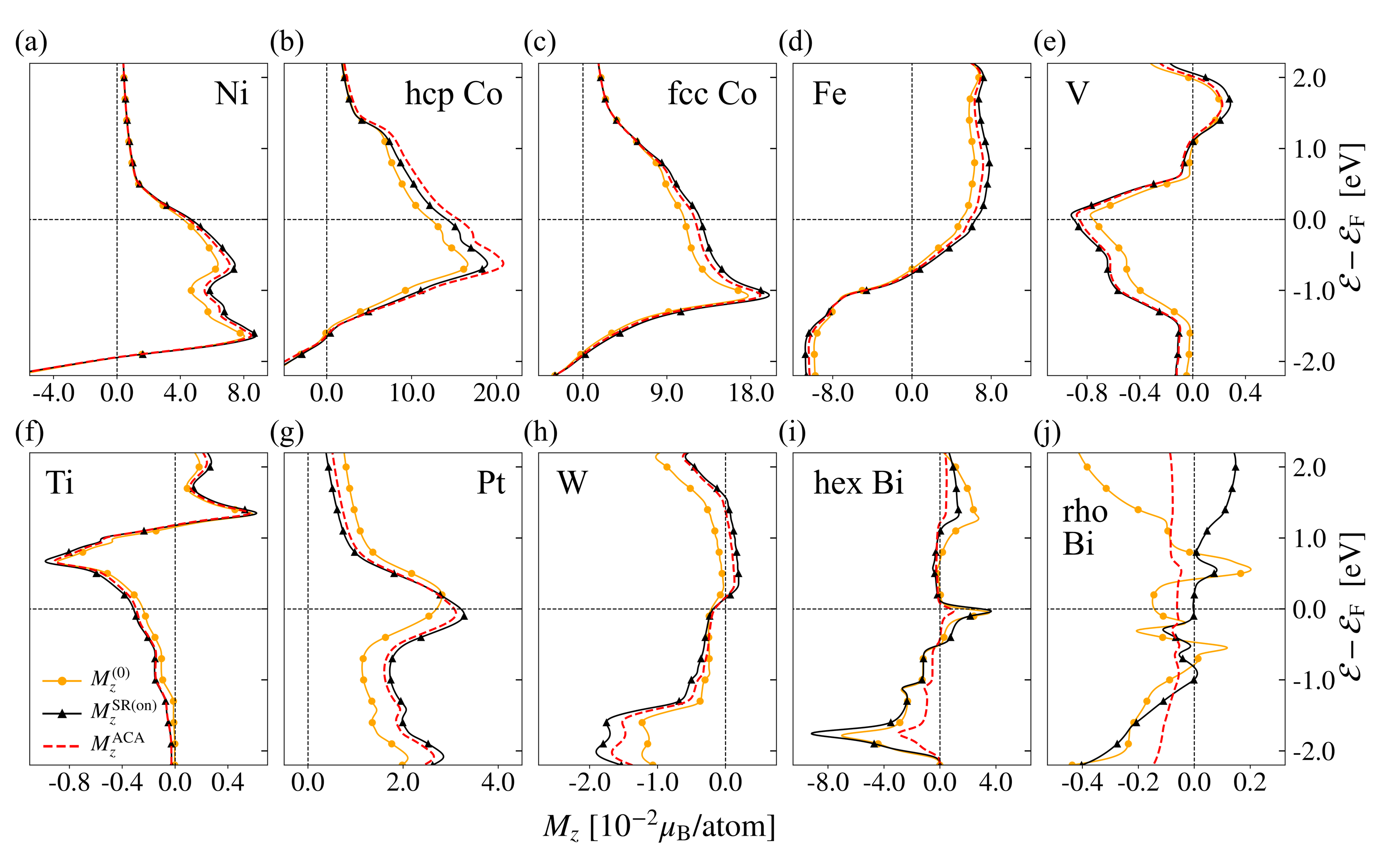}
\caption{\label{figA1}\textbf{$M_{\gamma}^{\text{SR(on)}}$ vs ACA.} (a) Ni with spin-quantization axis [111], (b) hcp Co [0001], (c) fcc Co [111], (d) bcc Fe [001], (e) bcc V [001], (f) fcc Ti [111], (g) fcc Pt [111], (h) bcc W [001], (i) hex Bi [001], and (j) rho Bi [001]. The results of $M_{z}^{\text{ACA}}$ (red dashed line), $M_{z}^{(0)}$ (orange line), and $M_{z}^{\text{SR(on)}}$ (black line) are shown. The results of Ni, Co, and Fe are obtained by including DFT+U.}
\end{figure*}

\section{Derivation of orbital moment operator}\label{appendix:derivation_operator}

The modern theory of orbital magnetization was originally derived in terms of a trace. Recently, several studies~\cite{Pezo22PRB,Pezo23PRB,Lee25PRB,Busch23PRR,Gobel24PRL,Gobel25CP,Ghosh25SR,Cysne26arXiv} have attempted to construct an operator associated with the modern theory of orbital magnetization and use it to analyze various quantities related to orbital magnetism and OAM. Introducing such an operator offers two important advantages. First, it allows one to analyze the contribution of each individual band to the orbital magnetization (i.e., band-resolved analysis). Second, by explicitly presenting the off-diagonal terms, it opens a way to investigating orbital phenomena in nonequilibrium systems.

The attempts in previous studies relied on the single-band covariant derivative and thus produced operators that are covariant under a $U(1)$ gauge. In this paper, we define an orbital moment operator [Eq.~\eqref{eq:modern_operator_covariant_derivative}] that corresponds to the gauge-invariant formula of orbital magnetization [Eq.~\eqref{eq:modern3}]. Our operator is applicable beyond a simple model Hamiltonian, extending to real materials with complex band structures, and, in particular, it enables calculations based on Wannierization. In this section, we derive the orbital moment operator Eq.~\eqref{eq:modern_operator_covariant_derivative}.

First, we consider the occupied block, $\hat{P}_{\mathbf{k}}\hat{M}_{\gamma}(\mathbf{k})\hat{P}_{\mathbf{k}}$. Its derivation has already been presented in Appendix~\ref{appendix:derivation_orbital_magnetization}. Note that one must multiply by $i$ instead of taking the imaginary part in Eq.~\eqref{eq:imaginary_covariant_FGK} to properly account for the off-diagonal components. Taking the imaginary part is valid only when the trace is taken. Next, we consider the unoccupied block, $\hat{Q}_{\mathbf{k}}\hat{M}_{\gamma}(\mathbf{k})\hat{Q}_{\mathbf{k}}$. We assume that the contributions from states with energies far above $\mathcal{E}_{\text{F}}$ to the physical quantity evaluated with the operator are negligible. The states outside the reduced Hilbert space are not included in our operator. The energy spectrum of states lying outside the inner window does not coincide with the real one and therefore the corresponding values are unreliable. Thus, the unoccupied space in which we expect our operator to function reasonably is the range between $\mathcal{E}_{\text{F}}$ and the inner window (i.e., $\bar{\mathcal{H}}^{\text{in}} - \bar{\mathcal{H}}^{\text{occ}}$, where $\bar{\mathcal{H}}^{\text{occ}}$ denotes the occupied space). To treat this region, we replace $\partial_{\alpha}\hat{P}$ in Eq.~\eqref{eq:building_blocks} with $\partial_{\alpha}\hat{Q}$. We also extend the covariant derivative to the states in $\bar{\mathcal{H}}^{\text{in}} - \bar{\mathcal{H}}^{\text{occ}}$. According to the discussions in Appendix~\ref{appendix:gauge-invariant-objects}, we practically define the \textit{occupation-weighted covariant derivative} as follows:
\begin{align}\label{eq:deriv_cov_deriv}
    \ket{\bar{D}_{\alpha}u_{n\mathbf{k}}} &=
    \begin{cases}
        \hat{Q}_{\mathbf{k}} \ket{\partial_{\alpha}u_{n\mathbf{k}}}, \quad \text{for} \; \bar{\mathcal{E}}_{n\mathbf{k}} \leq \mathcal{E}_{\text{F}}
        \\[2mm]
        (\hat{\mathbb{Q}}_{\mathbf{k}} + \hat{P}_{\mathbf{k}}) \ket{\partial_{\alpha}u_{n\mathbf{k}}}. \quad \text{for} \; \bar{\mathcal{E}}_{n\mathbf{k}} > \mathcal{E}_{\text{F}}
    \end{cases}
\end{align}
In Hamiltonian gauge, it is expressed by Eq.~\eqref{eq:occupation-weighted}. Finally, we consider the mixing block between the occupied states and the unoccupied states. It is obtained by evaluating Eq.~\eqref{eq:building_blocks} with $\partial_{\alpha} \hat{P}$ for the occupied states and with $\partial_{\alpha} \hat{Q}^{\text{I}}$ for the unoccupied states, using in each case the covariant derivative defined in Eq.~\eqref{eq:deriv_cov_deriv}. Combining all contributions yields Eq.~\eqref{eq:modern_operator_covariant_derivative}. The gauge and space dependence of the operator is discussed in Appendices~\ref{appendix:gauge_selection} and \ref{appendix:space_selection}. The comparison with the other formulations, the validity and the limitation of this operator are discussed in Appendix~\ref{appendix:operator_discussion}.

\section{Discussion of orbital moment operator}\label{appendix:operator_discussion}

The orbital moment operator [Eq.~\eqref{eq:modern_operator_covariant_derivative}] is obtained by converting the gauge-invariant formula for the orbital magnetization, originally introduced in Ref.~\cite{Ceresoli06PRB} and extended in Ref.~\cite{Lopez12PRB} in a form suitable for Wannierization, to an operator form while preserving gauge covariance. This construction enables a band-resolved analysis for the states within the inner window of the reduced Hilbert space. On the other hand, Ref.~\cite{Pezo22PRB} constructed an equilibrium OAM operator in solids by evaluating $\hat{\mathbf{L}} \propto \text{Sym}[\hat{\mathbf{r}} \times \hat{\mathbf{v}}] = (\hat{\mathbf{r}} \times \hat{\mathbf{v}} - \hat{\mathbf{v}} \times \hat{\mathbf{r}})/2$ in the cell-periodic part of the Bloch eigenstates as follows:
\begin{align}\label{eq:OAM_operator_Pezo}
     \mathbf{L}_{nm}&= \frac{e\hbar^{2}}{4g_{L}\mu_{B}}\text{Im}\sum_{l \neq n,m} \left( \frac{1}{\mathcal{E}_{ln}} + \frac{1}{\mathcal{E}_{lm}}\right) (\mathbf{v}_{nl} \times \mathbf{v}_{lm}),
\end{align}
where $\mathbf{L}_{nm} = \braket{u_{n\mathbf{k}}|\hat{\mathbf{L}}|u_{m\mathbf{k}}}$, $g_{L}$ is the orbital $g$ factor, $\mu_{B} = e\hbar / 2m_{e}$ is the Bohr magneton, $\mathbf{v}_{pq} = \braket{u_{p\mathbf{k}}|\mathbf{v}|u_{q\mathbf{k}}}$, and $\mathcal{E}_{pq} = \mathcal{E}_{p\mathbf{k}} - \mathcal{E}_{q\mathbf{k}}$. However, since off-diagonal elements are generally complex, this expression gives correct values only for diagonal elements. As pointed out in Ref.~\cite{Busch23PRR}, this should be modified to the following form
\begin{align}\label{eq:OAM_operator_Busch}
    \mathbf{L}_{nm} &= -\frac{ie\hbar^{2}}{4g_{L}\mu_{B}}\sum_{l \neq n,m} \left( \frac{1}{\mathcal{E}_{ln}} + \frac{1}{\mathcal{E}_{lm}}\right) (\mathbf{v}_{nl} \times \mathbf{v}_{lm}).
\end{align}
This corrected expression of OAM operator has been widely used to evaluate the OAM of electrons~\cite{Busch23PRR,Gobel24PRL,Cysne25NPJS,Lee24PRB,Lee25PRB,Ghosh25SR} and magnons~\cite{Go24NanoLett,Lee26NanoLett}.

Equation~\eqref{eq:OAM_operator_Busch} yields the correct values for the diagonal matrix elements, but it misses certain contributions in the off-diagonal terms. Since the position operator does not commute with the momentum operator, it inherently contains a $\mathbf{k}$-nonlocal contribution, which gives rise to group-velocity-related terms. The importance of this contribution was first pointed out by Cysne \textit{et. al.}~\cite{Cysne26arXiv}. They proposed the following expression, which includes the missing contribution:
\begin{align}\label{eq:OAM_operator_Cysne}
    \mathbf{L}_{nm} &= -\frac{ie\hbar^{2}}{4g_{L}\mu_{B}} \left( \sum_{l \neq n} \frac{(\mathbf{v}_{nl} \times \mathbf{v}_{lm})}{\mathcal{E}_{ln}} + \sum_{l \neq m}\frac{(\mathbf{v}_{nl} \times \mathbf{v}_{lm})}{\mathcal{E}_{lm}} \right) .
\end{align}
The $\mathbf{k}$-nonlocal contribution can be encoded in the diagonal part of the velocity matrix elements, which is precisely what give rise to the difference between Eqs.~\eqref{eq:OAM_operator_Busch} and \eqref{eq:OAM_operator_Cysne}. To make this more transparent, we revisit the representations of the position and velocity operators in solids. In a crystalline solid, the position operator cannot be represented by a $\mathbf{k}$-local matrix. For an infinite periodic lattice, the Bloch representation of the position operator is expressed by
\begin{align}\label{eq:position_operator_CMR}
    \braket{\psi_{n\mathbf{k}}|\hat{r}_{\alpha}|\psi_{m\mathbf{k^{\prime}}}} &= -i\partial_{k^{\prime}_{\alpha}}\delta (\mathbf{k} - \mathbf{k}^{\prime}) \delta_{nm} + A_{\alpha ,nm}(\mathbf{k}) \delta(\mathbf{k} - \mathbf{k}^{\prime}),
\end{align}
where $\ket{\psi_{n\mathbf{k}}} = e^{i\mathbf{k} \cdot \hat{\mathbf{r}}}\ket{u_{n\mathbf{k}}}$ is the Bloch state and $A_{\alpha,nm}(\mathbf{k}) = i\braket{u_{n\mathbf{k}}|\partial_{\alpha}|u_{m\mathbf{k}}}$ is the non-Abelian Berry connection. We note that the first term of the right hand side reflects the $\mathbf{k}$-nonlocal nature of $\hat{\mathbf{r}}$, that is, it cannot be expressed solely be a well-defined matrix at a point $\mathbf{k}$. In a finite sample, this nonlocality generates $\mathbf{k}$-off-diagonal matrix elements~\cite{Si25EPL}. Thus, the matrix representation of the velocity operator, $\hat{\mathbf{v}} = [\hat{\mathbf{r}},\hat{H}]/i\hbar$, contains the group velocity, $\partial_a \mathcal{E}_n$, as follows:
\begin{align}\label{eq:building_block_velocity}
    v_{\alpha,nm} &=\frac{1}{\hbar} \left( \partial_{\alpha}\mathcal{E}_{n}\delta_{nm} - i\mathcal{E}_{mn}A_{\alpha ,nm}\right).
\end{align}
The expression above applies to a nondegenerate case. More generally, it can be written as
\begin{align}\label{eq:building_block_velocity_general}
    v_{\alpha,nm} &=\frac{1}{\hbar} \left( \partial_{\alpha}H_{nm} - i[\hat{A}_{\alpha},\hat{H}]_{nm}\right).
\end{align}
Meanwhile, the matrix element of the position operator between two cell-periodic Bloch eigenstates is written as follows~\cite{Cysne26arXiv}:
\begin{align}\label{eq:building_block_position}
    \braket{n|\hat{r}_{\alpha}|m} &= \braket{n|iD_{\alpha}|m} = i\braket{n|\partial_{\alpha}|m} - \delta_{nm}A_{\alpha ,m}.
\end{align}

Now, we use these building blocks [Eqs.~\eqref{eq:building_block_velocity} and \eqref{eq:building_block_position}]to evaluate the representation of $\text{Sym}[\hat{\mathbf{r}} \times \hat{\mathbf{v}}]$ between two nondegenerate cell-periodic Bloch states
\begin{widetext}
\begin{align}\label{eq:OAM_operator_Cysne_derivation_nondegenerate}
    \braket{u_{n\mathbf{k}}|\text{Sym}[\hat{\mathbf{r}} \times \hat{\mathbf{v}}]_{\gamma}|u_{m\mathbf{k}}}&= \frac{\varepsilon_{\alpha \beta \gamma}}{2}(\hat{r}_{\alpha ,nl}\hat{v}_{\beta ,lm} + \hat{v}_{\beta ,nl}\hat{r}_{\alpha ,lm}) \notag
    \\[2mm]
    &=\frac{\varepsilon_{\alpha \beta \gamma}}{2\hbar} \left[ (A_{\alpha ,nl} - \delta_{nl}A_{\alpha ,n})(\partial_{\beta}\mathcal{E}_{m}\delta_{lm} - i\mathcal{E}_{ml}A_{\beta ,lm}) + (\partial_{\beta}\mathcal{E}_{n}\delta_{nl} - i\mathcal{E}_{ln}A_{\beta ,nl})(A_{\alpha ,lm} - \delta_{lm}A_{\alpha ,m})\right] \notag
    \\[2mm]
    &= -\frac{i}{2\hbar} \left[ \sum_{l \neq n, m} \left(\mathcal{E}_{ml}A_{\alpha ,nl}A_{\beta ,lm} + \mathcal{E}_{nl}A_{\alpha ,nl}A_{\beta ,lm}\right) - (\alpha \leftrightarrow \beta) \right] \notag
    \\[2mm]
    &\quad + \frac{1}{2\hbar} \left[ A_{\alpha ,nm}\partial_{\beta}(\mathcal{E}_{n} + \mathcal{E}_{m})(1 - \delta_{nm}) - (\alpha \leftrightarrow \beta)\right].
\end{align}
\end{widetext}
Eq.~\eqref{eq:OAM_operator_Cysne_derivation_nondegenerate} makes explicit the separation between the contribution originating from the $\mathbf{k}$-nonlocal nature and the remaining contributions. Then, Eq.~\eqref{eq:OAM_operator_Cysne} is obtained by replacing the Berry connections with the corresponding velocity matrix elements [Eq.~\eqref{eq:connection-naive}] and multiplying by the appropriate factors.
The first term of Eq.~\eqref{eq:OAM_operator_Cysne_derivation_nondegenerate} yields Eq.~\eqref{eq:OAM_operator_Busch}, while the second term is the ``$g^{\text{II}}$ contribution'' introduced in Ref.~\cite{Cysne26arXiv}.

Although Eq.~\eqref{eq:OAM_operator_Cysne} was derived in Ref.~\cite{Cysne26arXiv} only for the nondegenerate cases, it can be generalized to the degenerate cases by using Eqs.~\eqref{eq:building_block_position} and \eqref{eq:building_block_velocity_general}:
\begin{align}\label{eq:building_blocks_degenerate}
    A_{\alpha ,nl} - \delta_{nl}A_{\alpha ,n} &\rightarrow \hat{Q}_{\mathcal{D}_{n}}A_{\alpha ,nl}\notag
    \\[2mm]
    \partial_{\alpha}\mathcal{E}_{n} \delta_{nl} - i\mathcal{E}_{ln}A_{\alpha ,nl} &\rightarrow \partial_{\alpha}H_{nl} - i[\hat{A}_{\alpha},\hat{H}]_{nl}.
\end{align}
It should be noted, however, that it provides reasonable values only in a smooth Hamiltonian gauge. In particular, it must be evaluated with great care in the vicinity of band-crossing points. To recover the information lost during the restriction to a reduced Hilbert space, we use Eq.~\eqref{eq:Berry_connection} to evaluate $A_{\alpha}$, and $\mathbb{B}$ to evaluate the velocity as follows:
\begin{align}\label{eq:building_block_velocity_general_B}
    \hat{\mathbb{P}}v_{\alpha ,nm}\hat{\mathbb{P}} &= \frac{1}{\hbar} \left\{ \hat{\mathbb{P}}\partial_{\alpha} \mathbb{H}_{nm}\hat{\mathbb{P}} + i(\mathbb{B}_{\alpha ,nm} - \mathbb{B}^{\dagger}_{\alpha ,nm})\right\}\notag
    \\[2mm]
    &= \frac{1}{\hbar} \left\{ \partial_{\alpha} \mathbb{H}_{nm} + i(\widetilde{\mathbb{B}}_{\alpha ,nm} - \widetilde{\mathbb{B}}^{\dagger}_{\alpha ,nm} - [\mathbb{A}_{\alpha},\mathbb{H}]_{nm})\right\} ,
\end{align}
and
\begin{align}
    \hat{\mathbb{Q}}v_{\alpha ,nm}\hat{\mathbb{P}} &= -\frac{i}{\hbar} \widetilde{\mathbb{B}}^{\dagger}_{\alpha, nm} = -\frac{i}{\hbar} (\mathbb{B}^{\dagger}_{\alpha, nm} - \mathbb{A}_{\alpha ,nm}\mathbb{H}_{nm}).
\end{align}

On the other hand, if the same formalism is evaluated using the occupation-weighted covariant derivative, the corresponding OAM operator can be obtained by expanding the terms such as
\begin{align}
    \hat{\mathbb{P}}\hat{r}_{\alpha}\hat{v}_{\beta}\hat{\mathbb{P}} &= \{ \hat{P}\hat{r}_{\alpha}(\hat{Q}^{\text{I}} + \hat{\mathbb{Q}}) + \hat{Q}^{\text{I}}\hat{r}_{\alpha}(\hat{P} + \hat{\mathbb{Q}}) \} \hat{v}_{\beta}(\hat{P} + \hat{Q}^{\text{I}})
\end{align}
The derivation is straightforward. As discussed in Appendix~\ref{appendix:gauge-invariant-objects}, introducing the occupation-weighted covariant derivative greatly improves numerical stability, but it may miss interband contributions within the chosen active space. By contrast, evaluating Eq.~\eqref{eq:OAM_operator_Cysne_derivation_nondegenerate} using the covariant derivative for an $N$-fold degeneracy [Eq.~\eqref{eq:building_blocks_degenerate}] does not suffer from this omission. However, it is less robust for Wannierization and numerically less stable, and it requires a carefully chosen smooth Hamiltonian gauge.

We now compare these OAM operators [Eq.~\eqref{eq:OAM_operator_Cysne}] with the orbital moment operator in Eq.~\eqref{eq:modern_operator_covariant_derivative}. Taking the trace over the occupied states, the SR contribution obtained from the OAM operator coincides with that from the orbital moment operator. However, for off-diagonal matrix elements the two operators generally give different results. This discrepancy stems from the $\partial_{\alpha} H_{nm}$ and the covariant treatment of the velocity operator. The orbital moment operator is constructed so as to enable a band-resolution of the orbital magnetization, and thus it can reproduce the full orbital magnetization exactly. However, when one evaluates $\mathbf{r} \times \mathbf{v}$ directly by Bloch states, this operator misses certain contributions. By contrast, the OAM operator is an operator representation of $\mathbf{r} \times \mathbf{v}$ in terms of cell-periodic Bloch eigenstates and thus reproduces $\mathbf{r} \times \mathbf{v}$ accurately. However, it cannot capture the CM contribution and does not, by itself, fully reproduce the orbital magnetization. Moreover, within this approach there is, to date, no established way to further obtain the LC and IC contributions. Finally, we note that, unlike the spin operator, the OAM or orbital moment operator is not independent of the velocity operator, which implies that special care is required when using it to treat the orbital current~\cite{Culcer25PRL}.

\begin{figure*}[t!]
\includegraphics[width=500pt]{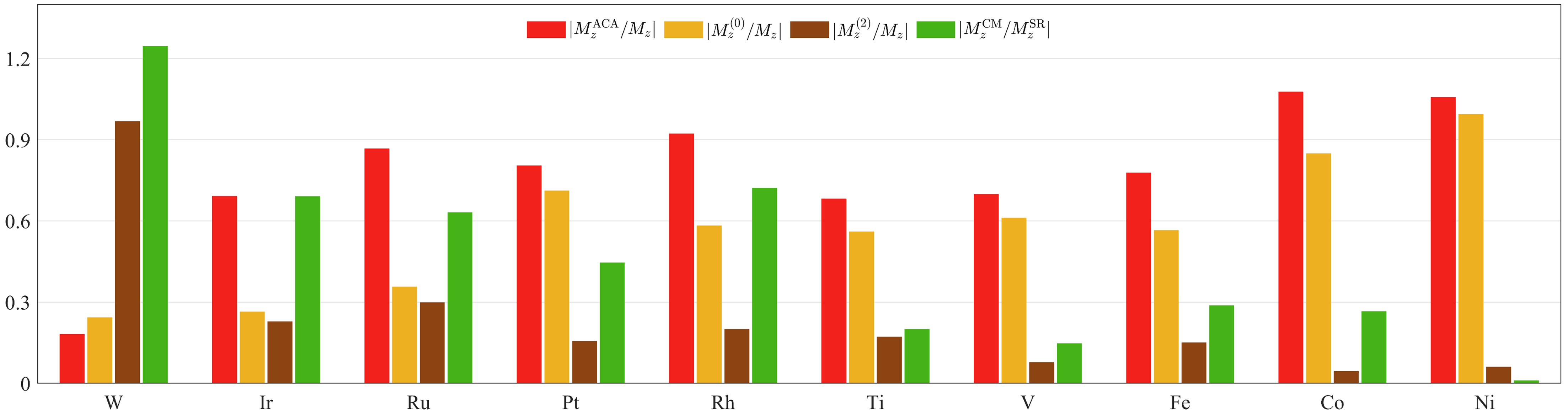}
\caption{\label{figA2}\textbf{Modern theory and ACA for orbital magnetization in various transition metals.} Ratios $\vert M_{z}^{\text{ACA}}$/$M_{z}\vert$ (red), $\vert M_{z}^{(0)}$/$M_{z}\vert$ (orange), $\vert M_{z}^{\text{CM}}$/$M_{z}^{\text{SR}}\vert$ (brown), and $\vert M_{z}^{(2)}$/$M_{z}\vert$ (green) for the selected $d$-transition metals are shown. The results of Ni, hcp Co, and Fe are obtained by including DFT+U.}
\end{figure*}

\section{Linear forms to occupation matrix}\label{appendix:linear_forms}

In this section, we show that Eq.~\eqref{eq:modern3} can be expressed as the product of $\hat{P}_{\mathbf{k}}$ and a matrix independent of the occupation matrices introduced in Eq.~\eqref{eq:orbital_moment_operator_total}. First, we show that the Berry curvature can be represented in a linear form to the occupation matrix. It suffices to show that
\begin{align}\label{eq:AF1}
    i\varepsilon_{\alpha \beta \gamma}\text{tr} \left[ \hat{P}A_{\alpha}\hat{P}A_{\beta}\hat{P} \right] &= -i\varepsilon_{\alpha \beta \gamma }\text{tr} \left[ \hat{P}A_{\beta} \hat{P} A_{\alpha}\hat{P} \right] =0.
\end{align}
Here, we use the cyclic property of the trace and the antisymmetric property of the permutation tensor. Note that it also implies that the anomalous Hall conductivity can also be expressed in the linear form to the occupied matrix. Next, we consider $i\varepsilon_{\alpha \beta \gamma}\text{tr} [\hat{\mathcal{G}}_{\alpha \beta} + \hat{\mathcal{K}}_{\alpha \beta}]$. To show that it can be represented in a linear form as the occupation matrix, it suffices to verify that
\begin{align}\label{eq:AF2}
    i\varepsilon_{\alpha \beta \gamma}\text{tr} \left[ \hat{P}A_{\alpha}\hat{P}\mathbb{H}\hat{P}A_{\beta}\hat{P} \right] &= -i\varepsilon_{\alpha \beta \gamma }\frac{1}{2}\text{tr} \left[ \hat{P} \{ \mathbb{H}, A_{\alpha} \hat{P} A_{\beta} \} \hat{P} \right]
\end{align}
and 
\begin{align}\label{eq:AF3}
    i\varepsilon_{\alpha \beta \gamma}\text{tr} \left[ \hat{P}A_{\alpha}\hat{P}\mathbb{B}_{\beta}\hat{P} \right] &= i\varepsilon_{\alpha \beta \gamma}\text{tr} \left[ \hat{P}A_{\alpha}\hat{P}\mathbb{H}\mathbb{A}_{\beta}\hat{P} \right] .
\end{align}
To show that Eq.~\eqref{eq:AF2} holds, we use the cyclic property of the trace and the antisymmetric property of the permutation tensor. Eq.~\eqref{eq:AF3} is obtained from the identity
\begin{align}\label{eq:AF4}
    f^{\text{H}}_{n}\mathbb{B}^{\text{H}}_{nn^{\prime}} &= if^{\text{H}}_{n}\braket{u^{\text{H}_{n}}|\hat{H}|\partial_{\beta}u_{n^{\prime}}^{\text{H}}} = \bar{\mathcal{E}}_{n}\mathbb{A}^{\text{H}}_{\beta, nn^{\prime}}.
\end{align}
Therefore, all the terms of Eq.~\eqref{eq:modern3} can be rewritten in a linear form to the occupation matrix in the trace operation. Although this proof is presented for the ground-state projector $P_{\mathbf{k}}$, it remains valid even if $P_{\mathbf{k}}$ is replaced by an arbitrary projection operator $P_{\mathcal{D}\mathbf{k}}$ constructed from a subset of the occupied states (that is, $\mathcal{D}$ is a subspace of the occupied space).

Finally, we discuss the \textit{band additivity} of orbital magnetization~\cite{Liu23arXiv}. Suppose that the active subspace $\mathcal{H}_{A}$ consists of nondegenerate energy eigenstates, and consider a physical quantity $\mathcal{O}$ that is constructed from derivatives of the states, such as the orbital magnetization or Berry curvature. Let $\mathcal{O}_{\mathcal{H}_{A}}$ be the quantity evaluated within $\mathcal{H}_{A}$ using the covariant derivative defined with $\hat{Q}_{\mathcal{H}_{A}}$, and let $\mathcal{O}_{n}$ be the corresponding quantity obtained using the single-band covariant derivatives $\hat{Q}_{n}$ for each band. If these two quantities satisfy
\begin{align}\label{band_additive}
    \mathcal{O} = \sum_{n \in \mathcal{H}_{A}}\mathcal{O}_{n},
\end{align}
then $\mathcal{O}$ is said to be \textit{band additive}. If $\mathcal{O}$ can be written in a linear form to occupation matrix (see Sec.~\ref{subsec:orbital_moment_operator}), then it is band additive. Therefore, the above proof, which shows that both the orbital magnetization and the Berry curvature can be expressed in linear form, implies that they are band additive. In contrast, $M_{\gamma}^{\text{SR}}$ and $M_{\gamma}^{\text{CM}}$ are neither expressible in linear form nor band additive,

\section{$M_{\gamma}^{\text{SR(on)}}$ vs ACA}\label{appendix:SR_onsite}

In Sec.~\ref{subsec:term-by-term_analysis}, we introduce $M_{\gamma}^{\text{SR(on)}}$ to extract the contribution corresponding to the ACA by calculating the interacell part of self-rotation of atomic orbital basis. In this section, we evaluate how much $M_{\gamma}^{\text{SR(on)}}$ contributes to $M_{\gamma}^{(0)}$ and its consistency with $M_{\gamma}^{\text{ACA}}$. Figure~\ref{figA1} compares $M_{\gamma}^{\text{SR(on)}}$, $M_{\gamma}^{(0)}$, and $M_{\gamma}^{\text{ACA}}$ results.

\begin{table*}[t!]
\begin{center}
\small
\begin{tabular}{ @{\; \;} c @{\; \;} | @{\; \;} c @{\; \;} | @{\; \;} c @{\; \;} | @{\; \;} c @{\; \;} | @{\; \;} c @{\; \;} | @{\; \;} c @{\; \;} | @{\; \;} c @{\; \;} }
\hline
\hline
\rule{0pt}{2.5ex}Materials & Lattice constant [$a_{0}$] & $R_{\text{MT}}$ [$a_{0}$] & $K_{\text{max}}$ [$a_{0}^{-1}$] & $l_{\text{max}}$ & DFT $\mathbf{k}$-mesh & Interpolation $\mathbf{k}$-mesh \\
\hline
\rule{0pt}{2.2ex}Fe (bcc) & $a = 5.42$~\cite{Go24PRB} & 2.29 & 4.5 & 12 & 16 $\times$ 16 $\times$ 16 & 150 $\times$ 150 $\times$ 150 \\[1mm]
\rule{0pt}{2.2ex}Co (hcp) & $a = 4.74$, $c = 7.69$~\cite{Go24PRB} & 2.30 & 4.5 & 12 & 16 $\times$ 16 $\times$ 16 & 150 $\times$ 150 $\times$ 150 \\[1mm]
\rule{0pt}{2.2ex}Co (fcc) & $a = 6.67$ \cite{Schoen17PRB} & 2.30 & 5.0 & 12 & 16 $\times$ 16 $\times$ 16 & 150 $\times$ 150 $\times$ 150 \\[1mm]
\rule{0pt}{2.2ex}Ni (fcc) & $a = 6.59$~\cite{Kirkin15NPJCM} & 2.30 & 4.5 & 12 & 16 $\times$ 16 $\times$ 16 & 150 $\times$ 150 $\times$ 150 \\[1mm]
\rule{0pt}{2.2ex}Ti (fcc) & $a = 7.75$~\cite{Choi23NAT} & 2.60 & 4.5 & 12 & 16 $\times$ 16 $\times$ 16 & 150 $\times$ 150 $\times$ 150 \\[1mm]
\rule{0pt}{2.2ex}Pt (fcc) & $a = 7.49$~\cite{Kirkin15NPJCM} & 2.56 & 4.5 & 12 & 16 $\times$ 16 $\times$ 16 & 150 $\times$ 150 $\times$ 150 \\[1mm]
\rule{0pt}{2.2ex}Ir (fcc) & $a = 7.31$~\cite{Kirkin15NPJCM} & 2.50 & 5.0 & 12 & 16 $\times$ 16 $\times$ 16 & 150 $\times$ 150 $\times$ 150 \\[1mm]
\rule{0pt}{2.2ex}Ag (fcc) & $a = 7.77$~\cite{Kirkin15NPJCM} & 2.55 & 5.0 & 12 & 16 $\times$ 16 $\times$ 16 & 150 $\times$ 150 $\times$ 150 \\[1mm]
\rule{0pt}{2.2ex}Au (fcc) & $a = 7.82$~\cite{Kirkin15NPJCM} & 2.65 & 5.0 & 12 & 16 $\times$ 16 $\times$ 16 & 150 $\times$ 150 $\times$ 150 \\[1mm]
\rule{0pt}{2.2ex}Rh (fcc) & $a = 7.23$~\cite{Kirkin15NPJCM} & 2.47 & 5.0 & 12 & 16 $\times$ 16 $\times$ 16 & 150 $\times$ 150 $\times$ 150 \\[1mm]
\rule{0pt}{2.2ex}Al (fcc) & $a = 7.63$~\cite{Kirkin15NPJCM} & 2.40 & 5.0 & 12 & 16 $\times$ 16 $\times$ 16 & 150 $\times$ 150 $\times$ 150 \\[1mm]
\rule{0pt}{2.2ex}V (bcc) & $a = 5.73$~\cite{Go24PRB} & 2.42 & 4.5 & 12 & 16 $\times$ 16 $\times$ 16 & 150 $\times$ 150 $\times$ 150 \\[1mm]
\rule{0pt}{2.2ex}W (bcc) & $a = 5.96$~\cite{Go24PRB} & 2.52 & 4.5 & 12 & 16 $\times$ 16 $\times$ 16 & 150 $\times$ 150 $\times$ 150 \\[1mm]
\rule{0pt}{2.2ex}Bi (hex) & $a = 8.14$~\cite{Fang18SCA} & 2.40 & 5.0 & 12 & 16 $\times$ 16 $\times$ 1 & 400 $\times$ 400 $\times$ 1 \\[1mm]
\rule{0pt}{2.2ex}Bi (rho) & $a = 9.07$, $\alpha = {57.423^{\circ}}$~\cite{Cucka62ACTA} & 2.40 & 5.0 & 12 & 16 $\times$ 16 $\times$ 16 & 150 $\times$ 150 $\times$ 150 \\[1mm]
\rule{0pt}{2.2ex}Ru (hcp) & $a = 5.15$, $c=8.13$~\cite{Go24PRB} & 2.46 & 4.5 & 12 & 16 $\times$ 16 $\times$ 16 & 150 $\times$ 150 $\times$ 150 \\[1.5mm]
\rule{0pt}{2.2ex}MoS$_{2}$ (1H) & $a = 6.03$~\cite{He16NRL} & \makecell[c]{Mo: 2.43\\[1mm] S: 1.90} &
5.0 & 12 & 16 $\times$ 16 $\times$ 1 & 400 $\times$ 400 $\times$ 1 \\[3.5mm]
\rule{0pt}{2.2ex}WTe$_{2}$ (T$_{d}$) & $a = 6.57$, $b=11.81$~\cite{Soluyanov15NAT} & W, Te: 2.44 & 5.0 & 12 & 16 $\times$ 16 $\times$ 1 & 400 $\times$ 400 $\times$ 1 \\[1mm]
\hline
\hline
\end{tabular}
\end{center}
\caption{Lattice structures and parameters used for the DFT calculations within the FLAPW method. For bcc and fcc structures, the lattice constant $a$ is presented. For hexagonal lattice, the in-plane lattice constant $a$ is presented. For hcp structures, the out-of-plane lattice constants $c$ is also presented. All lattice constants are in units of the Bohr radius $a_{0}$. $R_{\text{MT}}$ denotes the muffin-tin radius, $K_{\text{max}}$ is the plane-wave cutoff, and $l_{\text{max}}$ is the maximum number of the harmonic expansion in the muffin-tin. The DFT $\mathbf{k}$-mesh and the interpolation $\mathbf{k}$-mesh refer to the grids used for self-consistent field calculations and for evaluating the orbital magnetization [Eq.~\eqref{eq:modern3}] in the Wannier representation, respectively.}\label{table:2}
\end{table*}

For $d$-transition metals, $M_{\gamma}^{\text{SR(on)}}$ dominates $M_{\gamma}^{(0)}$ and closely agrees with $M_{\gamma}^{\text{ACA}}$ [Figs.~\ref{figA1}(a)-\ref{figA1}(h)]. Especially, the discrepancies between $M_{\gamma}^{\text{SR(on)}}$ and $M_{\gamma}^{\text{ACA}}$ at $\mathcal{E}_{\text{F}}$, defined as $\vert (M_{\gamma}^{\text{ACA}} - M_{\gamma}^{\text{SR(on)}})/M_{\gamma}^{\text{ACA}}\vert$, are 3\% in Ni, 5\% in V, and 7\% in Ti. Even in W, where the electronic wave function is relatively delocalized among the $d$-transition metals, the discrepancy is only 15\% and $M_{\gamma}^{\text{SR(on)}}$ reproduces $M_{\gamma}^{\text{ACA}}$ well. This result clearly demonstrates that the modern theory includes the ACA as a part of it, thereby clarifying the relationship between the two approaches.

In contrast, in $sp$ metals [Figs.~\ref{figA1}(i) and \ref{figA1}(j)] $M_{\gamma}^{\text{SR(on)}}$ fails to reproduce $M_{\gamma}^{\text{ACA}}$: the discrepancy is 278\% in hex Bi and 94\% in rho Bi. In particular, in rho Bi, the $\mathcal{E}$ dependence of $M_{\gamma}^{\text{SR(on)}}$ presents a completely different trend from both $M_{\gamma}^{\text{ACA}}$ and $M_{\gamma}^{(0)}$. This results indicate that in $sp$ metals, where the electrons are strongly delocalized, the contribution from the interstitial region to the self-rotation of electrons becomes significant, so that the ACA fails to properly capture even the magnetization arising from the self-rotation.

\section{Localization and orbital magnetism}\label{appendix:localization}

In this section, we examine the correlation between the localization of electronic wave functions and orbital magnetism by comparing the magnitudes of each term at $\mathcal{E}_{\text{F}}$. Figure~\ref{figA2} summarizes the magnitude of the ratio of each terms. First, we focus on the magnitude of the ratio of the CM to self-rotation contribution, $\vert M_{z}^{\text{CM}}/M_{z}^{\text{SR}} \vert$, shown as brown bars in Fig.~\ref{figA2}. In general, for materials with strongly localized electrons (e.g., $d$-ferromagnets, Ti, V), $M_{z}^{\text{SR}}$ dominates over $M_{z}^{\text{CM}}$. In particular, Ni exhibits $\vert M_{z}^{\text{CM}}/M_{z}^{\text{SR}} \vert$ = 1\%. As the degree of delocalization increases, the CM contribution tend to become more important. In the case of W, especially, $\vert M_{z}^{\text{CM}} \vert$ exceeds $\vert M_{z}^{\text{SR}} \vert$. In $sp$ metal rho Bi, the ratio reaches to 6.1\% (Table~\ref{table:1}). Although the overall trend of $\vert M_{z}^{\text{CM}}/M_{z}^{\text{SR}} \vert$ follows the tendency of electron localization, its exact value is determined by the detailed band structure of each material. Comparing $M_{z}^{\text{SR}}$ with $M_{z}$, we find that in $d$-transition ferromagnets, V, and Ti, $M_{z}^{\text{SR}}$ exhibits a similar $\mathcal{E}$ dependence to $M_{z}$, differing only by a slightly larger magnitude. In contrast, in Pt, Ir, and W, both the $\mathcal{E}$ dependence and the magnitude are substantially different from those of $M_{z}$. In particular, Pt shows a pronounced $M_{z}^{\text{SR}}$ peak at about 0.4 eV below $\mathcal{E}_{\text{F}}$ [Fig. \ref{fig6}(d)], while a similar peak appears at about 0.5 eV above $\mathcal{E}_{\text{F}}$ in Ir.

The ratios $\vert M_{z}^{(0)}/M_{z} \vert$ and $\vert M_{z}^{(2)}/M_{z} \vert$ are shown in Fig.~\ref{figA2} as orange and green bars, respectively. The value of $\vert M_{z}^{(0)}/M_{z} \vert$ tends to increase with stronger electron localization. Ni exhibits $\vert M_{z}^{(0)}/M_{z} \vert$ = 95\%, and the $\mathcal{E}$ dependence of $M_{z}^{(0)}$ closely follows that of $M_{z}$ [Fig.~\ref{fig4}(b)]. In contrast, W exhibits a value of 24\%, and even the sign of $M_{z}^{(0)}$ differs from that of $M_{z}$. Especially, the ratio decreases to 7\% in rho Bi. Conversely, the ratio $\vert M_{z}^{(2)}/M_{z} \vert$ tends to increase with electron delocalization. While Ni shows only 6\%, Ir and Ru exhibit values exceeding 20\%. In W, the ratio even reaches to 97\%. In the case of rho Bi, $\vert M_{z}^{(2)}/M_{z} \vert$ = 98\%.

The ratio $\vert M_{z}^{\text{ACA}}/M_{z} \vert$ is shown as red bars in Fig.~\ref{figA2}. In $d$-transition metals, the ACA reproduces the modern theory well. Except for W, the ACA captures more than about 70\% of the total orbital magnetization for the $d$-transition metals presented in Fig.~\ref{figA2}.

\section{Computational details}\label{appendix:computational_details}

The parameters for the materials used for the DFT calculation are summarized in Table~\ref{table:2}. For Fe, Ni, Co, and T$_{d}$-WTe$_{2}$ monolayer, we applied the LDA+$U$ method within the self-consistent DFT cycle~\cite{Shick99PRB}. The on-site Coulomb interactions $U$ and the intra-atomic exchange interactions $J$ are set to $U = 1.2$ eV and $J = 1.0$ eV for Fe~\cite{Sasioglu11PRB}, $U = 1.6$ eV and $J = 0.9$ eV for Co~\cite{Tung12PRB}, and $U = 1.9$ eV and $J = 1.2$ eV for Ni~\cite{Yang01PRL}. For the T$_{d}$-WTe$_{2}$ monolayer, we use $U = 5.5$ eV and $J = 0$ eV to obtain a band gap of $\approx$ 55 meV~\cite{Tang17NatPhys}.

To transform the Bloch states obtained from the DFT calculation into Wannier states, we construct 18 Wannier functions per atom for each $d$-block element and 8 for each $p$-block element. The number of Bloch states used for this construction is twice the number of resulting Wannier states. Atomic orbitals were used as the initial projections. The inner windows (frozen energy windows) are set from 5 to 12 eV above $\mathcal{E}_{\text{F}}$. The BZ integration is performed using a Monkhorst-Pack $\mathbf{k}$-mesh~\cite{Monkhorst76PRB} (see Table ~\ref{table:2}). The temperature is set to $T = 300$ K, and a constant level broadening $\eta = 0.0259$ eV was introduced to calculate $J^{\text{H}}_{\alpha}$ [Eq. \eqref{eq:J_Hamiltonian_gauge}].

\begin{verbatim}
\end{verbatim}

\nocite{*}

\bibliographystyle{apsrev4-2}
\bibliography{Paper}

\end{document}